\documentclass[12pt,preprint]{aastex}
\usepackage{graphicx}
\usepackage{fancyhdr}
\usepackage{multicol}
\usepackage{comment}
\usepackage{morefloats}
\usepackage{float}

\pdfoutput=1

\def\AaA{{\em Astr.~Astrophys.}}

\def\AIPC{{\em Amer.~Inst.~Phys.~Conf.}}
\def\AJ{{\em Astr.~J.}}
\def\AN{{\em Astron.~Nachr.}}
\def\ApJ{{\em Astrophys.~J.}}

\def\ApJS{{\em Astrophys.~J.~Suppl.}}
\def\APSS{{\em Astrophys.~Sp.~Sci.}}
\def\ARAA{{\em Ann.~Rev.~Astr.~Astrophys.}}
\def\ASPC{{\em ASP~Conf.~Series}}

\def\ExA{{\em Exp.~Astr.}}

\def\IJMP{{\em Intl.~Jour.~Mod.~Phys.}}
\def\MN{{\em Mon.~Not.~R.~astr.~Soc.}}
\def\Nat{{\em Nature}}
\def\NuPhS{{\em Nuc.~Phys.~B.~Proc.~Supp.}}
\def\PASJ{{\em Publ.~Astr.~Soc.~Japan}}
\def\PASP{{\em Publ.~Astr.~Soc.~Pacif.}}

\def\PhRvL{{\em Phys.~Rev.~Lett.}}
\def\PR{{\em Phys.~Reports}}

\def\RMdAA{{\em Rev.~Mex.~Astr.~Astrophys.}}

\def\SPIE{{\em Soc.~Pho.~Inst.~Eng.}}
\def\SSR{{\em Space Sci.~Rev.}}
\def\etal{{et al.\thinspace}}

\def\spose#1{\hbox to 0pt{#1\hss}}

\def\multleft#1{\hbox to size{\vbox {\halign {\lft{##}\cr #1}}\hfill}\par}
\def\multright#1{\hbox to size{\vbox {\halign {\rt{##}\cr #1}}\hfill}\par}

\def\degmark{^\circ}
\def\boxit#1{\vbox{\hrule\hbox{\vrule\kern3pt\vbox{\kern3pt
          #1 \kern3pt}\kern3pt\vrule}\hrule}}

\def\cm{{\rm\thinspace cm}}

\def\erg{{\rm\thinspace erg}}
\def\eV{{\rm\thinspace eV}}

\def\keV{{\rm\thinspace keV}}
\def\km{{\rm\thinspace km}}

\def\m{{\rm\thinspace m}}

\def\Msun{\hbox{$\rm\thinspace M_{\odot}$}}

\def\s{{\rm\thinspace s}}
\def\ks{{\rm\thinspace ks}}

\def\ergcmps{\hbox{$\erg\cm\s^{-1}\,$}}

\def\ergpcmsqps{\hbox{$\erg\cm^{-2}\s^{-1}\,$}}

\def\kmps{\hbox{$\km\s^{-1}\,$}}

\def\pcmsq{\hbox{$\cm^{-2}\,$}}

\newcommand{\squishlist}{
 \begin{list}{$\bullet$}
  { \setlength{\itemsep}{0pt}
     \setlength{\parsep}{3pt}
     \setlength{\topsep}{3pt}
     \setlength{\partopsep}{0pt}
     \setlength{\leftmargin}{1.5em}
     \setlength{\labelwidth}{1em}
     \setlength{\labelsep}{0.5em} } }

\newcommand{\squishend}{
  \end{list}  }

\begin{document}
\title{Measuring Supermassive Black Hole Spins in Active Galactic Nuclei}
\author{Laura Brenneman\altaffilmark{1} \\
{\it SpringerBrief DOI:} 10.1007/978-1-4614-7771-6 \\
http://link.springer.com/book/10.1007/978-1-4614-7771-6/page/1}
\altaffiltext{1}{Harvard-Smithsonian Center for Astrophysics, 60
  Garden St., MS-67, Cambridge, MA  02138  USA,  lbrenneman@cfa.harvard.edu}

\begin{abstract}
Measuring the spins of supermassive black holes (SMBHs) in active galactic
nuclei (AGN) can inform us about the relative role of gas accretion vs. mergers
in recent epochs of the life of the host galaxy and its AGN.  Recent
advances in theory and observation have enabled spin measurements for
a handful of SMBHs thus far, but this science is still very much in its infancy.
Herein, I discuss how and why we seek to measure black hole spin in AGN, using recent
results from long X-ray observing
campaigns on three radio-quiet AGN (MCG--6-30-15, NGC~3783 and
Fairall~9) to illustrate this
process and its caveats.  I then present our current knowledge of the
distribution of SMBH spins in the
local universe.  I also address prospects for improving the accuracy, precision and
quantity of these spin constraints in the next decade and beyond with
instruments such as {\it NuSTAR, Astro-H} and a future generation large-area
X-ray telescope.
\end{abstract}


\section{Introduction}
\label{sec:intro}

Black holes represent the ultimate frontier in
astrophysics: the one-way passage to the unknown and the unknowable.
These exotic objects
are defined by a characteristic radius known as the event horizon: the
radius from the central collapsed remnant (or singularity) at which the escape
velocity of the black hole equals the speed of light.  Black holes
therefore emit no light themselves, and we can only observe
them indirectly by analyzing the electromagnetic (e/m)
radiation released from the gas they accrete.  This accretion
typically takes the 
form of a geometrically-thin, optically-thick disk (Shakura \& Sunyaev 1973) for
black holes which are actively accreting gas ($L_{\rm X}/L_{\rm Edd} \geq 0.001$, Miller 2007).
The finite value of
the speed of light renders all material and spacetime within the event
horizon causally separated from the Universe in which we live; at
present, there is no known way to access information from beyond the
event horizon.\footnote{Quantum mechanical ``tunneling''
  theoretically enables black
holes to emit thermal radiation at a very slow rate.
This is
known as ``Hawking Radiation,'' (Hawking 1974) and can eventually evaporate a black
hole.  However, a black hole with solar mass would take $\sim10^{66}$
years to evaporate via this process.  A supermassive black hole
would take considerably longer.}  Due to this limitation, all of our knowledge of black hole
systems comes from Einstein's Special and General theories of 
Relativity, and from e/m observations of accretion disks around known or suspected black hole
systems which are, almost invariably, bright and/or nearby.

In spite of their enigmatic nature, black holes are arguably the simplest
objects in the Universe, possessing only three fundamental properties
by which they can be completely defined: (1) mass, (2) spin, and (3)
electric charge.  In practice, the electric charge of a black hole in any
environment other than a pure vacuum is assumed to be negligible, as
the black hole would rapidly accrete oppositely charged particles and
neutralize itself.  Mass and spin ---or angular momentum--- are thus the
only two meaningful properties that describe an astrophysical black hole (Kerr 1963).
The mass of a black hole determines the degree to which the spacetime
in which it resides is warped (as in the classic ``bowling ball on a
trampoline'' analogy), whereas spin determines the degree to which
that spacetime is twisted (much like beaters in thick batter). 

The masses of stellar-mass black holes within our own galaxy are typically determined by examining the
orbital and radiative properties of their companion stars.  Measuring the
masses of SMBHs can be more difficult, however.  Several methods have been developed to
estimate the masses of SMBHs: e.g., reverberation mapping (Blandford \& McKee 1982), stellar
velocity dispersion (Ferrarese \& Merritt 2000), tracing of stellar orbits (Ghez
\etal 2000, Genzel \etal 2000), maser
observations (Watson \& Wallin 1994), and gravitational lensing (Silvestro 1974).  Most of these methods
rely on measuring radiation emitted relatively far from the black hole.
Observationally, black holes range in size from $\sim3 - 10^{10} \Msun$, with
most stellar-mass black holes clustered in the $5-20 \Msun$ range and most
SMBHs with masses of $10^6-10^8 \Msun$.  Over the last decade there
has been some
evidence to support the existence of intermediate mass black holes with masses
of order $10^2-10^4 \Msun$ (e.g., Miller \& Colbert 2004).

Though black hole mass is by no means trivial to measure, spin is the more
challenging property to ascertain.  In contrast to constraining mass, measuring spin
requires probing the nature of the spacetime
within a few
gravitational radii of the event horizon (where the gravitational radius is defined as
$r_{\rm g} \equiv GM/c^2$; $G$ is Newton's constant, $M$ is the mass
of the black hole and $c$ is the speed of light).  The angular momentum of a black hole only
manifests through Lense-Thirring precession, also known as
frame-dragging, which occurs only in the innermost part of the
accretion disk where relativistic effects cause the spacetime in this region to
become twisted in the same direction that the black hole is rotating.  To observe this
effect, observations of the innermost disk must be made in X-rays, given the
energetic processes at
work in the cores of actively-accreting black holes.  Current X-ray
telescopes lack the spatial resolution necessary to resolve the
innermost regions of the accretion disk, even in bright, nearby AGN.
As such, X-ray spectra taken from the core of the AGN are the tool of
choice for examining the spacetime of the inner disk.

Spin (in dimensionless form) is defined as $a \equiv cJ/GM^2$, where cosmic
censorship within the framework of General Relativity dictates that $-1 \leq a
\leq +1$ (negative spin values represent retrograde
configurations in which the black
hole spins in the opposite direction to the disk, positive values
denote prograde spin configurations, and $a=0$ implies a non-spinning black
hole), and $J$
represents the black hole angular momentum (Bardeen, Press \& Teukolsky 1972,
Thorne 1974).  If spin is known to within $\Delta a
\leq 10\%$, then meaningful correlations can be drawn
between spin and other environmental variables, e.g., the history of the
accretion flow and the presence and power of relativistic jets in the system.

Supermassive black holes are particularly interesting to examine,
given that their masses and spins have likely evolved considerably
in the billions of years since their formation.  SMBHs grow by either
merging with other black holes or accreting gas, most often by a combination
of the two processes (e.g., Volonteri \etal 2005).  Additionally, as a SMBH
grows, it can go through periods
where it produces powerful outflows of kinetic and radiative energy
through the production of winds and jets (Fabian 2012), seeding the surrounding
environment with matter and energy.  Such heating of the ambient
gas in and around the host galaxy may ultimately play a significant
role in regulating its rate of
star formation.  This type of
``feedback'' process has been cited as a potential explanation for the
famed $M-\sigma$ relation linking
the mass of the SMBH to the velocities of the stars in the central
bulge of its host
galaxy, as well as to the mass of the bulge itself (e.g., Ferrarese \&
Merritt 2000, G\"{u}ltekin \etal 2009).  Given that jets are thought to be
launched by the magnetic extraction of rotational energy from the ergosphere of the
black hole (Blandford \& Znajek 1977) when the black hole spin gets sufficiently
large ($a \geq 0.93$; Agol \& Krolik 2000), spin may play a siginficant role in
regulating galaxy growth on scales far beyond the gravitational sphere of
influence of the black hole. 

Put simply, measurements of the spins of SMBHs in AGN can contribute to our
understanding of these complex and energetic
environments in three principal ways:
\begin{itemize}
\item{They offer a rare probe of the nature of the spacetime proximal to the event
horizon of the black hole, well within the strong-field gravity regime
(Fabian \etal 1989, Laor 1991);}
\item{They can shed light on the relation of a black hole's angular momentum
to its outflow power in the form of jets (e.g., Narayan \& McClintock
2012, Steiner \etal 2012 for stellar-mass black holes);}  
\item{They can also inform us about the relative role of gas accretion vs. mergers in recent
epochs of the life of the host galaxy and its AGN (Berti \& Volonteri 2008).}
\end{itemize}
For these reasons,
developing a theoretical and observational framework in which to measure black
hole spin accurately and precisely is of critical importance to our understanding of how galaxies form
and evolve over cosmic time.  

Advances in theoretical
modeling as well as observational sensitivity in the {\it Chandra/XMM-Newton/Suzaku}
era are finally producing robust constraints on the spins of a handful of
SMBHs.  Computationally, new algorithms developed within the past decade
(Dov\v{c}iak \etal 2004, Beckwith \& Done 2005, Brenneman \& Reynolds 2006,
Dauser \etal 2010, 2013) have made it possible to perform fully relativistic
ray-tracing of photon paths emanating from the accretion disk close to the black hole,
keeping the black hole spin as a variable parameter in the model.  When such models are
fit to high signal-to-noise (S/N) X-ray spectra from the innermost accretion disk,
they yield vital physical information about the black hole/disk system, including
constraints on how fast ---and in what direction--- the black hole is rotating.      

In this work, I discuss our current
knowledge of the distribution of SMBH spins in the local universe and future
directions of black hole spin research.  I begin
in \S2 with an examination of the spectral modeling techniques used to measure
black hole spin, focusing on those most effective in constraining spin in AGN.
I then discuss the models involved, reviewing the caveats that must be
considered in the process
in \S3.  In \S4 I demonstrate the
application of these techniques to deep observations
of the nearby, type 1 AGN MCG--6-30-15, NGC~3783 and Fairall~9.  I examining our current knowledge of the spin
distribution of local SMBHs in \S5, along with its implications.
Future directions for this field of research are presented in \S6.

\section{Measuring Black Hole Spin}
\label{sec:methods}

\begin{figure}[hp]
\centerline{
\includegraphics[width=1.0\textwidth]{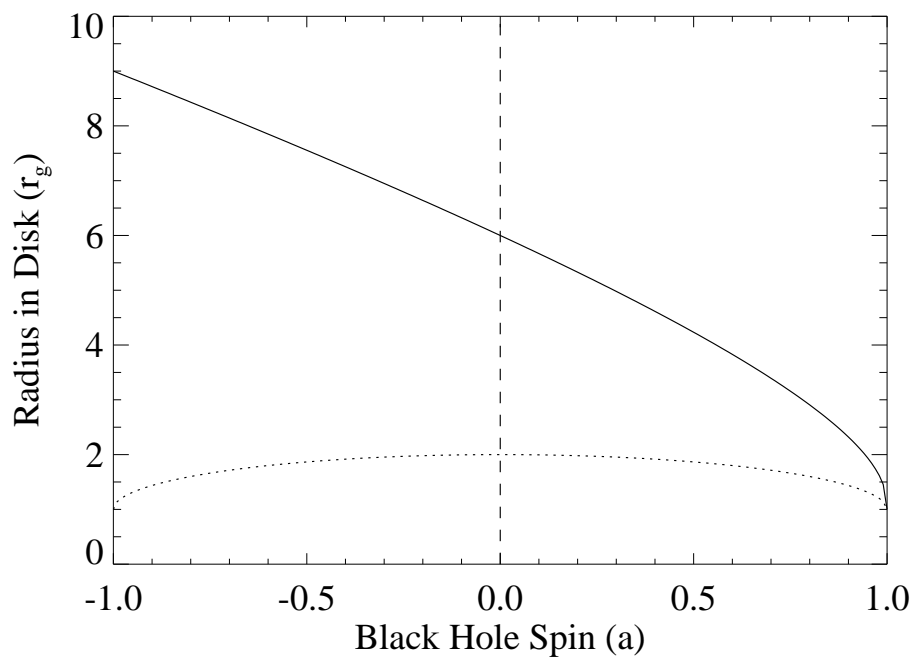}}
\vspace{-10cm}
\caption{\small{Radius of the ISCO (solid line) and event horizon (dotted line) as a
  function of black hole spin.  Spin values to the left of the dashed line indicate a black hole
spinning in the opposite direction relative to the accretion disk (retrograde), while spins
to the right of the dashed line indicate a black hole spinning in the
  same direction as the disk (prograde).}}
\label{fig:isco}
\end{figure}

In principle, there are at least five ways that spin can be measured for a
single (i.e., non-merging) SMBH.  All
five are predicated on the assumption that General Relativity provides the
correct description of the spacetime near the black hole, and that there is an
easily-characterized, monotonic relation between the radius of the innermost
stable circular orbit (ISCO) of the accretion disk and the black hole spin (see
Fig.~\ref{fig:isco}).  The disk is also assumed to remain geometrically thin,
optically thick and radiatively efficient down to the ISCO boundary, and to
truncate relatively rapidly therein (see \S\ref{sec:systematics}).  

\begin{figure}[hp]
\centerline{
\includegraphics[width=1.0\textwidth]{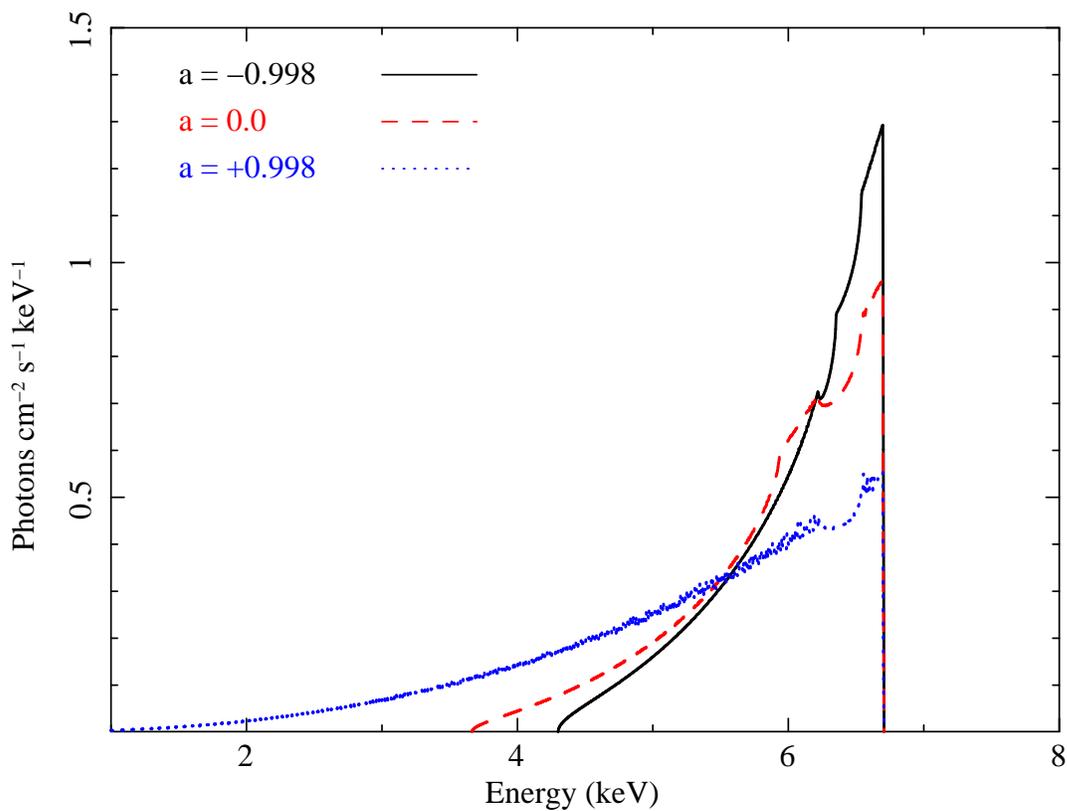}}
\caption{\small{Change in the shape of the Fe K$\alpha$ line as a function of black hole spin.
The black solid line represents $a=+0.998$, the red dashed line shows $a=0.0$ and the blue
dotted line shows $a=-0.998$.  Note the enhancement in the breadth of the red wing of the
line as the black hole spin increases.  The {\tt relline} code of Dauser \etal (2010)
was used to plot the three lines.}}
\label{fig:FeK}
\end{figure}

\begin{figure}[hp]
\centerline{
\includegraphics[width=1.0\textwidth]{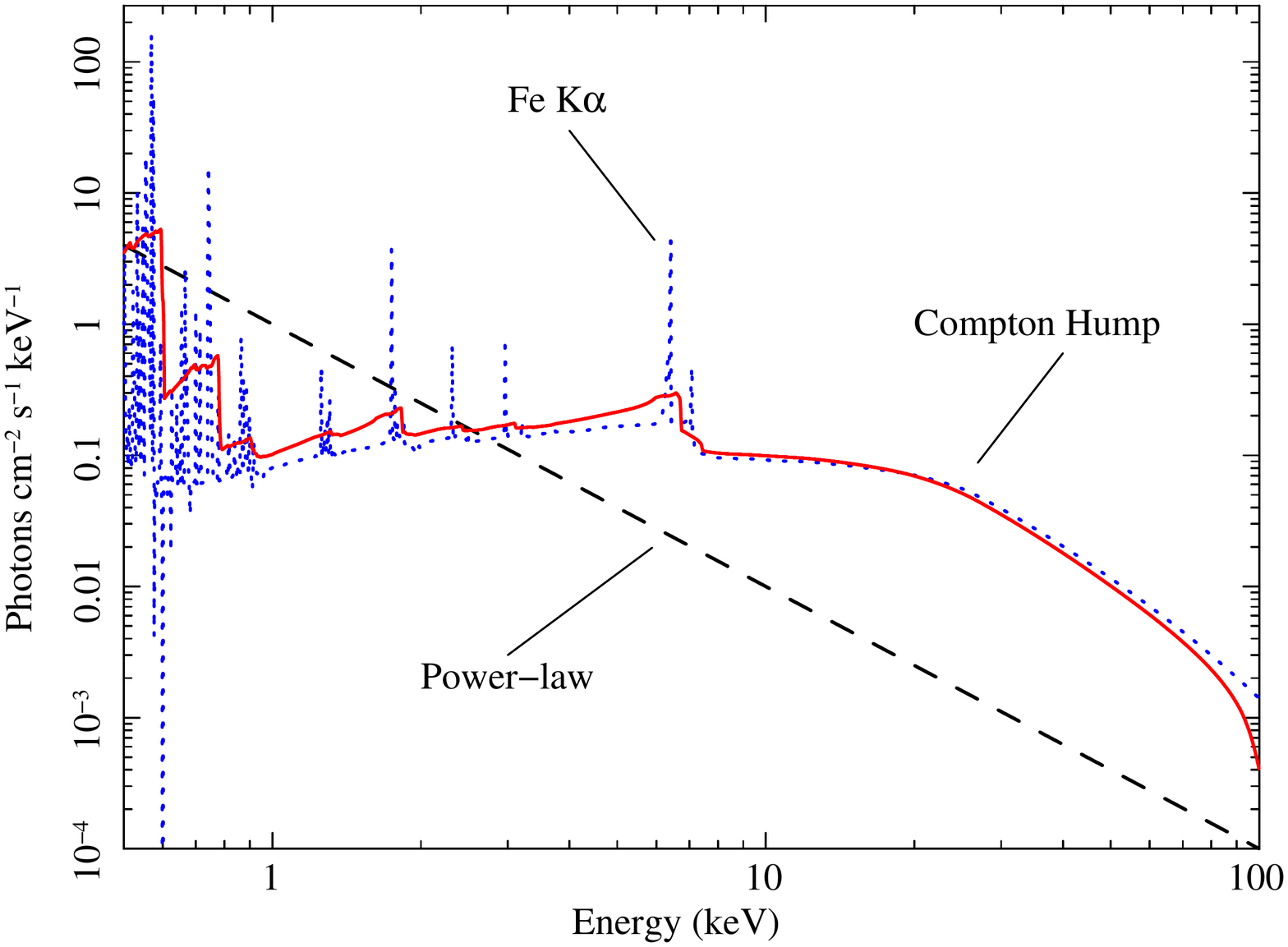}}
\caption{\small{The reflection spectrum for a neutral accretion disk with
    solar iron abundance, irradiated by a power-law continuum of
    $\Gamma=2$.  The power-law component is shown as a dashed black line
    while the static disk reflection spectrum is the dotted blue
    line.  When the disk is convolved with Doppler and relativistic
    effects from the spacetime around even a non-spinning black hole extending
    down to the ISCO, we see the blurred spectrum in solid red.  Note that
    the fluorescence lines are affected in shape more than the
    continuum, and that the Fe K$\alpha$ line and Compton hump are
    especially prominent relative to the power-law.  The static reflection spectrum is
    modeled by the {\tt xillver} code of Garcia \& Kallman (2010).  The blurred
    reflection spectrum has been convolved with the {\tt relconv} model of
    Dauser \etal (2010).}}
\label{fig:refl_spec}
\end{figure}

The five methods for measuring spin are listed below.
\begin{itemize}
\item{{\bf Thermal Continuum Fitting} (e.g., Remillard \& McClintock 2006)
  treats the inner accretion disk 
  as a modified blackbody, and the radius of the ISCO is computed via the
  Stefan-Boltzmann law, by
  measuring the peak temperature and flux of this blackbody ($F
  d^2 \propto R^2 T^4 \cos (i)$, where $F$ is the disk blackbody flux,
  $d$ is the distance to the source, $R$ is the radius of the ISCO,
  $T$ is the peak blackbody temperature of the disk, and $i$ is the
  inclination angle of the disk to our line of sight).  
The physics behind this method is
  straightforward, much like the method one would use to measure the
  photospheric radius of a star.  When dealing with an accretion disk, however, caveats
  include the degree to which the disk emission is Comptonized by the highly
  ionized disk atmosphere (``spectral hardening,'' as per Davis \etal 2006),
  which can be difficult to quantify
  precisely.  The disk luminosity must also fall within a range roughly
  $1-30\%$ of the Eddington value in order to ensure that the blackbody emission
dominates over the Comptonized, power-law component.  Because this method also relies on
precise, accurate, independent measurements of the distance to and mass of the black hole, as well as
its disk inclination angle, the thermal continuum fitting method can only viably
be used to measure spin in stellar-mass black holes (for which there are $14$ published
spin constraints at the time of this writing).  Moreover, the temperature of the
accretion disk goes as $T \propto M^{-1/4}$ (Frank, King \& Raine 2002), so the
blackbody peak for AGN disks is in the UV, whereas stellar-mass black hole disks peak in
soft X-rays.  The prevalence of absorption in the UV band can present serious complications
for accurately measuring the thermal disk emission in AGN.}

\item{{\bf Inner Disk Reflection Modeling} (e.g., Brenneman \&
  Reynolds 2006; hereafter BR06) assumes that the high-energy X-ray
  emission ($\geq 2 \keV$) is dominated by thermal, UV disk emission which has been
  Comptonized by hot electrons in a centrally-concentrated ``corona.''  This structure
  may represent the disk atmosphere, the base of a jet or some alternative
  geometry (e.g., Markoff, Nowak \& Wilms 2005).  Some of the Comptonized photons
  will scatter out of the system and form the
  power-law continuum characteristic of typical AGN in X-rays.  A certain
  percentage of the photons, however, will be scattered back down (``reflected'') onto the
surface of the disk again.  Provided that the disk is not completely ionized, the
irradiation from the continuum power-law photons will excite a series of
fluorescent emission lines of
various atomic species at energies $\leq 7 \keV$, along with a ``Compton hump'' shaped
by the Fe K absorption edge at $\sim 7.1 \keV$ and by downscattering at $\sim 20-30
\keV$ (see Fig.~\ref{fig:refl_spec}). 
The most prominent of the fluorescent lines produced is Fe K$\alpha$ at
a rest-frame energy of $6.4 \keV$, due largely to its high abundance and fluorescent yield.
As such, the Fe K$\alpha$ line is the most important diagnostic feature of the
inner disk reflection spectrum; its shape is altered from the typical near-delta
function profile
expected in a laboratory, becoming highly broadened and skewed due to the
combination of Doppler and relativistic effects close to the black hole (See Fig.~\ref{fig:FeK}).  The
energy at which the ``red'' wing (i.e., low-energy tail) of this line truncates is directly linked to
the location of the ISCO, and therefore the spin of the black hole (see Reynolds \&
Nowak 2003 and Miller 2007 for comprehensive reviews of the reflection modeling
technique).  This method
does not require {\it a priori} knowledge of the mass, distance or inclination
of the black hole system in question, and can therefore be applied to any black hole system.
However, the principal caveat for the reflection method is that the effects of
disk ionization (especially in stellar-mass black holes with hotter disks, e.g.,
Davis \etal 2006) and/or
complex absorption from gas along the line of sight to the system (particularly within
AGN systems; e.g., Halpern 1984, Reynolds 1997) can make the
  determination of the low-energy bound of the red wing
challenging.}

\item{{\bf High Frequency Quasi-Periodic Oscillations} (e.g.,
  Strohmayer 2001), in which the X-ray power
  density spectrum of the emission from the inner accretion disk is
  characterized by $1-2$ pulses at certain harmonic frequencies (e.g., in a 3:2
  ratio), indicative of some type of resonance or regular oscillation within the
accretion flow in this region.  Such phenomena have been reported commonly in
actively accreting stellar-mass black holes (e.g., P\'{e}tri 2008), but only once in an AGN
(Gierli\'{n}ski \etal 2008).  The physical mechanism for producing these HFQPOs is
not yet known, but if the frequencies at which they are observed are related in
a fundamental way to the frequency of the ISCO (the Fourier transform
  of its period of rotation around the black hole), then the radius of
  the ISCO could potentially be derived from HFQPO
observations, and the black hole spin thereby inferred.}

\item{{\bf X-ray Polarimetry} (e.g., Tomsick \etal 2009), in which the reflected emission from the
  inner accretion disk is expected to be polarized if the disk is indeed a
  geometrically-thin, optically thick slab of gas, as expected (i.e., Shakura
  \& Sunyaev 1973).  The degree and angle of the observed polarization
  vs. energy function would then have
a characteristic shape depending on the spin of the black hole, due to the
influence of the black hole spin on the shape of the spacetime immediately surrounding
the black hole, and the radius of the ISCO (Schnittman \& Krolik 2009).
Measurements of the degree and angle of polarization vs. energy would require a
sensitive X-ray polarimeter to be flow in space, however, and there are
currently no active or planned missions to incorporate such an instrument.}

\item{{\bf Imaging the Event Horizon Shadow} (e.g., Broderick \etal 2011), in which sub-mm Very Long Baseline
  Interferometry (VLBI) is used to obtain micro-arcsecond spatial resolution to
  produce the first-ever images of the innermost accretion disk surrounding a
  black hole.  By comparing these images with detailed models of the appearance
of the innermost disk, which incorporate the necessary relativity and light
bending as a function of physical parameters such as the radius of the ISCO, the
black hole spin can be constrained (Doeleman \etal 2008).  Both the modeling and VLBI measurement techniques
are a work in progress at present, however, and this method is currently only
able to achieve the spatial resolution necessary to image black holes with event
horizons of comparatively large angular size (tens of micro-arcseconds),
limiting the sample to Sgr A* and M87 at present.}
\end{itemize}

There are limitations in applying the last three methods listed above, and the
continuum fitting method has only been applied successfully to
stellar-mass black holes.  We are therefore
currently restricted to using only the reflection modeling method for
constraining the spins of SMBHs in AGN.  

\section{Applying the Reflection Model}
\label{sec:apply}

Though the reflection modeling method could, in principle, be used to measure
the angular momentum of an actively accreting black hole of any mass, we restrict our
focus here to determining spin in SMBHs.

\subsection{Requirements}
\label{sec:requirements}

An AGN must satisfy a few important requirements in order to be
a viable candidate for obtaining spin constraints.  
Firstly, it must be bright enough to achieve
the necessary S/N in X-rays to accurately separate the reflection
spectrum from (a) the
continuum, and (b) any intrinsic absorption within the host galaxy and its nucleus.
Typically one must obtain $\geq 200,000$ photons over the $2-10 \keV$
energy range (Guainazzi \etal 2006), though
in practice the required number of counts can be substantially higher in sources
with complex absorption.  

Secondly, the
AGN must possess a broad Fe K$\alpha$ line of sufficient strength relative to
the continuum to allow its red wing to be successfully located; usually this
corresponds to a line equivalent width of $EW \gtrsim 100 \eV$.  The first
and best-studied broad iron line in an AGN is that of MCG--6-30-15,
which was initially discovered in an {\it ASCA} observation by Tanaka
\etal (1995), largely because of its high equivalent width
($EW=330^{+180}_{-120} \eV$).
The strength and breadth of this feature have been confirmed in
subsequent {\it XMM-Newton} and {\it Suzaku} observations by many
authors, most recently Brenneman
\& Reynolds (2006), Miniutti \etal (2007), and Chiang \& Fabian (2011).  Not
all type 1 AGN have been observed to possess such features, however.  Recent surveys of
hundreds of AGN with {\it XMM-Newton} have concluded that broadened Fe K$\alpha$
lines are only present in $\sim 40\%$ of all bright, nearby type 1 AGN (Nandra
\etal 2007, de la Calle P\'{e}rez \etal 2010), and some broad iron lines have been
ephemeral, appearing and disappearing in the same object observed during
different epochs (e.g., NGC~5548, Brenneman \etal 2012).

Thirdly, the Fe K$\alpha$ line in question must be {\it relativistically} broad
in order to be used to constrain black hole spin; that is, it must have a measured inner
disk edge ---assumed to correspond to the ISCO--- of $r_{\rm in} \leq 9\,r_{\rm g}$.  
Because the measurement of spin is predicated on the
assumption that the inner edge of the disk truncates at the ISCO, the
traditional approach when employing spectral models for the iron line (and other
reflection features) that allow black hole
spin to be a free parameter in the fit is to fix $r_{\rm in}=r_{\rm ISCO}$.  
A valuable first step before applying such a model, however, is to assess the
location of the inner edge of the disk by first fitting the
iron line with a fixed-spin model such as {\tt diskline} ($a=0$; Fabian et
al. 1989) or {\tt laor} ($a=+0.998$; Laor 1991) and allowing the $r_{\rm in}$ to
fit freely.  If $r_{\rm in} \lesssim 9\,r_{\rm g}$, subsequent fitting with a
free-spin model will return believable spin constraints.  

Though line-only
models are useful as a tool to establish an initial starting point in parameter
space for spectral
fitting, it is important
to model the inner disk full reflection spectrum when making spin measurements,
not just the broad Fe K$\alpha$ line.  Separate, line-only models can be
misleading when attempting robust spin measurements because they do not take
into account the curvature of the reflection spectrum over the full energy range
of the data, e.g., from the Compton hump and associated iron edge absorption.
Only an accurate, holistic modeling approach that treats the continuum,
absorption and the entire reflection spectrum at once will yield robust
constraints on black hole spin.

Taking all these points into consideration, the potential sample size of
spin measurements for AGN in the local universe is $\sim30-40$ sources (Miller 2007).
Most of these AGN are type 1, lacking significant obscuration by dust and gas
along the line of sight to the inner disk.  
Though it is possible to obtain spin measurements for more heavily
absorbed type 2 sources, great care must be taken to properly account for all of the
absorption in the system.

\subsection{Spectral Models}
\label{sec:models}

The reflection spectrum from the inner disk can be
self-consistently reproduced by models such as {\tt reflionx}
(Ross \& Fabian 2005; see Fig.~\ref{fig:refl_spec}) or {\tt xillver}
(Garcia \& Kallman 2010).  These models
simulate not only the broad Fe K$\alpha$ line, but all other fluorescent
emission species at lower energies, as well as the Compton hump at higher
energies.  
Typical free parameters include the ionization of the disk and its
iron abundance (assumed, simplistically, to be constant values), the photon index of the
irradiating power-law continuum (usually tied to that of the power-law itself),
and the flux or normalization of the reflection spectrum.  
In order to
incorporate the effects of relativity and Doppler shift, this static reflection
spectrum must then be convolved with a smearing algorithm which computes the photon
trajectories and energies during transfer from the accretion disk to the
observer.  Several free-spin smearing
algorithms are currently available for use (see \S1).  The {\tt kerrconv} algorithm of
BR06 is the only one of these models that is currently
built into {\sc xspec}, though it limits black hole spin to prograde
values only.  A more recent improvement is the
{\tt relconv} model of Dauser \etal (2010, 2013), which generalizes the possible
spins to incorporate retrograde black holes (see Fig.~\ref{fig:FeK}).

Isolating the inner disk reflection spectrum from other X-ray spectral
signatures is often the greatest challenge in obtaining robust
constraints on black hole spin, even in deep ($\gtrsim100,000 \s$)
observations of bright AGN.  Due to the relatively poor spatial
resolution of X-ray telescopes, the spectra obtained from the cores of
AGN represent a superposition of emission and absorption from several
different physical processes within these regions.  The principal
components of AGN X-ray spectra (other than relativistic reflection) and their
physical origins are enumerated
below.  Most, though not all, of these components are present in any
given AGN.

\begin{enumerate}
\item{{\bf Continuum Emission:} As discussed in \S2, the source of the
continuum in X-rays is Compton upscattering of thermal photons from
the accretion disk.  At present very little is known about the
origin, geometry and location of the hot electrons responsible for this
scattering, but the timescales of variability ($\sim$hours) for the power-law
component in AGN spectra suggest that this corona is compact and
likely also close to the accretion disk (e.g., Markoff, Nowak \& Wilms 2005).
Magnetic fields almost certainly play a critical role in its formation, perhaps
facilitating the heating and/or flaring of certain regions of the electron
plasma (Di Matteo 2001).
Though this corona is readily approximated by a power-law with a high-energy
cutoff (a proxy for the electron temperature), more physical models have been
created which include
free parameters for optical depth, electron temperature, seed photon
temperature, compactness, thermal vs. non-thermal electron population, etc.  Three
of the most popular of these models are {\tt compTT} (Titarchuk 1994), its
successor, {\tt compPS} (Poutanen \& Svensson 1996), and
{\tt eqpair} (Coppi 1999).}

\item{{\bf Intrinsic Cold (Neutral) Absorption:} The putative molecular torus of
AGN unification schemes (Antonucci 1993, Urry \& Padovani 1995) manifests in two
forms in X-rays: as the source of the primary neutral absorbing column within
the nucleus, and as a scattering medium for the continuum emission, forming the
distant reflection signatures discussed in item $\#4$ below.  This reservoir of
gas is a
  relatively cold, neutral, optically-thick medium thought to reside on the order of $10^4-10^5 \,
  r_{\rm g}$ from the black hole.  Its gas appears to be
anisotropically distributed, though its origins are still a topic of active
research.  The radius of this neutral gas from the black hole, with respect to the broad emission line region
(BELR), may vary from object to object, however.  For example, NGC~1365 shows
evidence for a clumpy absorber at a radius $r \leq 2 \times 10^{15} \cm$ from
the black hole, well within the BELR, in which the clumps of gas eclipse the inner disk/corona region
(e.g., Risaliti \etal 2005a, Maiolino \etal 2010, Brenneman \etal
2013).  Recent work suggests that radiation
pressure may play a prominent role in forming these collections of neutral gas
(Elvis 2012).  Whatever its origin, this cold absorbing gas is best modeled with a
simple photoelectric absorption component whose low-energy cutoff is determined
by the column density of the gas (e.g., {\tt phabs}, within {\sc
  xspec}, or {\tt tbabs}, from
Wilms, Allen \& McCray 2000), or a
partial-covering variant of these models with an additional parameter for the
covering fraction of the gas over the continuum source (e.g., {\tt
  pcfabs}, within {\sc xspec}).}

\item{{\bf Intrinsic Warm (Ionized) Absorption:} Some of the first signatures of
  ionized absorbing gas within AGN were originally reported by Halpern (1984),
  though they were not detected commonly in X-rays until the {\it ASCA} era
  (e.g., Reynolds 1997).  Early CCD resolution could only detect the two most
  obvious manifestations of these features,
  the O\,{\sc vii} and O\,{\sc viii} absorption edges at $0.74$ and $0.87 \keV$,
  respectively.  But with the advent of grating spectroscopy in the {\it
    Chandra} and {\it XMM-Newton} era, warm absorbers are now known to harbor a
  rich forest of lines and edges from various species of gas and dust
  (Lee 2010).
  The presence of these
  features can extend up into the Fe K band in some cases
  where the ionization of the gas is high enough, and are sometimes
  associated with outflows having speeds of up to $v_{\rm out} \sim 0.4c$ based on their
  blueshifted absorption features (e.g., Tombesi \etal
  2010).  Most warm absorbers are now
  thought to incorporate several ``zones'' of material, distinct in their
  kinematic properties as well as their column densities and ionizations, but
  maintained in pressure balance (e.g., NGC~3783; Krongold \etal 2003).  Though
  they can be modeled with individual absorption lines and edges, the most
  self-consistent way to model these features over an entire spectrum is with tables produced by
  spectral synthesis codes (including radiative transfer) such as {\sc
  cloudy} (Ferland \etal 2013) or {\sc xstar} (Kallman \& Bautista 2001).}

\item{{\bf Distant Reflection:}  When irradiated by the X-ray power-law continuum
  emission, the outer disk/torus produces a reflection spectrum much like that of
  the inner
  accretion disk, only without the convolved relativistic effects due to its
  distance from the black hole (George \& Fabian 1991, Matt \etal 1992).  The narrow Fe
  K$\alpha$ line and Compton hump are the two most prominent features of this
  distant reflection spectrum; in fact, the narrow Fe K$\alpha$ line appears to
  be nearly ubiquitous in Seyfert 1 galaxies (Yaqoob \& Padmanabhan 2004).
  Though the Compton hump can be modeled
  adequately with a {\tt pexrav} component (Magdziarz \& Zdziarski 1995), one
  must include additional Gaussian emission lines to model the discrete Fe K
  features.  Alternatively, the {\tt pexmon} model (Nandra \etal 2007) includes
  the Fe K$\alpha$, K$\beta$ and Ni K$\alpha$ lines as well as the Compton
  shoulder of the Fe K$\alpha$ line self-consistently with the {\tt
  pexrav} reflected continuum.  There are two important caveats to
  keep in mind when applying these models, however: (1)
  {\tt pexrav} and {\tt pexmon} are designed to
  simulate the spectrum produced from the irradiation of a thin disk
  rather than a puffy or toroidal structure,
  so this model may not be an accurate representation of the system;
  (2) neither model accounts for the contribution of the cold
  reflected emission at energies $E \lesssim 3 \keV$.  To consider
  the contribution from emission below this energy, one can also
use a static {\tt reflionx} model to simulate the contribution from
  the torus across all energies,
though this model also assumes a disk geometry.  The {\tt
    MYTorus} model of Murphy \& Yaqoob (2009) is a more physical alternative
  that takes the geometry of the reprocessing medium into account, as
  well as its low-energy emission, though at present it is unclear how well the
  opening angle of the reprocessor can be constrained, and the model does not
  allow the iron abundance of the reprocessor to vary freely.}

\item{{\bf Soft Excess Emission:} Some of the first observational evidence for a
``soft excess'' in X-rays above the power-law continuum was observed by {\it EXOSAT} in Mrk~841 (Arnaud \etal
  1985).  Though this emission can be well fit with a blackbody or modified disk
  blackbody ({\tt diskbb}) component, however, the typical temperature ($\sim0.2
  \keV$) is too high to be thermal emission from the disk in an AGN, as a rule.
  Advanced CCD and grating spectra made possible in the {\it
    Chandra/XMM-Newton/Suzaku} era have since demonstrated both the ubiquity of
  this feature in AGN with moderate-to-high accretion rates, and that the typical
  temperature of this component does not appear to change with source luminosity, posing
  theoretical problems
  for many thermal and single-zone Comptonization models (Bianchi \etal 2009, Done \etal 2012).
  Other possible origins for the soft excess include a contribution from inner
  disk reflection, photoionized or extended thermal plasma emission (e.g., from
  a circumnuclear starburst region), scattering of continuum photons, or
  emission from the base of a jet (see
  Lohfink \etal 2013a and references therein).  Because the soft excess typically manifests as a smooth feature,
an adequate fit can be equally well achieved with any (or more than one) of the above
modeling components, leading to uncertainty about its physical origin.  The
best-fit model component(s) of the soft excess also seem to vary between AGN.
Effective modeling of the soft excess is critical to constraining the physical
parameters of the AGN spectrum, however, because
X-ray detectors tend to have higher collecting areas at lower energies where
this component dominates (i.e., $\leq 2 \keV$).  The fit statistic will
therefore be dominated by this region of the spectrum as well.}
\end{enumerate}

\subsection{Time-averaged vs. Time-resolved Spectra}
\label{sec:TAvTR}

An examination of the time-averaged spectrum maximizes the S/N
of the observation and allows one to assess which physical components
are present in the data.  By contrast, time-resolved spectroscopy is
critical for identifying and properly
modeling the various physical components of a typical AGN system.
Many of these components may vary substantially during the
observation, and those variations may appear averaged-out and provide
misleading information about the source when viewed through the
time-averaged spectrum alone.  

Whenever possible (i.e., given enough S/N in
reasonable time bins, $\sim10,000$ counts per bin), a time-resolved spectral
analysis should be undertaken in addition to the time-averaged
analysis.  Data from different time intervals and/or flux states within an observation
(or indeed, data of the same source taken during multiple epochs) 
can be analyzed jointly in order to assess the physical nature and
variability of all of the components in a given X-ray spectrum, often yielding a
more physical picture of the system whose changes during an observation are
washed out in a time-averaged spectrum.  

For example, the power-law and inner disk reflection are expected to vary on timescales
as short as hours in a typical AGN if the continuum emission and
reflection are centrally concentrated (e.g., Miniutti \& Fabian 2004,
Uttley, McHardy \& Vaughan 2005).  By contrast, distant reflection from the
outer disk/torus region typically varies on the order of
$\sim$days-weeks (McHardy, Papadakis \& Uttley 1999), and warm
absorbers can show changes in their column
densities and/or ionizations on timescales of $\sim$weeks-months
(Krongold \etal 2005).  Both of these components take longer to
respond to changes in the continuum emission than the inner disk radii
due to their relatively large distances from the corona.
The soft excess
in many AGN is often unrelated to source luminosity and can be constant over
years-long timescales (Crummy \etal 2006).  By jointly analyzing
spectra from different time or flux intervals in a given AGN, one can
tie certain model parameters together between intervals if they are
not expected to vary during or between observations, while leaving the
other model parameters free to vary.  Doing so effectively increases
the S/N of the data and yields more accurate constraints on slowly- or
non-varying parameters of these systems (e.g., black hole spin, disk
inclination, iron abundance).  

Time-resolved spectroscopy is particularly useful for
disentangling the effects of complex absorption from the properties of
the continuum and inner disk reflection, because the majority of AGN show
evidence for warm absorption in their spectra.  Some researchers have even
proposed that, in many AGN with purported broad Fe K$\alpha$ lines, these apparent
reflection features are actually artifacts of improperly modeled absorption.
For example, Miller, Turner \& Reeves (e.g., 2008, 2009; MTR) argue that the archetypal
broad iron line AGN, MCG--6-30-15, actually shows no relativistic
inner disk reflection, but instead has five layers (or ``zones'') of absorbing gas
covering a wide range in column density, ionization parameter and covering
fraction.  The superposition of spectral features created by these
absorbing structures mimics the appearance of relativistic reflection features.
The MTR absorber's incorporation of partial covering, especially, is
what allows the model to achieve a goodness-of-fit comparable to that of the
relativistic reflection model, which does include complex absorption,
but with fewer zones.  The debate
is ongoing regarding which model is a more plausible physical representation of
the system, and a combination of a broad spectral bandpass and time-resolved
and/or multi-epoch spectral analysis are needed to definitively address this
question (see \S5).  When viewed holistically in this manner, the relativistic
reflection model (e.g., that of BR06, Miniutti \etal 2007
and Chiang \& Fabian 2011) will vary in ways that have no analog in the MTR
absorber model, and vice versa.  

In the following sections, I describe the practicalities of using the
relativistic reflection model (employing a {\tt reflionx} disk
reflection model and either a {\tt kerrconv} or {\tt
  relconv} smearing algorithm) to measure the spins of the SMBHs in 
three well-known AGN using {\it XMM-Newton, Chandra} and {\it Suzaku}
observations: MCG--6-30-15, NGC~3783 and Fairall~9.

\section{Case Studies: MCG--6-30-15, NGC~3783 and
  Fairall~9}

\subsection{MCG--6-30-15}
\label{sec:mcg6}

The type 1 AGN MCG--6-30-15 ($z=0.0077$) was the first galaxy in which a broad Fe
K$\alpha$ line was observed, using {\it ASCA} spectra (Tanaka \etal
1995).  This feature still stands as the broadest line of
its kind to date, with a red wing extending down to $\sim3 \keV$,
and its strength measured at $\sim200 \eV$ (Fabian \etal 2002).  As
such, MCG--6-30-15 is one of the most observed AGN in X-rays, with
numerous pointings from {\it Chandra, XMM-Newton} and {\it Suzaku} over the
past decade in the public archives.\footnote{NASA/GSFC maintains the
  High-Energy Astrophysics Science Archive Research Center, or
  HEASARC: http://heasarc.gsfc.nasa.gov/.}

The first measurement of the SMBH spin in MCG--6-30-15 using the technique
described above in \S\ref{sec:apply} was made by
BR06 using {\it XMM-Newton} data from the long
2001 observation first reported by Fabian \etal (2002).  BR06
constrained the spin to $a \geq +0.98$ to 90\% confidence with 
a model incorporating a {\tt kerrconv} smearing kernel acting on a
{\tt reflionx} disk reflection
spectrum, a three-zone, dusty warm absorber and a soft excess that was
modeled equally well with either a blackbody or {\tt compTT} component.
The limited spectral range of {\it XMM-Newton} ($0.3-12 \keV$) posed
difficulties for modeling the Compton hump, however, since this feature lies
mostly outside of the telescope's energy band.  

\begin{figure}[hp]
\centerline{
\hspace{-2cm}
\includegraphics[width=0.8\textwidth]{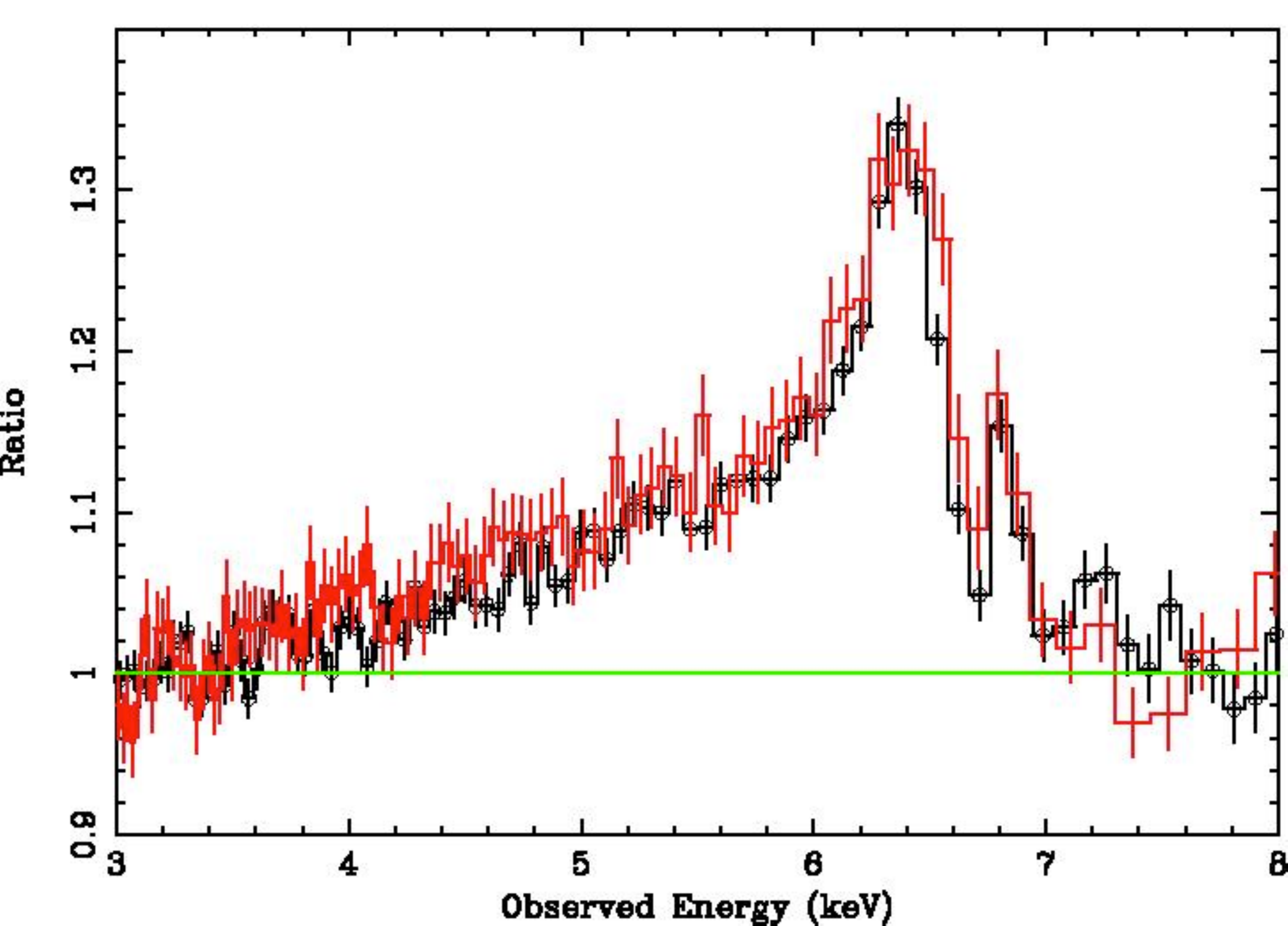}}
\caption{\small{{\it Suzaku} (black, 2006) and {\it XMM-Newton} (red, 2001) data
    showing the broad Fe K$\alpha$ line in MCG--6-30-15 ratioed
    against a power-law continuum.  The feature shows little variation in
    strength and breadth between epochs.  Solid black and red lines are meant to
    guide the eye and do not represent a model.  The solid green line
    shows a data-to-model ratio of unity.  Figure is from Miniutti
    \etal, 2006, \PASJ, 59S, 315.  Reprinted by permission of \PASJ.}}
\label{fig:miniutti_xmm_suz}
\end{figure}

The launch of {\it
  Suzaku} in 2005 enabled the $0.3-12 \keV$ energy range to be
complemented by simultaneous data up to $\sim60 \keV$ for bright AGN,
using {\it Suzaku}'s XIS and PIN instruments in tandem.  Miniutti
\etal (2007) examined the {\it Suzaku} spectrum of MCG--6-30-15 for
the first time, using $\sim330 \ks$ of data obtained over two weeks in
January 2006.  The flux of MCG--6-30-15 at this time was measured at $4.0 \times
10^{-11} \ergpcmsqps$, yielding $1.98 \times 10^6$ counts in the XIS detectors
and $1.50 \times 10^5$
counts in the PIN detector. 
Miniutti \etal noted the striking
similarity between the
{\it Suzaku} and {\it XMM-Newton} data in the Fe K band (see
Fig.~\ref{fig:miniutti_xmm_suz}).  They then restricted their energy range to $\geq3
\keV$ in order to avoid most of the spectral complexities associated
with the warm absorber and used a model similar to that of BR06 to derive a
black hole spin of $a \geq +0.92$.  

\begin{figure}[hp]
\centerline{
\includegraphics[width=0.8\textwidth]{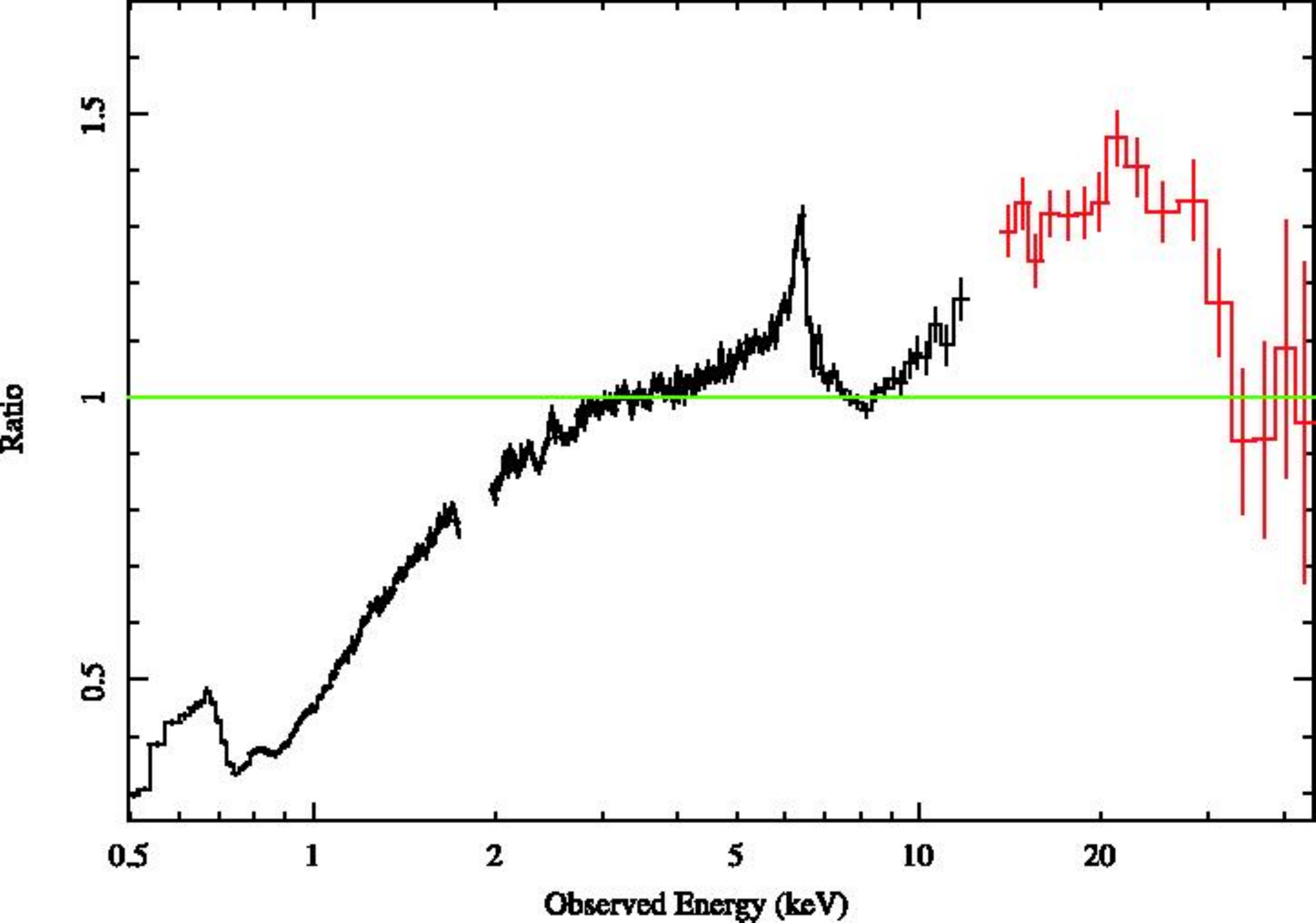}}
\caption{\small{{\it Suzaku}/XIS (black) and PIN (red) data from the
    2006 observation of MCG--6-30-15, here shown ratioed against a
    power-law.  The black and red solid lines connect the data points
    and do not represent a model.  The solid green line
    shows a data-to-model ratio of unity.  Figure is from Miniutti
    \etal, 2006, \PASJ, 59S, 315.  Reprinted by permission of \PASJ.}}
\label{fig:miniutti_suz}
\end{figure}

One of the most comprehensive analyses to date of MCG--6-30-15 was performed
by Chiang \& Fabian (2011; CF11).  These authors re-examined the {\it
  XMM-Newton+BeppoSAX} (2001), {\it Chandra} (2004) and {\it Suzaku} (2006)
data jointly in order to characterize the nature and variability of
the warm absorber, continuum and reflection components of the source
holistically.  The {\it Suzaku} spectrum is shown in
Fig.~\ref{fig:miniutti_suz} as a ratio to the
power-law continuum and Galactic photoabsorbing column in order to
illustrate the various residual spectral features present.  A very
strong Compton hump is visible above $10 \keV$ and rolls over at
$\sim20 \keV$.  At lower energies, the narrow Fe K$\alpha$ line is
prominent at $6.4 \keV$ along with an absorption line of Fe\,{\sc xxv}
at $\sim6.7 \keV$ and a small Fe K$\beta$ emission line at $\sim7
\keV$.  The broad line is quite prominent, extending down to $\sim3
\keV$ on the red wing.  Below this energy, the spectrum takes on a
concave shape due to the presence of absorption by gas at lower
ionization states.  A weak soft excess is evident below $\sim0.8
\keV$.

The goal in
modeling the spectrum over the entire energy range is to achieve the
best possible statistical fit with the lowest possible number of
parameters using a physically self-consistent approach.  CF11
assumed that the basic components of
the fit are the same as those seen in all other type 1 AGN, as
described in \S3.2: power-law continuum, distant and inner disk
reflection, complex absorption and a soft excess.  No {\it a priori}
constraints were placed on
the physical nature of the
soft excess emission, as this is still a topic of debate within the
community and may vary among AGN.  The inner disk was assumed to extend
radially from $r_{\rm in}=r_{\rm ISCO}$ to $r_{\rm out}=400\,r_{\rm g}$. 
CF11 modeled
the spectra from {\it Chandra, XMM+BeppoSAX} and {\it Suzaku} accordingly,
beginning with the power-law continuum modified by Galactic
photoabsorption and progressively adding new components as warranted
by an improvement in the fit statistic.  

Absorption modifies an entire spectrum in a
multiplicative sense ($\propto e^{-\tau}$, where $\tau$ is the optical
depth of the absorbing gas), meaning that it can affect the shape of the
overall spectrum across a significant fraction of the energy band.  As
such, the absorption should be addressed early during modeling because
it will have a significant effect on the parameters of the continuum
(most notably the slope of the power-law).  CF11 employ {\sc xstar}
tables to model the multi-zone warm absorber in MCG--6-30-15, allowing
the column density and ionization to be free parameters in the fit.
Primarily informed by the high-resolution {\it Chandra}/HETG data, the
authors find that three zones of ionized absorbing gas intrinsic to
the AGN are required to properly model its spectral curvature.  

\begin{figure}[hp]
\vspace{-4in}
\centerline{
\includegraphics[width=3.0\textwidth]{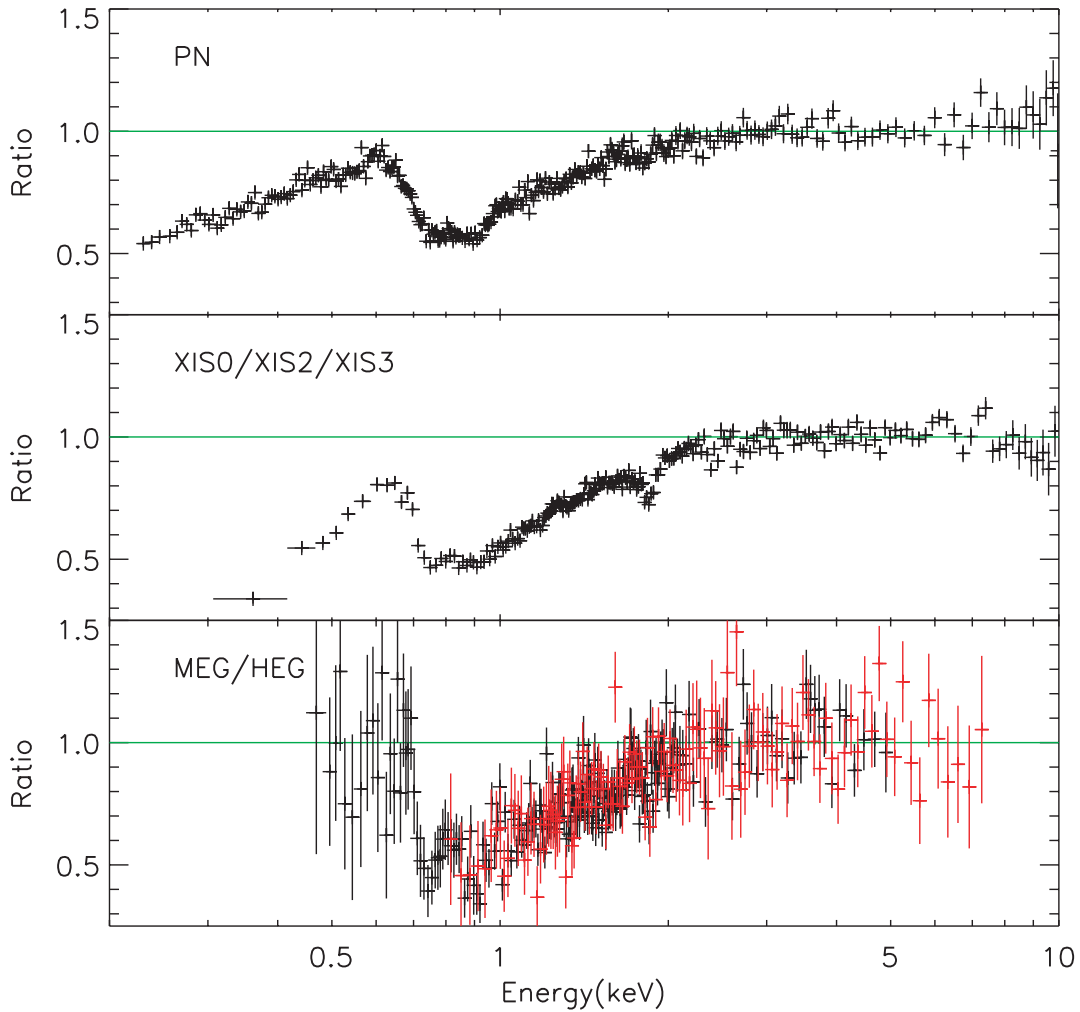}}
\vspace{-4.5in}
\caption{\small{The difference spectra from top to bottom are
    extracted from
    {\it XMM-Newton}/PN, {\it Suzaku}/XIS-FI and {\it Chandra}/HETG,
    respectively.  The absorption structure around $1.8 \keV$ in the
    middle panel is an uncalibrated silicon feature produced by he
    detector.  Note the similarity in spectral shape between the three
    observations.  Figure is from Chiang \& Fabian, 2011, \MN, 414,
    2345.  Reprinted with permission from Oxford University Press.}}
\label{fig:chiang_diff}
\end{figure}

To illustrate that
these three absorption components are both necessary and sufficient,
CF11 consider difference spectra created by subtracting the spectrum
of the low-flux state of the source from that of the high-flux state
of the source in each observation (see Fig.~\ref{fig:chiang_diff}).
Although examining a difference spectrum is a common
technique used to assess the contribution of additive components
(e.g., individual emission lines) to a
spectrum at difference times, the multiplicative components (e.g.,
absorption) also manifest because they cannot be subtracted out.  The
difference spectra drawn from the three satellites are shown here as a
ratio to a simple power-law model modified by Galactic absorption.
They are remarkably similar at energies below $3 \keV$.  An obvious
drop below $2 \keV$ is seen and goes to nearly the same depth in all
difference spectra.  This suggests that the low energy spectra can be
represented by the same model, and that the warm absorber does not change
significantly between the three observations.  The $3-10 \keV$ ($3-7.5 \keV$
for {\it Chandra}) difference spectra can be fitted by a simple
power-law, implying that the warm absorber causes little curvature
above $3 \keV$ and that the (additive) reflection signatures are
largely unchanged between epochs. 

The warm absorber does not appear to mimic the red wing structure seen
in the Fe K band, nor the excess emission seen at higher energies that
is commonly attributed to the Compton hump also produced by reflection.  This
contrasts with the model suggested by Miller \etal (2008, 2009) which
incorporates two additional partial-covering clumpy absorbing zones to
model the high-energy spectrum, in particular.  One of these zones
mimicked the shape of broadened Fe K$\alpha$ line, and the other
partially covered the continuum in order to explain the hard excess.
The main difference between the absorption-only model and the
reflection-dominated model is that the former has no distortions due
to relativistic effects.  

Miller \etal (2009) have claimed that the
reflection model fails to interpret the hard excess.  However, CF11
showed that the hard excess can be simply explained by a
relativistically blurred reflection component of super-solar iron
abundance {\it without} any additional, partial-covering absorption.
However, it should be noted that, though the Miller \etal model does
not address the difference spectra
of MCG--6-30-15, both the reflection and absorption-only models do 
provide adequate fits to the spectrum and variability seen in
MCG--6-30-15 in these data.  

\begin{figure}[hp]
\centerline{
\includegraphics[width=0.4\textwidth,angle=90]{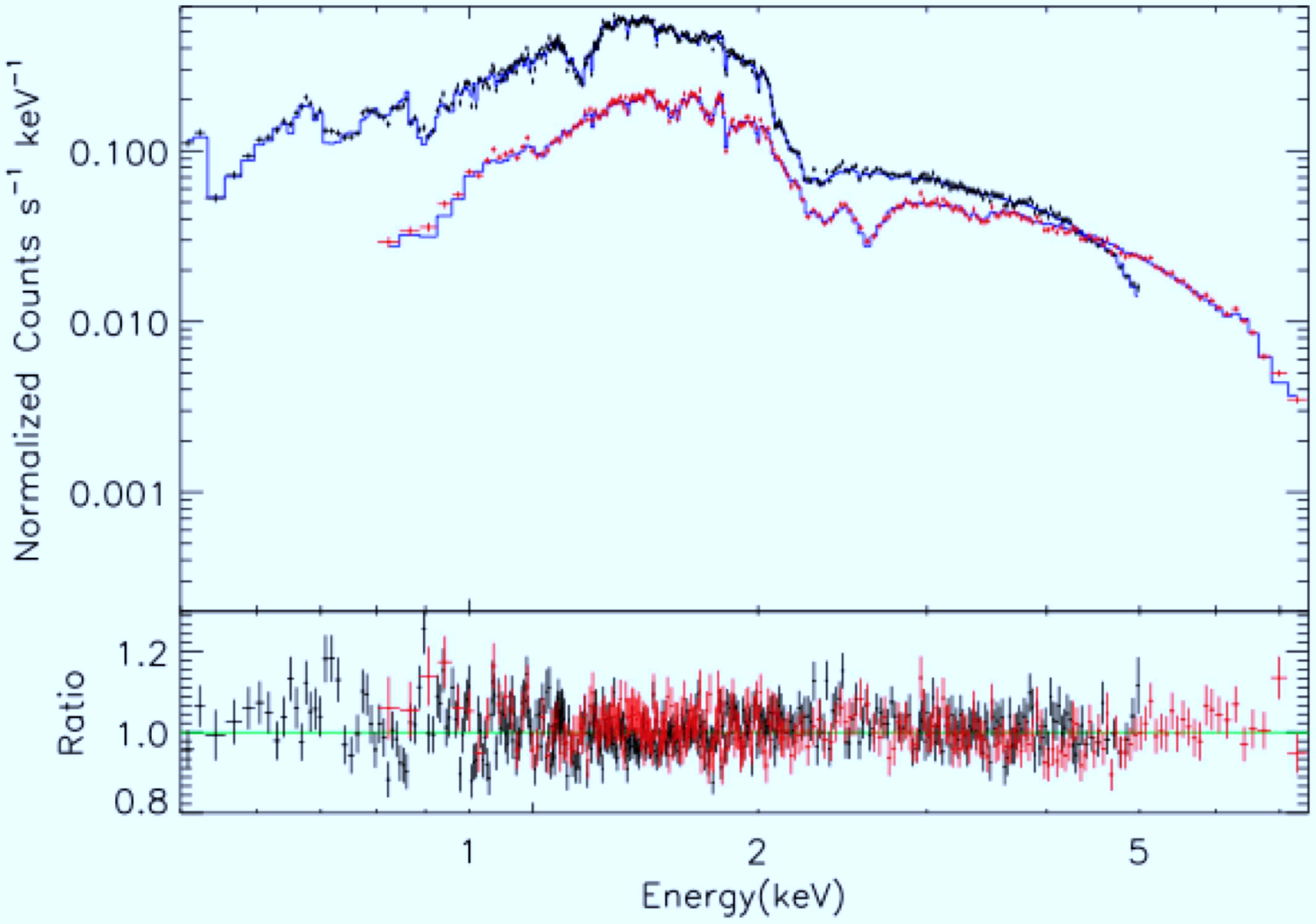}
}
\vspace{-1.0cm}
\centerline{
\includegraphics[width=0.4\textwidth,angle=90]{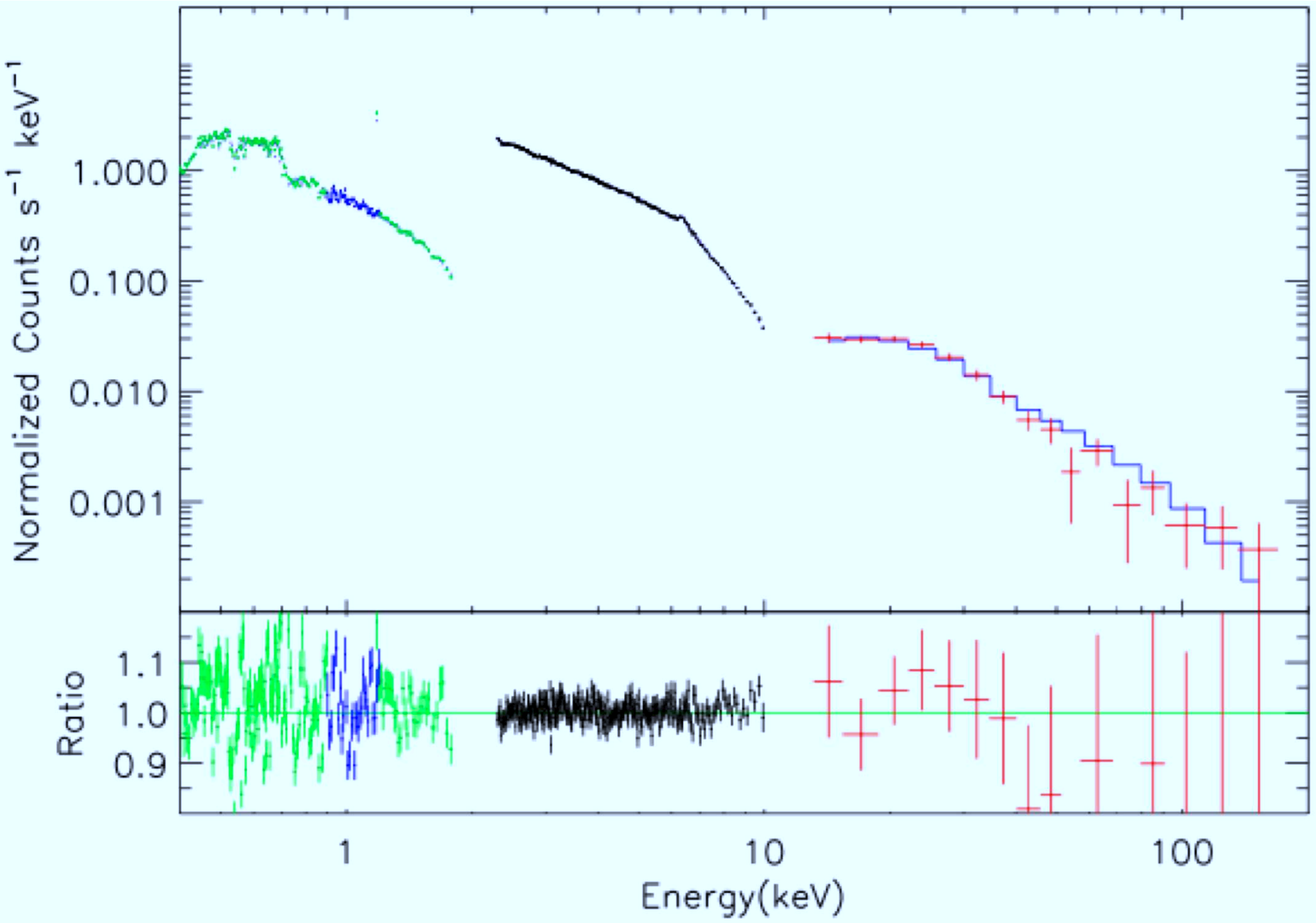}
}
\vspace{-1.0cm}
\centerline{
\includegraphics[width=0.4\textwidth,angle=90]{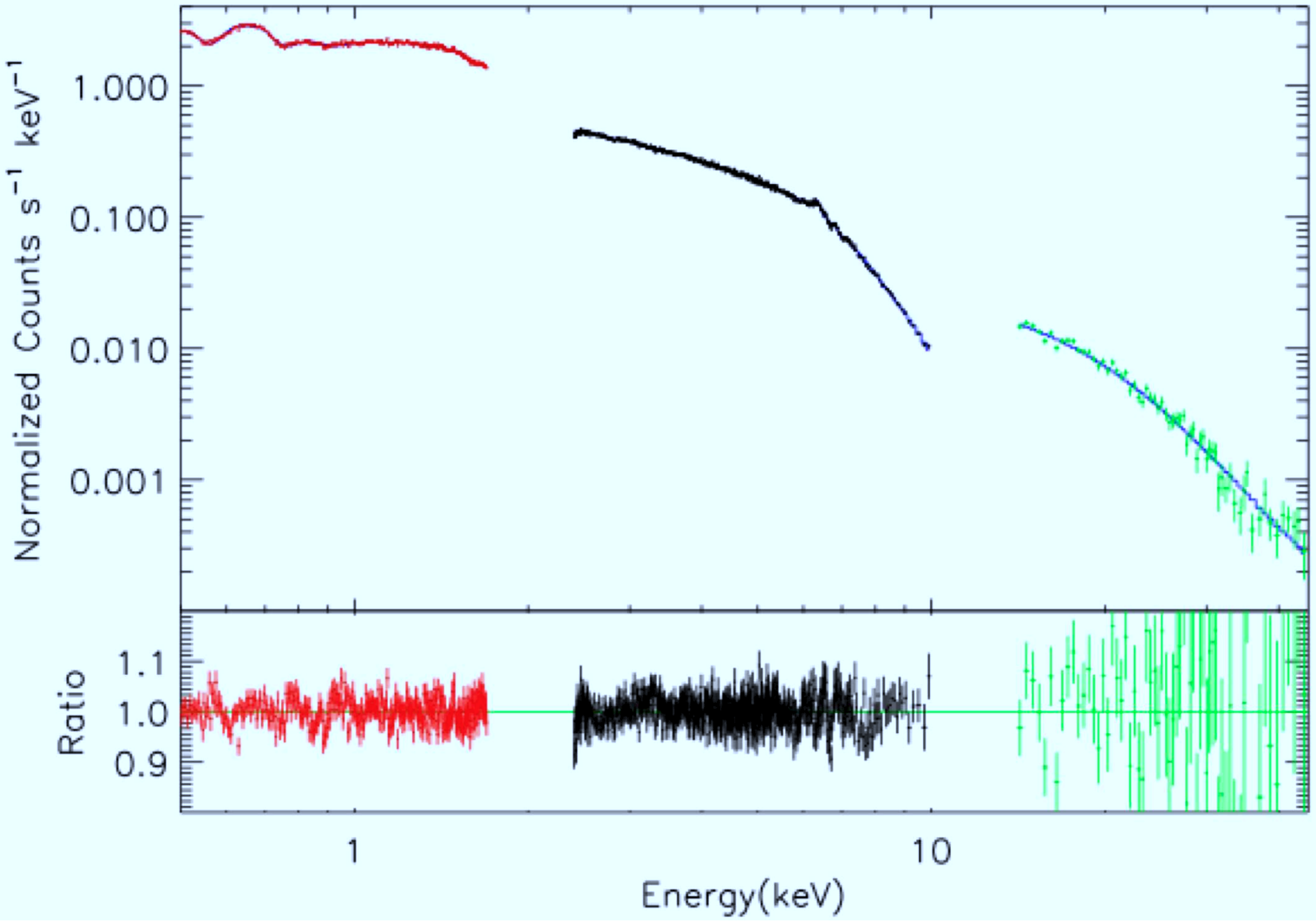}
}
\vspace{-1.0cm}
\caption{\small{The best-fit model of CF11 applied to data from {\it
      Chandra}/HETG $\pm 1$ order MEG (top; black points show +1, red show -1),
    {\it XMM-Newton}+{\it BeppoSAX} (middle; green/blue points show {\it XMM-Newton}/RGS, black
    show {\it XMM-Newton}/PN, red show {\it BeppoSAX}/PDS) and {\it Suzaku}/XIS+PIN
(bottom; red points show the XIS-BI, black show the co-added XIS-FI and green
    show the PIN).  Models fit to each dataset are shown as solid lines in the
    top panels.  The bottom panels on each plot show the data-to-model ratio, where
    the solid green line indicates a theoretical perfect fit.  Figure
      is from Chiang \& Fabian, 2011, \MN, 414, 2345.  Reprinted with
      permission from Oxford University Press.}}
\label{fig:chiang_best}
\end{figure}

\begin{figure}[hp]
\centerline{
\includegraphics[width=1.0\textwidth]{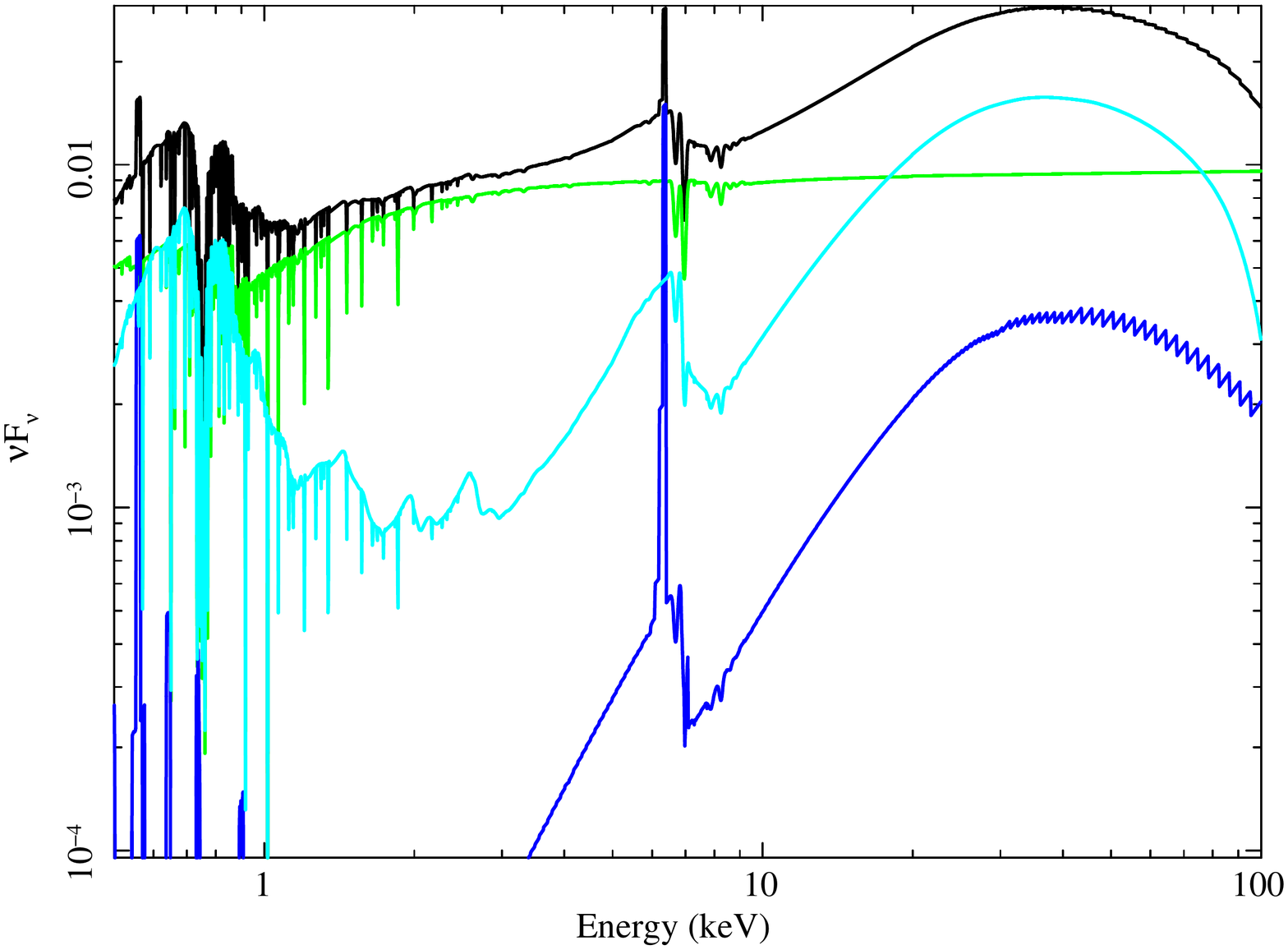}
}
\caption{\small{The best-fitting model components for the CF11 model as applied
    to the 2006 {\it Suzaku} data for MCG--6-30-15.  The total model is depicted
in black, the power-law in green, the distant reflector in dark blue and the
inner disk reflector in light blue.  The three-zone warm absorber modifies all
spectral components.  Figure is created based on the best-fit model
    from Chiang \& Fabian, 2011,
    \MN, 414, 2345.}}
\label{fig:chiang_best_eemo}
\end{figure}

Within the framework of the reflection model, both distant and inner
disk reflection components must be included in order to accurately
model the spectrum of MCG--6-30-15.  CF11
parametrize these features with two {\tt reflionx} models, convolving the
inner disk component with a {\tt kdblur} smearing kernel analogous to
the {\tt laor} relativistic line profile (i.e., spin fixed at
$a=+0.998$) while leaving the distant {\tt reflionx} component
unsmeared.  Though it is not possible to formally constrain black hole spin using the
approach these authors have chosen, the measurement of the inner disk
radius obtained via this method can provide some insight into the
magnitude and direction of the black hole's angular momentum.  The best fit
obtained is
$\chi^2/\nu=5059/3809\,(1.33)$ for {\it XMM+BeppoSAX},
$\chi^2/\nu=2418/2139\,(1.13)$ for {\it Chandra} and
$\chi^2/\nu=1685/1576\,(1.07)$ for {\it Suzaku}.  In each case, the
good quality of the fit is a strong indication that a
rapidly-spinning, prograde black hole resides in MCG--6-30-15.  This is
confirmed by the constraints placed on the inner edge of the disk,
particularly in the {\it XMM-Newton} observation, which has the highest S/N
in the Fe K band: $r_{\rm in} \leq 1.7\,r_{\rm g}$.  This equates to a
spin of $a \geq +0.97$, entirely consistent with the results of
BR06 and Miniutti \etal (2007).  The
reflectors are both approximately neutral,
with a measured iron abundance of ${\rm
  Fe/solar}=1.7^{+0.2}_{-0.1}$ and a disk inclination angle of
$i=(38^{+3}_{-2}) \degmark$ to the line of sight.  The soft excess is
represented adequately by the inner disk reflector in the CF11 model,
so no additional spectral component is required.  The best-fitting model to all
three datasets is shown in Fig.~\ref{fig:chiang_best}, while the best-fitting
model components are shown in Fig.~\ref{fig:chiang_best_eemo}.

Even if we assume that the reflection-dominated model is the most
physically realistic explanation for the spectrum and variability of
MCG--6-30-15, controversy over the derived reflection parameters
remains.  Patrick \etal (2011; hereafter P11) employ a very similar
spectral model to CF11 with a three-zone warm absorber, power-law
continuum and both distant and inner disk reflection, yet their measured spin
parameter is $a=0.61^{+0.15}_{-0.17}$, more than a $2\sigma$ off the value
measured by BR06 and Miniutti \etal (2007).  However, P11
also model the soft excess with a {\tt compTT} component that
is assumed from the start of their modeling, rather than fitted as a
remaining residual after the continuum, absorption and reflection have
been accounted for.  This {\tt compTT} component has a modest
temperature, optical depth and flux, in keeping with the
modest strength of the soft excess in this source ($kT=3.9 \keV$,
$\tau=0.8$, $F_{\rm 0.6-10}=7.2 \times 10^{-12} \ergpcmsqps$).

Another, and perhaps more critical difference of the P11 model from that of
CF11 is in the construction of the warm absorber tables.  CF11 use
iterative fitting to determine the turbulent velocities of the three warm
absorption zones: $v_{\rm turb}=500 \kmps$ for two zones, $v_{\rm turb}=100 \kmps$
  for the remaining zone.  P11, by contrast, keep the turbulent velocity of the
  two low-ionization zones at $v_{\rm turb}=200 \kmps$ and fix that of the
  high-ionization zone to $v_{\rm turb}=1000 \kmps$.  Additionally, CF11 allow
  the iron abundance of the warm absorber to vary (tying it to that of the
  distant and inner disk reflector), whereas P11 fix the iron abundance at the
  solar value.  These two differences have significant effects on the appearance
  of the warm absorber, giving it broader spectral lines and requiring higher
  column densities to model iron features than would be required if super-solar
  iron abundance were allowed (as found in CF11, using {\it Chandra}/HETG data).
  These differences are likely responsible for the different column densities
  and ionizations measured for the warm absorber zones by the two groups.
  Finally, P11 employ an additional neutral, high column, partial-covering absorber to account for a
  hard excess above $10 \keV$ above the dual reflector model.  This absorber has
  $N_{\rm H}=3.4 \times 10^{24} \pcmsq$ and $f_{\rm cov}=50\%$.  No such
  additional absorber is needed in the CF11 analysis; the hard excess is well
  accounted for by the super-solar iron abundance allowed in the CF11 absorber
  tables.  The low-ionization absorber of P11 possesses much more curvature up into the Fe K
  band, while the partial-covering absorber takes up the hard excess emission
  that would otherwise be modeled by the Compton hump in the CF11 model.  This
  combination results in the lower spin derived by P11.

Disregarding the ongoing debate about which model is a more physical
representation of the system, the fact remains that a good statistical fit to
the spectral data can be
achieved by the reflection-dominated model of BR06 and
CF11, the absorption-only model of Miller \etal (2008, 2009) and the ``hybrid''
model of P11 that combines features of both.  Breaking the degeneracy between
these models will require high S/N over a broad bandpass in X-rays,
and good spectral resolution over the $\sim0.5-10 \keV$ range,
especially in the Fe K band (i.e., $\sim100 \eV$ resolution at $6
\keV$).  These
capabilities will enable the continuum, absorption, reflection and any remaining
soft or hard excess emission to be accurately and simultaneously characterized
based on their discrete and broad-band features.  

\begin{figure}[hp]
\centerline{
\includegraphics[width=0.6\textwidth]{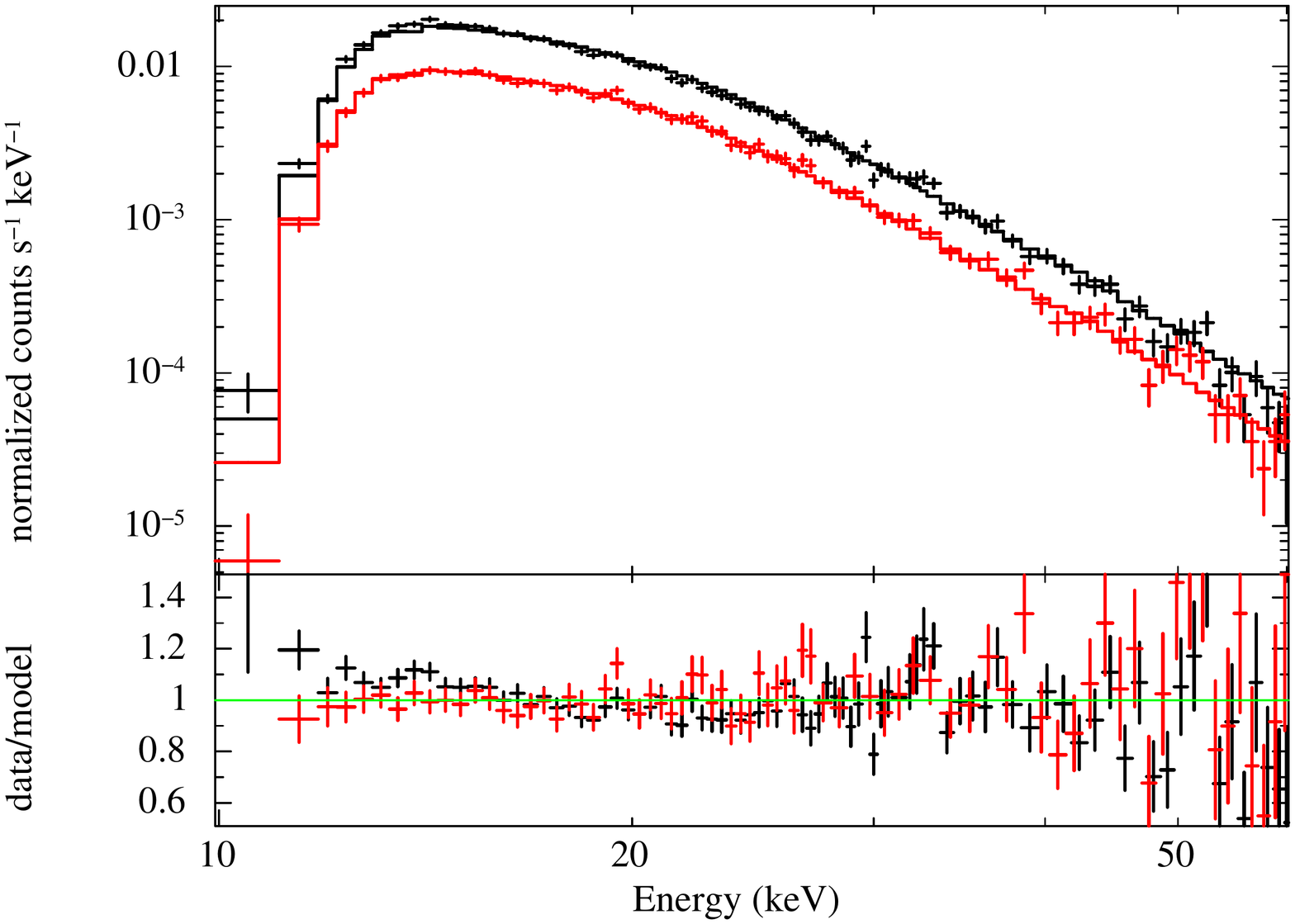}
}
\centerline{
\includegraphics[width=0.6\textwidth]{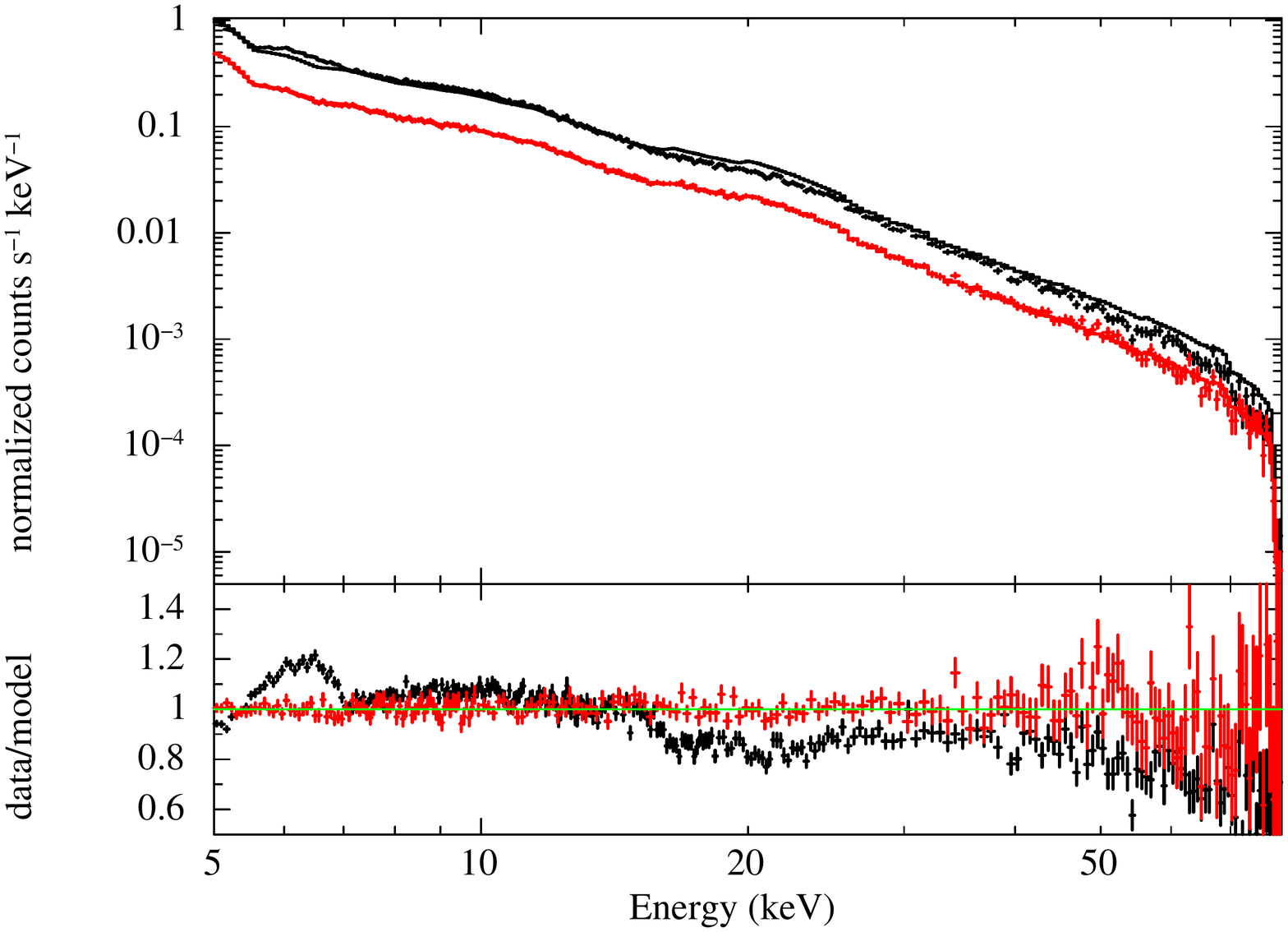}
}
\caption{\small{{\it Top:} The reflection-dominated model of Brenneman \&
    Reynolds (2006) (black
    points) and the absorption-only model of Miller \etal (2008; red points)
    for MCG--6-30-15 simulated with the {\it Suzaku}/PIN response for $150
    \ks$.  Data points are offset in amplitude for clarity.  The black line
    in the top panel represents the absorption model applied to the simulated reflection
    data while the red line shows the absorption model applied to the absorption
data.  The data-to-model ratio in the bottom panel shows that the two models are
not noticeably different to the eye; the residuals from both simulated datasets
are clustered about the green line (representing a perfect fit) in approximately
the same way.  $\chi^2/\nu=1.48$ for the joint fit.  {\it Bottom:} The same plot, but with the two
models simulated through the {\it NuSTAR} response for $150 \ks$.  In this case,
applying the absorption model to both simulated datasets results in a much more
divergent fit above $10 \keV$ both to the eye and statistically:
$\chi^2/\nu=3.55$ for the joint fit.}}
\label{fig:mcg6_refl_abs}
\end{figure}

\begin{figure}[hp]
\centerline{
\includegraphics[width=0.7\textwidth,angle=0]{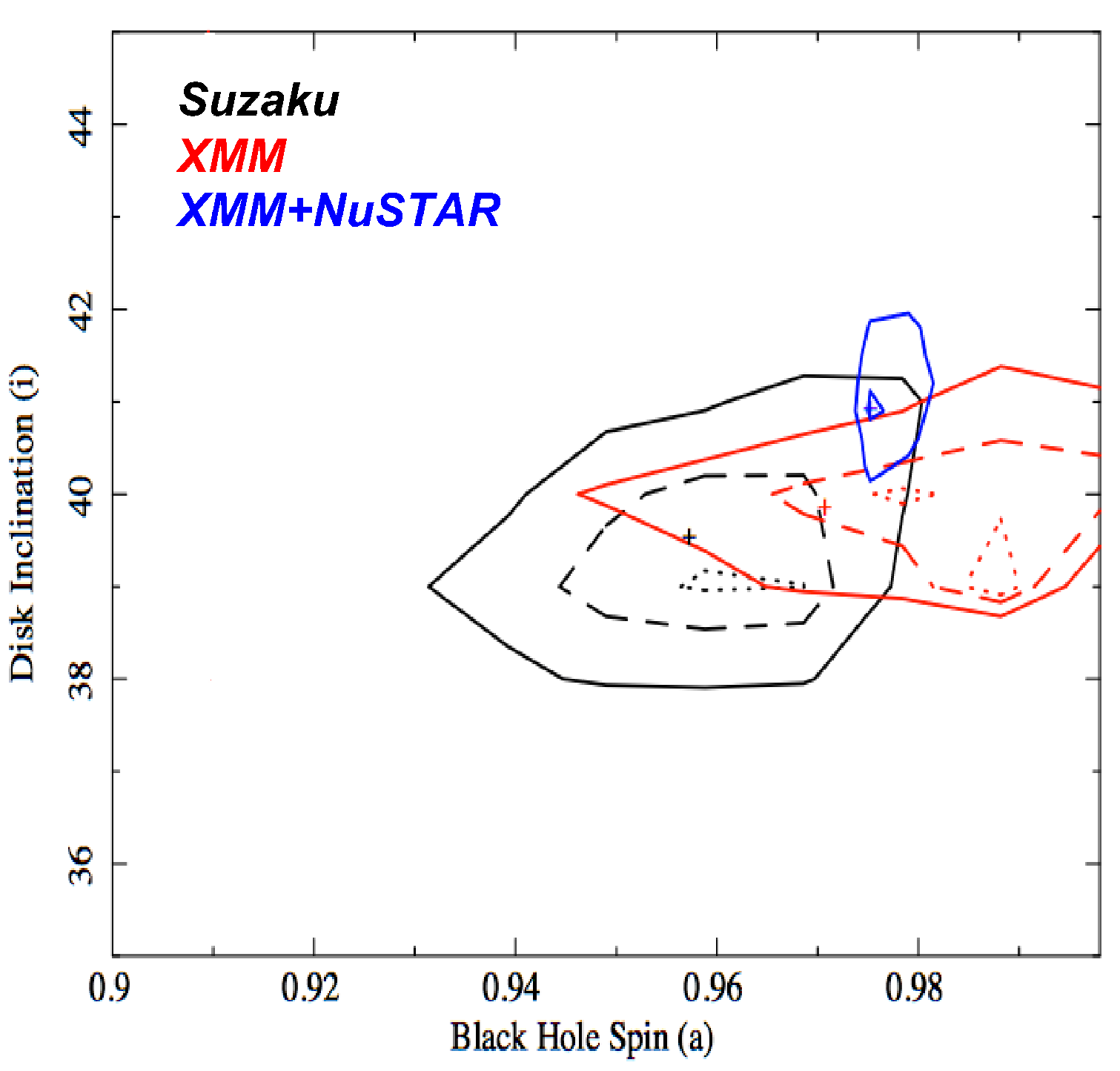}
}
\caption{\small{Contour plot of the constraints placed on black hole spin and disk
    inclination angle in MCG--6-30-15 with
    {\it Suzaku} (black), {\it XMM-Newton} (red) and {\it XMM+NuSTAR} (blue).
    Solid lines show the $67\%$ confidence region, dashed lines show $90\%$
    confidence, and dotted lines show $99\%$ confidence.  These contours are
    derived from simulations of the BR06 model run
    through the response of each detector, then refit with the model using a
    Markov Chain Monte Carlo algorithm in order to assess the confidence
    intervals for each parameter.  Simultaneous {\it NuSTAR/XMM-Newton} or {\it
    NuSTAR/Suzaku} data will improve the precision of the spin
    constraint by a factor of $\sim10$ for MCG--6-30-15, while also
    improving the accuracy of the measurement.}}
\label{fig:mcg6_spin}
\end{figure}

The recently launched {\it NuSTAR} telescope (Harrison \etal 2013) will provide
the best S/N above $10
\keV$ ever achieved owing to its large collecting area, its unique
focusing optics in this energy range, and its low background.
These capabilities will allow the differences between the
reflection-dominated and absorption-only models at higher energies to be constrained by the quality
of the data, enabling the correct model to be identified.  In a $150 \ks$
simulation of an observation of MCG--6-30-15, {\it NuSTAR} 
conclusively breaks the degeneracy between the two models, whereas the {\it
  Suzaku}/PIN instrument does not due to its higher background and lower
collecting area (see Fig.~\ref{fig:mcg6_refl_abs}).  
When used in tandem with instruments such as
{\it XMM-Newton} or {\it Suzaku}, {\it NuSTAR} data will also enable the most precise,
accurate constraints on black hole spin in AGN to be obtained by isolating the
reflection signatures from the other components in the spectrum more reliably
than ever before (see Fig.~\ref{fig:mcg6_spin}).  The launch of {\it Astro-H}
(Takahashi \etal 2010) in
2014 will further augment this
science by introducing the superior resolution of micro-calorimetry into the
broad-band spectrum, enabling the discrete features of the warm
absorber, soft excess and Fe K band emission lines to be conclusively identified
and modeled.  The contributions of {\it NuSTAR} and {\it Astro-H} to black hole spin
science will be further discussed in \S6.

\subsection{NGC~3783}
\label{sec:3783}

The type 1 AGN NGC~3783 ($z=0.0097$) was the subject of a $210 \ks$ {\it
  Suzaku} observation in 2009 as part of the {\it Suzaku} AGN Spin
  Survey Key Project (PI: C.~Reynolds, lead co-I: L.~Brenneman).  The
  source was observed with an average flux of $F_X=6.04 \times
  10^{-11} \ergpcmsqps$ from $2-10 \keV$
  during the observation, yielding a total of $\sim940,000$ photon counts over
  this energy range in the XIS instruments 
  the PIN instrument from $14-45 \keV$, 
 after background
  subtraction.  The results are reported in Brenneman \etal 2011 (hereafter B11).

\begin{figure}[hp]
\centerline{
\includegraphics[width=1.0\textwidth]{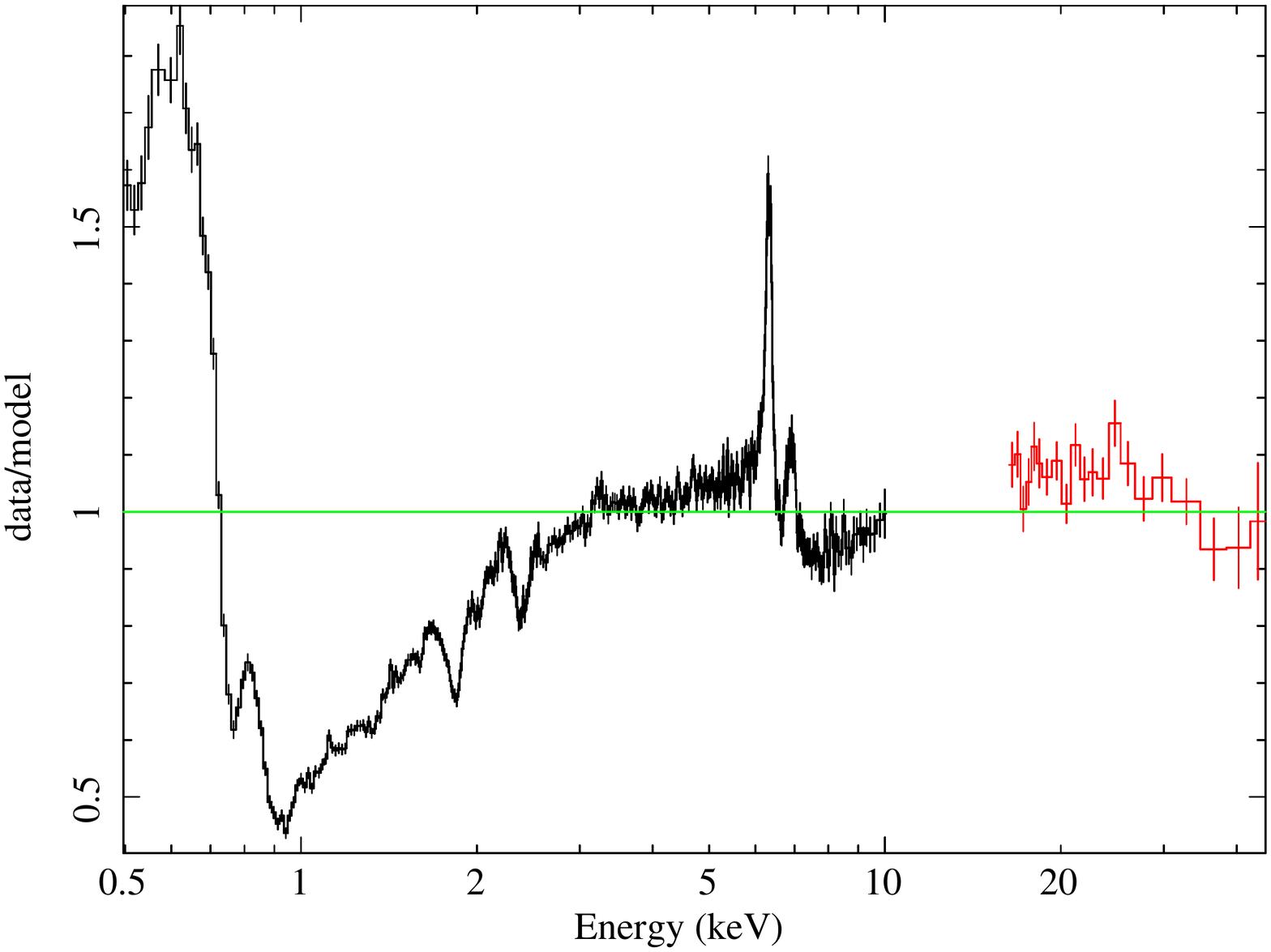}}
\caption{\small{{\it Suzaku} XIS-FI (front-illuminated; black crosses) and PIN (red crosses)
  data from the $210$ ks observation of NGC~3783 in 2009, ratioed
  against a simple power-law model for the continuum (fit over $2-4.5$
  and $7.2-10 \keV$) affected by
  Galactic photoabsorption.  Black and
  red solid lines connect the data points and do not represent a
  model.  The green line depicts a data-to-model ratio of unity.
  Data from the XIS back-illuminated CCD (XIS-BI) are not shown for
  clarity.  Figure is from Brenneman \etal 2011, \ApJ, 736, 103.
  Reproduced by permission of the AAS.}}
\label{fig:3783_po}
\end{figure}

\begin{figure}[hp]
\hspace{2cm}
\centerline{
\includegraphics[width=1.0\textwidth,trim=1.2cm 0cm 0cm 0cm,clip=true]{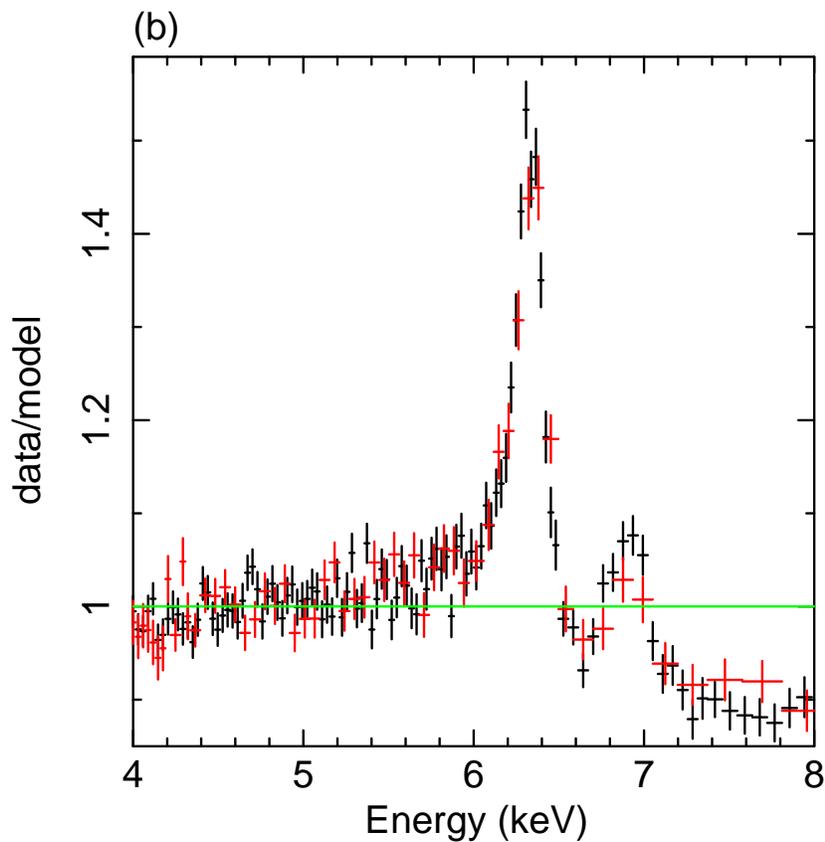}}
\caption{\small{A zoomed-in view of the Fe K region in the 2009 {\it Suzaku}
  observation of NGC~3783, ratioed against a simple power-law
  continuum.  Note the prominent narrow Fe K$\alpha$ emission line at
  $6.4 \keV$ and the blend of Fe K$\beta$ and Fe\,{\sc xxvi} at $\sim7
  \keV$.  XIS-FI is in black, XIS-BI in red.  Figure is from Brenneman \etal 2011, \ApJ, 736, 103.
  Reproduced by permission of the AAS.}}
\label{fig:zoom_narrow}
\end{figure}

\begin{figure}[hp]
\hspace{2cm}
\centerline{
\includegraphics[width=1.0\textwidth,trim=1.1cm 0cm 0cm 0cm,clip=true]{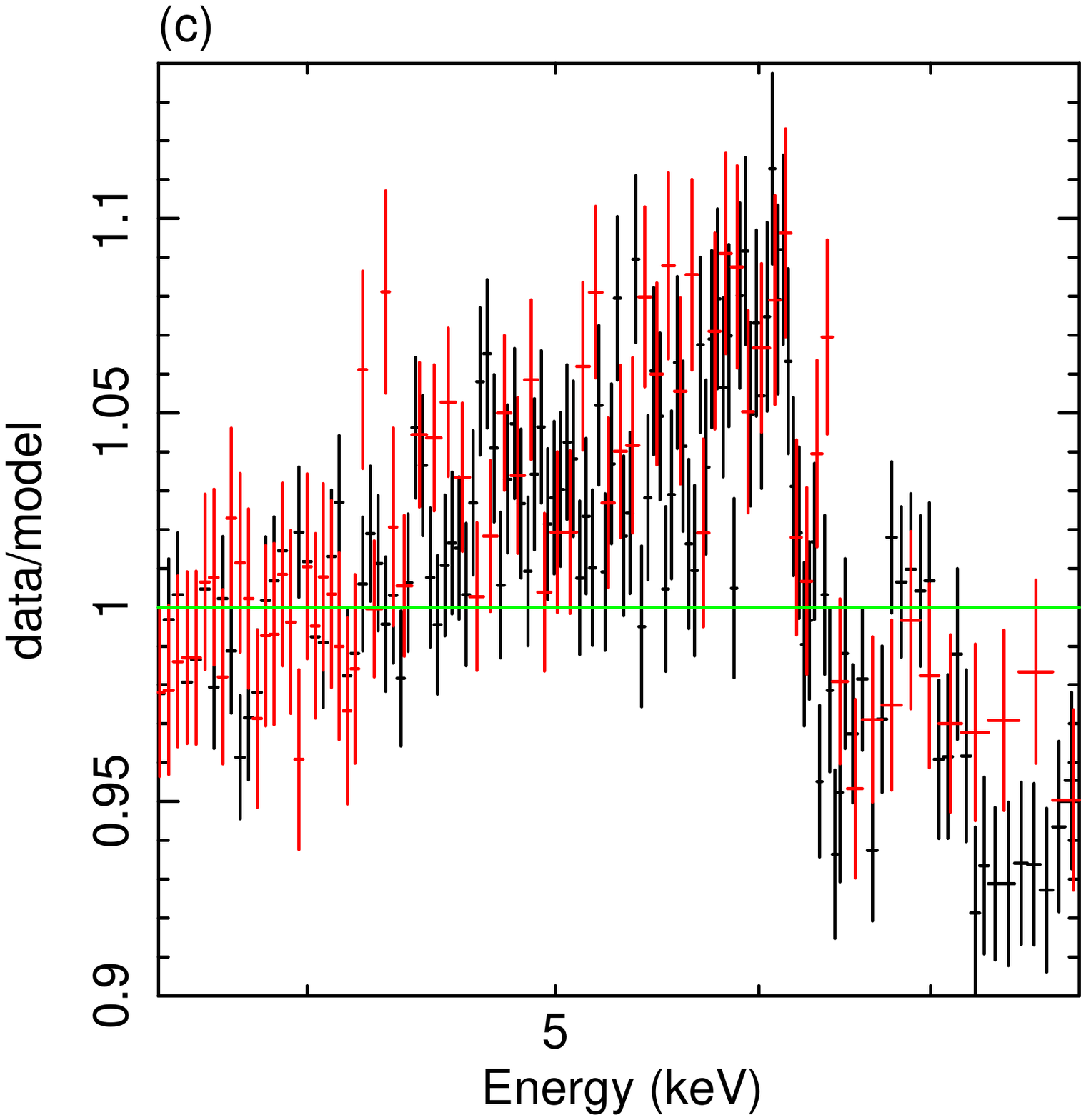}}
\caption{\small{The broad Fe K$\alpha$ line at $6.4 \keV$ becomes more obvious when the two
more prominent narrow emission lines are modeled out, in addition to
the power-law continuum.  Figure is from Brenneman \etal 2011, \ApJ, 736, 103.
  Reproduced by permission of the AAS.}}
\label{fig:zoom_broad}
\end{figure}

The spectrum ratioed against the power-law continuum is shown in
Fig.~\ref{fig:3783_po}.  The Compton hump is readily
apparent at energies $\geq 10 \keV$, though
its curvature is relatively subtle compared with more prominent
features of its kind (e.g., in MCG--6-30-15).  The $6-7 \keV$ band
the spectrum is dominated by narrow and broad Fe K features, including
a narrow Fe K$\alpha$ emission line at $6.4 \keV$ and a blend of Fe
K$\beta$ and Fe\,{\sc xxvi} emission at $\sim 7 \keV$.  The broad Fe
K$\alpha$ line manifests as an elongated, asymmetrical tail extending
redwards of the narrow Fe K$\alpha$ line to $\sim4-5 \keV$.  The Fe K region can
be seen in more detail in Figs.~\ref{fig:zoom_narrow}-\ref{fig:zoom_broad}.  At
energies below $\sim3 \keV$ the spectrum becomes concave due to the
presence of complex, ionized absorbing gas within the nucleus of the
galaxy; the gas is ionized enough that some contribution from this
absorber is seen at $\sim 6.7 \keV$ in an Fe\,{\sc xxv} absorption
line.  There is a rollover back to a convex shape below $\sim1 \keV$,
however, where the soft excess emission dominates.  

\begin{figure}[hp]
\hspace{2cm}
\centerline{
\includegraphics[width=1.5\textwidth]{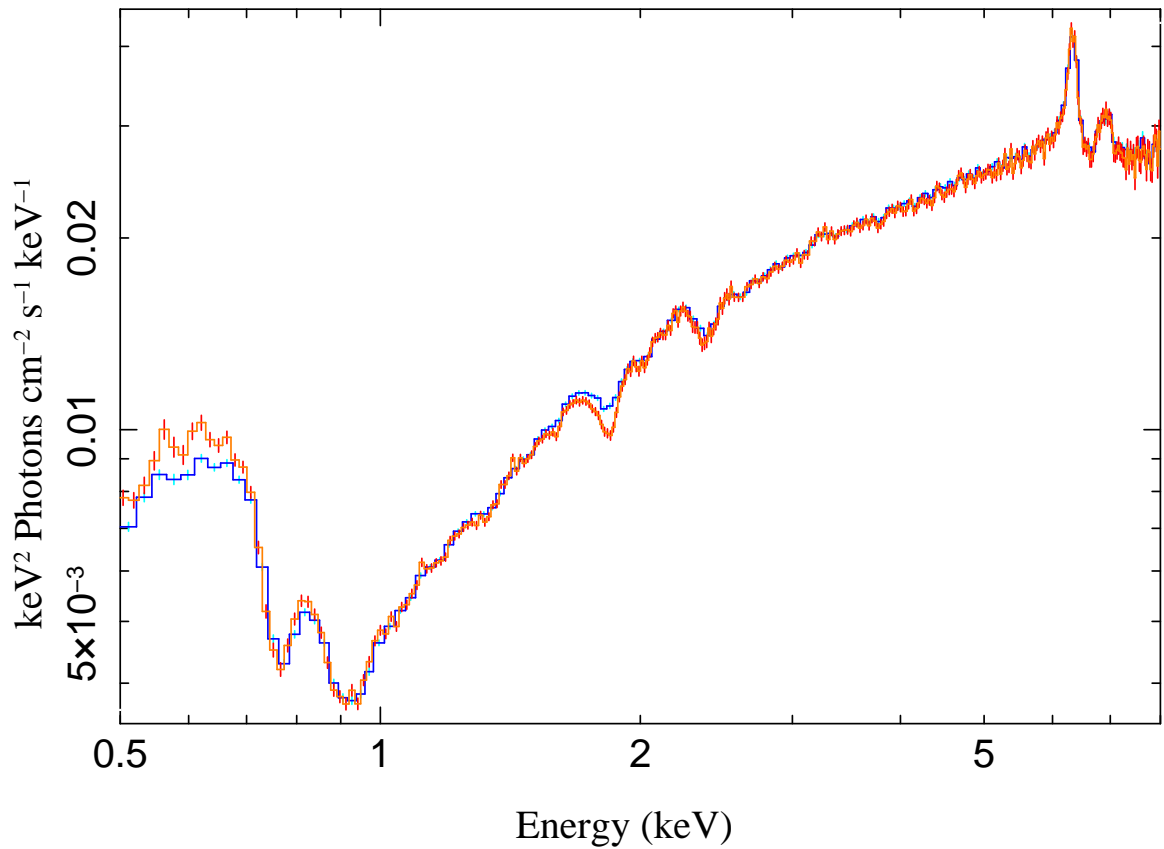}
}
\caption{\small{Plot of the best-fitting model from the 2001 {\it
      Chandra}/HETG data (red) on top of the 2009 {\it Suzaku}/XIS-FI
      data (blue)
      of NGC~3783.  No refitting has been performed, and the {\it
      Chandra} model has been folded through the {\it Suzaku}/XIS
      response.  Note the excellent agreement between the 2009 data
      and 2001 model, indicating that the warm absorber is in a very
      similar state in these two observations.  Figure is from Brenneman \etal 2011, \ApJ, 736, 103.
  Reproduced by permission of the AAS.}}
\label{fig:n3783_gratings}
\end{figure}

Brenneman \etal (2011) began their model fitting with the continuum power-law and Galactic
photoabsorption, then progressively added various model components to
represent the residual spectral features, provided that these added
components improved the fit statistically, according to the F-test.    
A 2001 {\it Chandra}/HETG observation of this AGN was used to inform the modeling of
the warm absorber, since {\it Suzaku}'s CCDs lack the resolution of
the {\it Chandra} gratings.  Though warm absorbers in AGN tend to vary
on timescales of $\sim$weeks-months (Krongold \etal 2005), the 2001 {\it Chandra} data were
a surprisingly good match for the 2009 {\it Suzaku} data in terms of
absorber appearance, enabling their use in this capacity (see Fig.~\ref{fig:n3783_gratings}).

\begin{figure}[hp]
\hspace{2cm}
\centerline{
\includegraphics[width=1.0\textwidth]{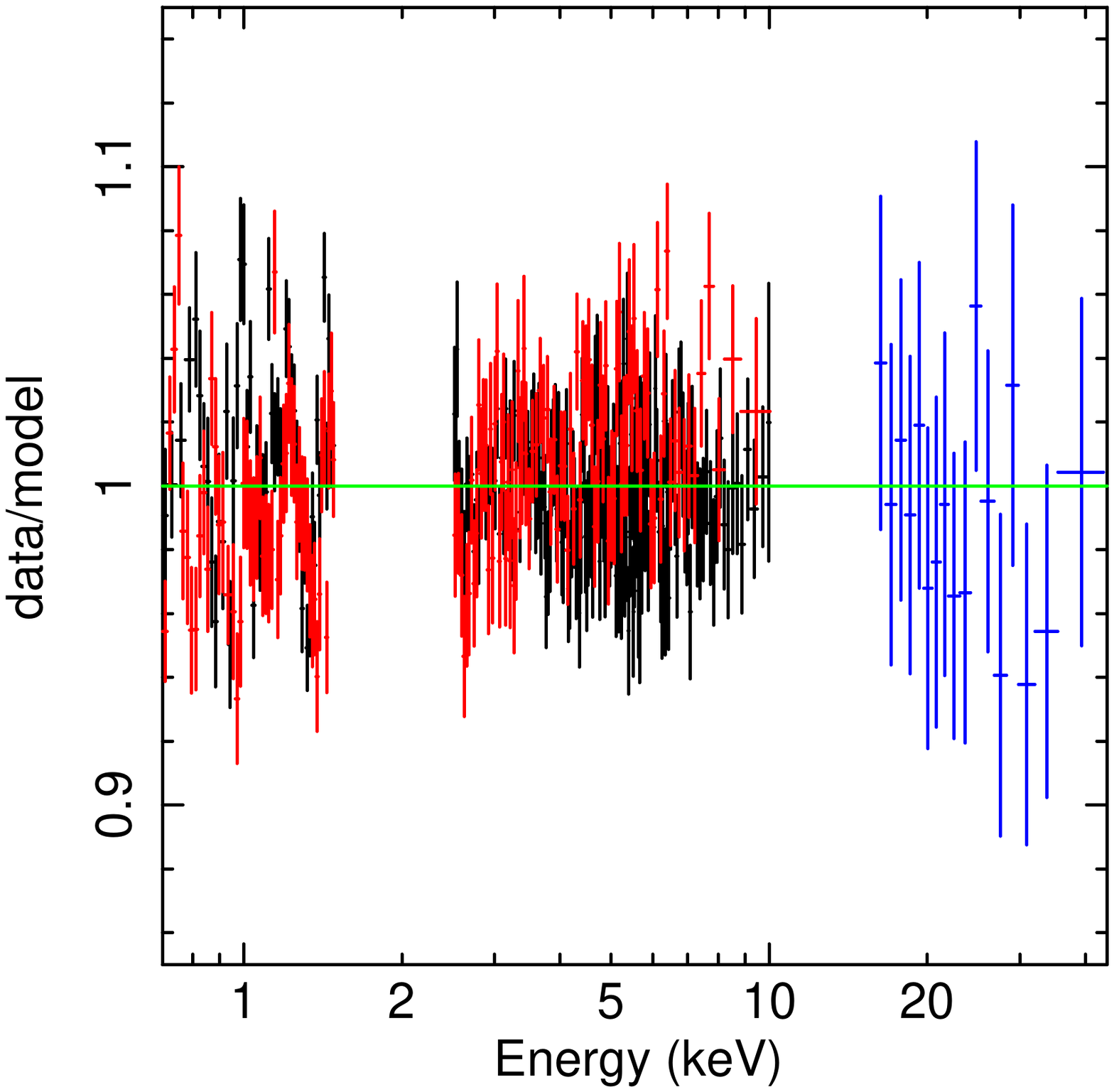}}
\caption{\small{Data/model ratio of the best-fitting model for the 2009 {\it
    Suzaku} observation of NGC~3783.  XIS-FI data are in black, XIS-BI
    data are in red and PIN data are in blue.  The green line
    represents a data/model ratio of unity.  Energies from $1.5-2.5
    \keV$ and $10-16 \keV$ are ignored due to calibration
    uncertainties.  Figure is from Brenneman \etal 2011, \ApJ, 736, 103.
  Reproduced by permission of the AAS.}}
\label{fig:ratio}
\end{figure}

\begin{figure}[hp]
\hspace{2cm}
\centerline{
\includegraphics[width=1.0\textwidth]{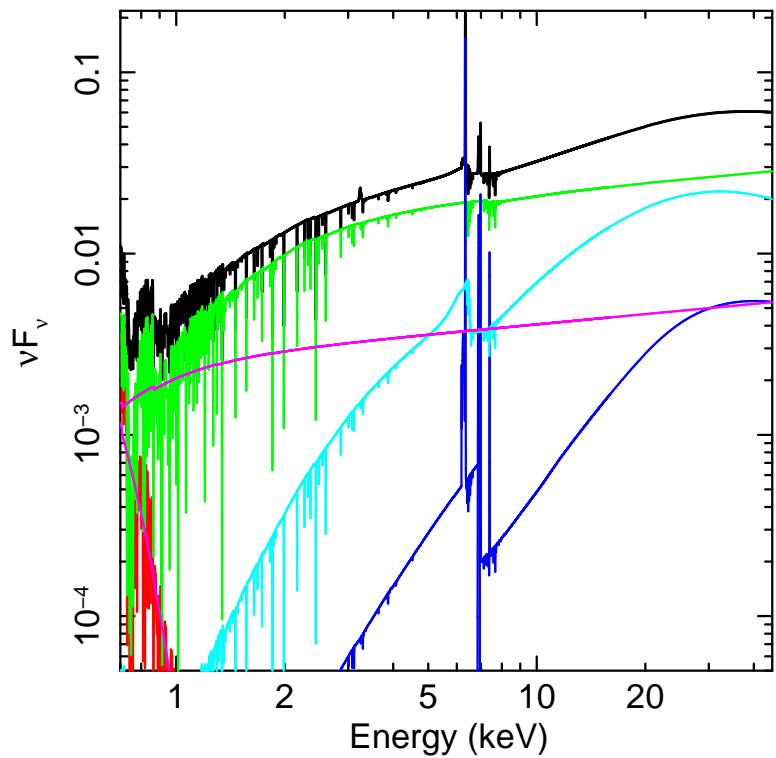}}
\caption{\small{The relative contributions of the various model components
  for the B11 best-fit to NGC~3783.  The black line represents the total
  model, the green shows the power-law continuum, red shows the
  blackbody soft excess, magenta shows the component of scattered
  emission, dark blue shows the distant reflection and light blue
  shows the inner disk reflection.  Figure is from Brenneman \etal 2011, \ApJ, 736, 103.
  Reproduced by permission of the AAS.}}
\label{fig:eemodel}
\end{figure}

Brenneman \etal (2011) used the models and methods described above in
\S3 to fit the
$0.7-45 \keV$ {\it Suzaku} spectrum of NGC~3783 with a statistical quality
of $\chi^2/\nu=917/664\,(1.38)$.  Most of the residuals
in the best-fit model manifested below $\sim3 \keV$ in the region dominated by
the warm absorber and soft excess, as is typical for type 1 AGN.
Because the S/N of the XIS detectors is highest at
lower energies due to the higher collecting area there, small residuals in the spectral
modeling of this region can have an exaggerated effect on the overall
goodness-of-fit.  Excluding energies below $3 \keV$ in
the fit, B11 achieved $\chi^2/\nu=499/527\,(0.95)$.  No significant
residuals remained.  
See Figs.~\ref{fig:ratio}-\ref{fig:eemodel} for the
best-fit data/model ratio and relative contributions of the various
model components.  The best-fit parameters of the black hole/inner disk system included a spin of $a
\geq +0.98$, a disk inclination angle of $i=22^{+3}_{-8}$$\degmark$ to
the line of sight, a disk iron abundance of ${\rm Fe/solar}=3.7 \pm
0.9$ and an ionization of $\xi \leq 8 \ergcmps$ (errors are quoted at
$90\%$ confidence for one interesting parameter).  These parameters
remained consistent, within errors, when energies $\leq 3 \keV$ were
ignored in the fit, negating the importance of the soft excess
emission in driving the fit to these parameter values.

\begin{figure}[hp]
\hspace{2cm}
\centerline{
\includegraphics[width=1.0\textwidth]{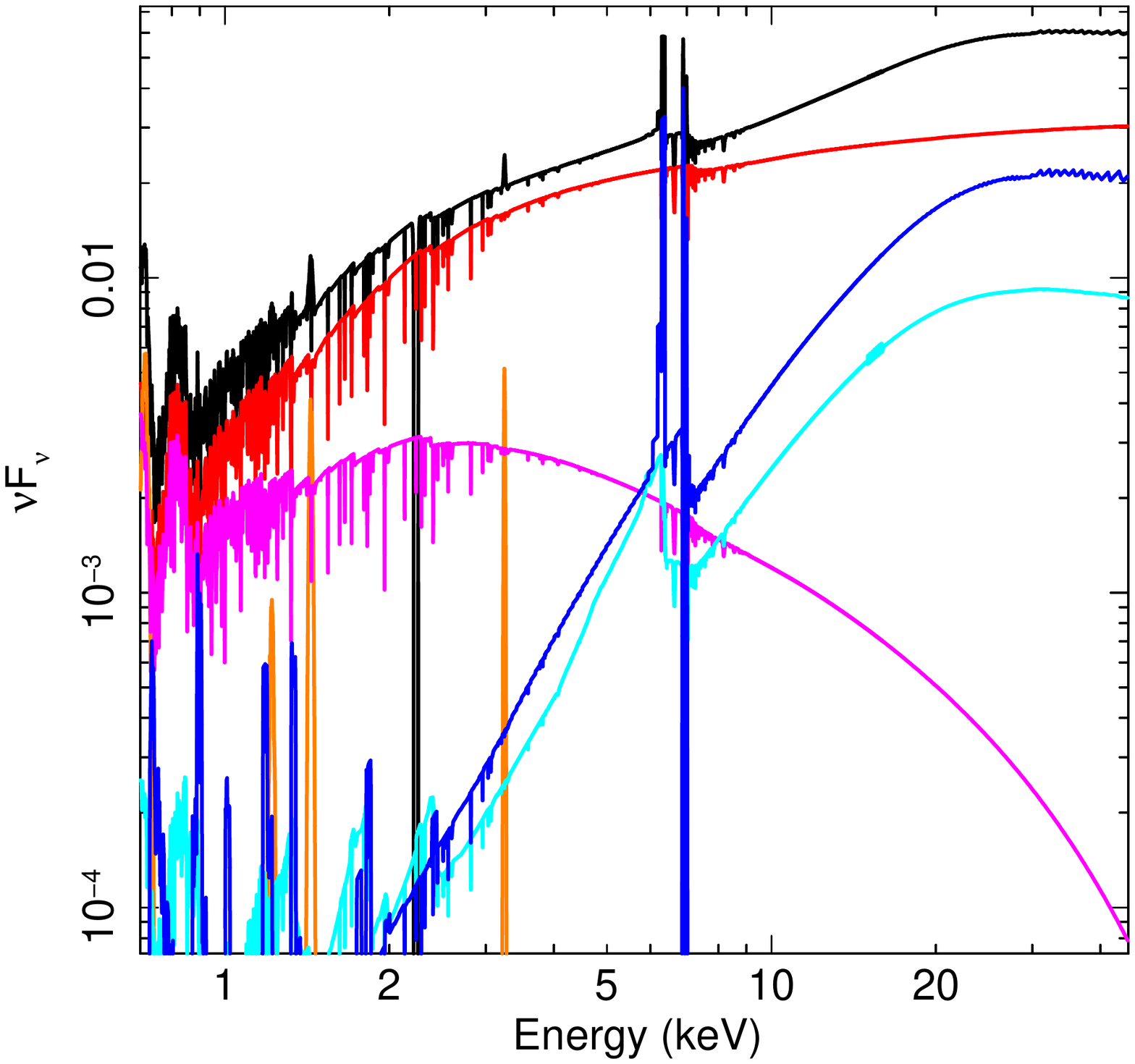}}
\caption{\small{The relative contributions of the various model components
  for the P11 best-fit to NGC~3783.  The black line represents the total
  model, the red line shows the power-law continuum, magenta shows the
  {\tt compTT} soft excess, dark blue shows the distant reflection and light blue
  shows the inner disk reflection.  Photoionized emission lines are in orange.
  Figure is created based on the
  best-fit ``dual reflector'' model presented in Patrick \etal
  2011, \MN, 416, 2725.}}
\label{fig:eemodel_patrick}
\end{figure} 

\begin{figure}[hp]
\vspace{-5cm}
\centerline{
\includegraphics[width=2.0\textwidth]{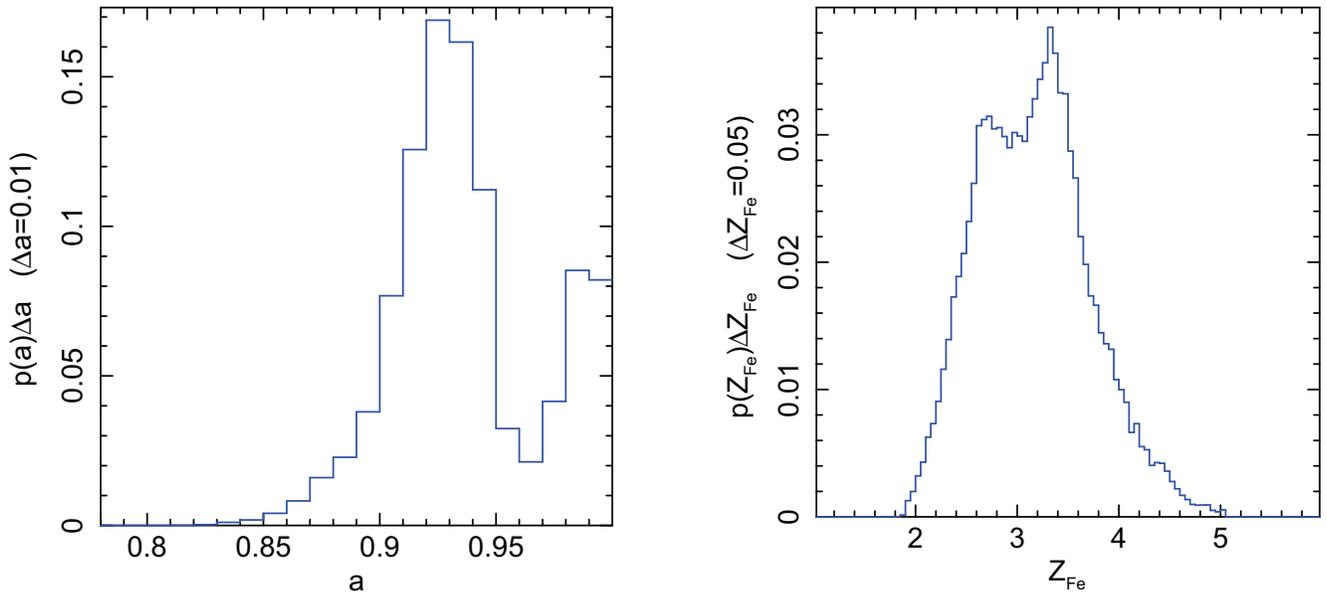}}
\vspace{-8cm}
\caption{\small{{\it Left:} The probability distribution of black hole spin
    for NGC~3783, based on an MCMC analysis (R12).  {\it Right:} The
    probability distribution for iron abundance in the inner disk,
    also from the MCMC analysis of R12.  Note the preference for high
    spin and iron abundance.  There is a positive correlation between
    these parameters; see R12 for a more complete discussion.  Figure
    is from Reynolds \etal 2012, \ApJ, 755, 88.  Reproduced by
    permission of the AAS.}}
\label{fig:mcmc}
\end{figure}

The results of B11 were corroborated by Reis \etal (2012), who examined
the temporal and spectral variability of NGC~3783 within the {\it
  Suzaku} observation, and by Reynolds \etal (2012, hereafter R12), who re-examined
the time-averaged data using a Markov Chain Monte Carlo (MCMC) analysis to
more closely probe the total available parameter space.  
These authors
especially noted the robustness of the rapid black hole spin and
super-solar iron abundance found by B11 (see Fig.~\ref{fig:mcmc}). 
The variability analysis of Reis \etal revealed that the spectrum is principally
composed of two elements: a variable soft component and a quasi-constant hard
component, similar in nature to that expected from reflection arising from the
inner parts of an accretion disk.  Further, difference spectra between different
flux states during the observation are all well fit by a simple power-law,
suggesting that the well-known warm absorber in this source is not variable
during the observation and
that the variability is due to changes in the power-law continuum flux.  An
excess of flux appears at energies $\geq 10 \keV$ in the later stages of the
observation.  This excess was shown to vary with time but not source flux, and
can be effectively accounted for by changes in the reflection strength and/or
ionization of the inner disk during the observation.

However, P11 analyzed the same data separately and reached
a strikingly different conclusion regarding the spin of the black hole in
NGC~3783: $a \leq 0.31$.  This discrepancy illustrates the importance
of assumptions and modeling choices in influencing the derived black hole
spin and other physical properties of the black hole/disk system.  P11 made three
critical assumptions that differed from B11: (1) that the iron abundance of the inner disk is 
fixed to the solar value; (2) that the warm absorber has a
high-turbulence ($v_{\rm turb}=1000$ km/s), high-ionization ($\xi \sim 7400 
\ergcmps$) component not reported by B11; (3) that the soft excess originates entirely through
Comptonization, with the Comptonizing medium at a temperature of $kT
\geq 9.5 \keV$ and an optical depth of $\tau=1.9 \pm 0.1$.  

Reynolds \etal (2012) demonstrated
that fixing the iron abundance at the solar value significantly
worsens the global goodness-of-fit in NGC~3783
when compared with allowing the iron abundance of the inner disk to fit
freely ($\Delta\chi^2=+36$).  B11 found no need to include a
high-turbulence component in their fit to the {\it Suzaku} data, and
noted no evidence for such a component in the higher-resolution 2001 {\it Chandra}/HETG
data.  Finally, R12 note that there is no physical reason to assume
that the soft excess originates entirely from Comptonization
processes, as other processes within the AGN might contribute (e.g.,
photoionized emission, scattering, thermal emission). 
R12 attempted several different model fits to the soft excess and found not only a much smaller
contribution to the overall model for the soft excess component than P11, but also no
statistical difference between fits using different models (e.g., blackbody
vs. {\tt compTT}).  It should be noted, however, that modeling the soft
excess with a Comptonization component of high temperature, high
optical depth and high flux, as P11 have done, requires
the {\tt compTT} component to possess significant curvature up into the
Fe K band, reducing the need for the relativistic
reflector to account for this same curvature seen in the data and thereby
eliminating the requirement of high black hole spin.  To illustrate
this, see
Fig.~\ref{fig:eemodel_patrick} for a plot of the relative importance
of the best-fit model components in the P11 analysis, as compared
with Fig.~\ref{fig:eemodel} for B11.  Clearly, different modeling
approaches can lead to vastly different conclusions regarding black hole spin and
careful consideration should be given to the models used and to their
allowed parameter ranges. 

\begin{figure}[hp]
\vspace{-10cm}
\centerline{
\includegraphics[width=3.0\textwidth]{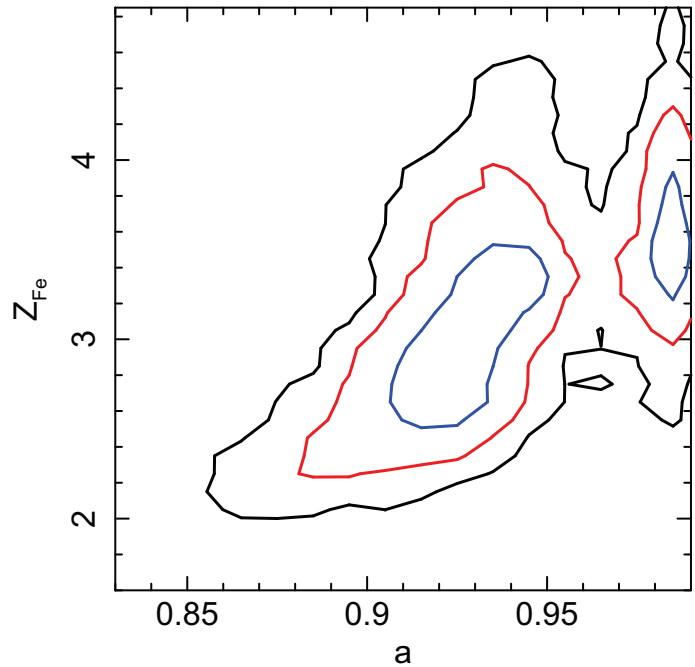}}
\vspace{-14cm}
\caption{\small{Probability density on the ($Z_{\rm Fe}$, $a$)-plane
    showing a positive correlation.
Contour levels are shown at p($Z_{\rm Fe}$, $a$) $= 1, 3.3, 10.0$
    (blue, red, black; i.e., inner, middle, outer lines),
    defined such that the probability of being in
the range $Z_{\rm Fe} \rightarrow Z_{\rm Fe} + \delta Z_{\rm Fe}$ and
$a \rightarrow a + \delta a$ is ${\rm p}(Z_{\rm Fe}, a) \delta Z_{\rm
  Fe} \delta a.$  Figure
    is from Reynolds \etal 2012, \ApJ, 755, 88.  Reproduced by
    permission of the AAS.}}
\label{fig:a-Z_correl}
\end{figure}

Degeneracies between model parameters can also be a factor
and should be carefully considered.  For example, R12 discuss the
positive correlation between black hole spin and iron abundance found through
their MCMC analysis of NGC~3783 (see Fig.~\ref{fig:a-Z_correl}).  Both have high values as
preferred in the best-fit model, and because increasing the amount of
iron in the disk will increase the strength of the reflection
features, a rapid spin is required in order to smooth those features
out enough to produce an adequate fit to the spectrum.  R12 note a worsening
of the fit when a fixed, solar iron abundance is adopted, however,
lending credence to the super-solar abundance measured. 

One possible explanation for the overabundance of iron detected in NGC~3783 and
other AGN (e.g., NGC 1365, Walton \etal 2010; 1H0707-495, Zoghbi \etal 2010) is
radiative levitation.  Previously discussed in the context of surface abundances
of white dwarfs (Chayer, Fontaine \& Wesemael 1995, Seaton 1996, Wassermann et
al. 2010), R12 applied the concept to AGN accretion disks for the first time.
Because disks are radiation-pressure dominated in their central regions and can
also possess fairly low ionization states of iron (Fe\,{\sc xvii} and below)
that have populated L- and M-shells, the radiation pressure acting on these
heavy metal ions can be much stronger than that acting on the surrounding,
fully-ionized plasma.  This outward force can also greatly exceed that of
gravity, causing the preferential upward drift of iron ions to the
disk surface, resulting in an enhancement of iron relative to other
elements in the disk atmosphere. 

\subsection{Fairall~9}
\label{sec:f9}

As we have seen in the cases of MCG--6-30-15 and NGC~3783, spectral complexities
like ionized
absorption intrinsic to the AGN can confuse our interpretation of the
spectrum.  The presence of such components can significantly affect
the derived physical parameters of the system, such as black hole spin.  It would
therefore be ideal to fit relativistic reflection models to a cleaner AGN system
without warm absorption as a kind of control case.

There exists a small sample of type 1 AGN, known as ``bare'' Seyferts, which seem
to lack any observable signatures of significant intrinsic absorption in
X-rays.  While most of these objects do display a soft excess, usually the flux
of this component is substantially smaller than that seen in NGC~3783, so the
exact model used to parametrize it will have a negligible effect on the
spectrum in the Fe K band and will not compromise the spin measured from the
broad Fe K$\alpha$ line and Compton hump.

Fairall~9 is one such bright, nearby ($z=0.047$), ``bare'' Seyfert with over
$160 \ks$ of data in the {\it XMM-Newton} archive and nearly $400 \ks$ in the
{\it Suzaku} archive.  Though a spectral analysis of the {\it XMM-Newton} data incorporating relativistic
reflection features was reported in Brenneman \& Reynolds (2009), the first
reported spin measurement for this AGN was published by Schmoll \etal (2009)
using a $167 \ks$ {\it Suzaku} observation from 2007.  The source had a flux of
$F_X=1.5 \times 10^{-11} \ergpcmsqps$ over $2-10 \keV$ at that time and was best fit using a
power-law continuum, distant reflection modeled with a {\tt pexrav} and narrow
Gaussians for the cores of the Fe K$\alpha$ and K$\beta$ lines, and ionized
inner disk reflection using a {\tt kerrconv} smearing algorithm convolved with
{\tt reflionx}.  The spin measured with this model was
$a=+0.65^{+0.05}_{-0.05}$, significantly less ($>6\sigma$) than the high spin values measured
for MCG--6-30-15 and NGC~3783, and perhaps indicative of a different galaxy and
SMBH evolution history in Fairall~9 than for the other two AGN considered thus
far.

In 2010, a deep {\it XMM-Newton} observation of Fairall~9 was obtained ($130 \ks$), and
from these data a weak spin constraint was established using the {\tt kerrconv}
model convolved with {\tt reflionx}: $a=+0.39^{+0.48}_{-0.30}$ (Emmanoulopoulos
\etal 2011).  Although formally consistent with the Schmoll \etal (2009) result
within errors, the disk inclination angle derived by these authors clashed
worryingly with that of Schmoll \etal: $i=(64^{+7}_{-9}) \degmark$ vs. $i=44
\pm 1 \degmark$.  

\begin{figure}[hp]
\centerline{
\includegraphics[width=0.7\textwidth,angle=270]{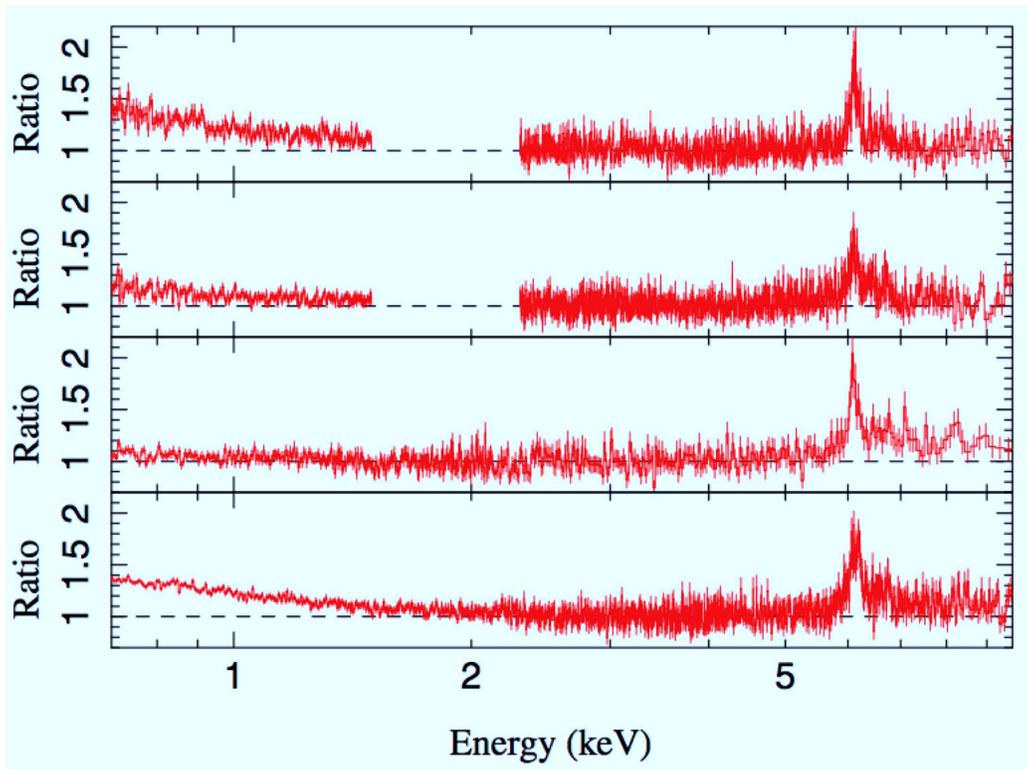}
}
\caption{\small{Data-to-model ratios of the Fairall~9 spectra to a simple
    power-law continuum modified by Galactic photoabsorption.  From top to
    bottom, the datasets represented are {\it Suzaku} 2007, {\it Suzaku} 2009,
    {\it XMM-Newton} 2009 and {\it XMM-Newton} 2000.  Figure is from Lohfink \etal
    2012, \ApJ, 758, 67.  Reproduced by permission of the AAS.}}
\label{fig:f9_ratio}
\end{figure}

\begin{figure}[hp]
\centerline{
\includegraphics[width=0.7\textwidth,angle=270]{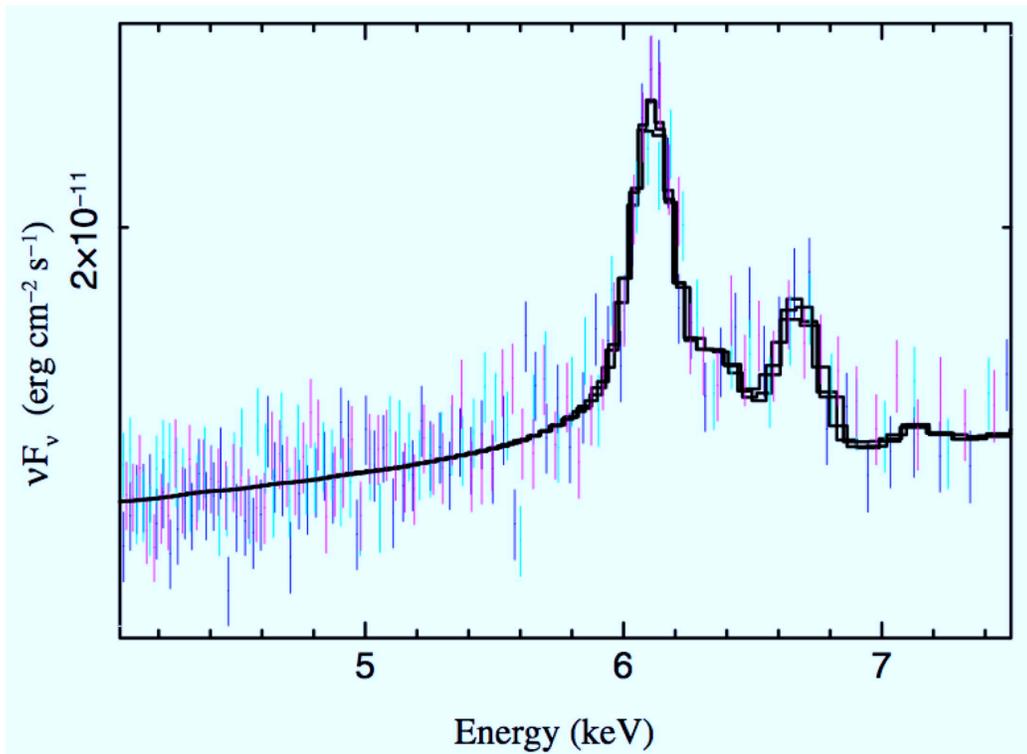}
}
\caption{\small{Unfolded spectrum (points) of the 2007 {\it Suzaku} data in the Fe K band
(using a diagonal response), fitted with the best-fitting reflection-only model
    of L12 (lines).  The data and best-fit models from all three XIS detectors
    are shown.  Figure is from Lohfink \etal
    2012, \ApJ, 758, 67.  Reproduced by permission of the AAS.}}
\label{fig:f9_FeK}
\end{figure}

A $229 \ks$ {\it Suzaku} observation of Fairall~9 was obtained in 2009 via the {\it
  Suzaku} AGN Spin Survey Project, and all four {\it XMM-Newton} and {\it Suzaku}
pointings have recently been analyzed jointly in Lohfink
\etal (2012; hereafter L12).  Both {\it Suzaku} pointings are also discussed in
P11.  By considering all four epochs of data, L12 note that the
  source has an average flux consistent with that of the 2007 {\it
  Suzaku} observation, but that the source varies
in flux by a factor of $\sim2$ over the $2-10 \keV$
  band during the 2010 {\it Suzaku} observation.  In
spite of the flux variation, the spectral shape remains very similar, with the
power-law and distant reflector evident, along with a broad
Fe K$\alpha$ line and a noticeable Compton reflection hump above $10 \keV$.
While most of the variation in the flux is evidently due to changes in the
power-law strength, a variable soft excess is also visible below $2 \keV$ (see
Fig.~\ref{fig:f9_ratio}).  L12 also note the presence of 
ionized emission lines
of Fe\,{\sc xxv} and Fe\,{\sc xxvi} in the 2009 {\it Suzaku} spectrum, which are
reported in the {\it XMM-Newton} observations (Brenneman \& Reynolds 2009) but not
robustly seen in the 2007
{\it Suzaku} pointing, according to Schmoll \etal (2009).  L12 do report these
features in the 2007 data, however; see
Fig.~\ref{fig:f9_FeK} for a close-up look at the Fe K region of this
observation.

\begin{figure}[hp]
\centerline{
\includegraphics[width=0.7\textwidth,angle=270]{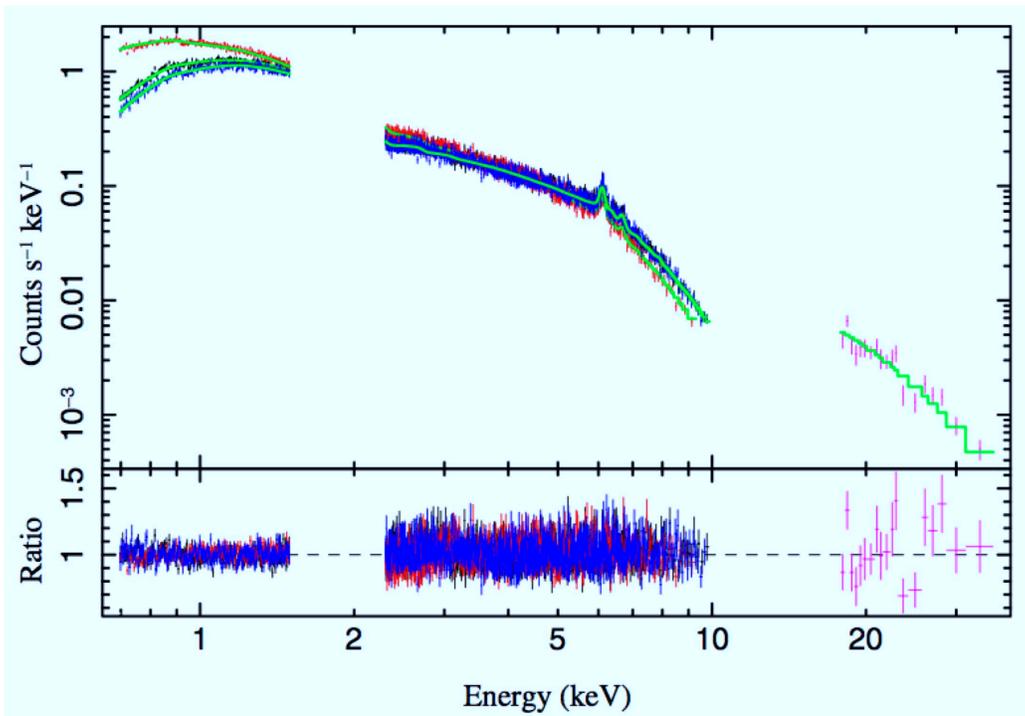}
}
\caption{\small{The 2007 {\it Suzaku} data fit with the reflection-only model of
L12.  XIS~0 data are the black points, XIS~1 data are in red, XIS~3 are in blue
and PIN data are in magenta.  The model is the solid green line.  The lower
panel depicts the data-to-model ratio, which is centered closely about unity.
Figure is from Lohfink \etal
    2012, \ApJ, 758, 67.  Reproduced by permission of the AAS.}}
\label{fig:f9_best}
\end{figure}

The best-fitting model obtained by L12 to the four datasets for Fairall~9, fit
jointly, requires the standard power-law continuum and near-constant distant
reflection (here modeled with a {\tt pexmon} component), plus reflection from an
inner accretion disk with sub-solar iron
abundance ${\rm Fe/solar}=0.67 \pm 0.08$, ranging in ionization from
$\xi=6^{+3}_{-4} \ergcmps$ (2007 {\it Suzaku}) to $\xi=1739^{+1143}_{-509}
\ergcmps$ (2009 {\it Suzaku}).  The inclination angle of the disk is measured at
$i=(37^{+4}_{-2}) \degmark$.  Additionally, a photoionized plasma is required
to explain the ionized iron emission lines seen in the spectrum; this plasma has
a loosely constrained ionization of $\xi \sim (0.02-10) \times 10^6 \ergcmps$.
The black hole spin is measured at $a=+0.71^{+0.08}_{-0.09}$, consistent with that
determined by Schmoll \etal (2009) from the 2007 {\it Suzaku} data alone.  This
model yields an excellent goodness-of-fit, with $\chi^2/\nu=5544/5276\,(1.06)$
(see Fig.~\ref{fig:f9_best}). 

There is still some controversy about the most physical model to use for
Fairall~9, however.  P11 consider the
{\it Suzaku} data from this source
with their dual reflector model, using a {\tt reflionx} component for both the
distant and relativistic reflection.  In so doing, they find that they must also
include a neutral absorber intrinsic to the source ($N_{\rm H}=4 \times 10^{23}
\pcmsq$) in order to remove the
contribution of the distant reflector from the soft excess emission and fit it
adequately with their {\tt compTT} component.  These authors also detect the
ionized emission from Fe\,{\sc xxv} and Fe\,{\sc xxvi}, choosing to parametrize
the lines with individual Gaussians rather than a photoionized emission model as
per L12.  There is no clear evidence to either support
or disprove the presence of the neutral absorber, which is not reported in any
other work on Fairall~9, but it is not necessary to include this component in order to
achieve a good fit: the $\chi^2/\nu=929/881\,(1.05)$ of P11 is comparable to
that of L12.  It should also be noted that P11 choose to fix the iron abundance
of Fairall~9 at ${\rm Fe/solar}=2$, contrary to their approach for the other
four AGN in their sample which have ${\rm Fe/solar}=1$.  

The {\tt compTT} soft excess used by P11 has a modest flux and optical depth,
with $F_{\rm 0.6-10}=(3.6 \pm
0.2) \times 10^{-11} \ergpcmsqps$ and $\tau=0.5^{+1.6}_{0.2}$, but the upper
limit on its temperature is quite high: $kT < 14.1 \keV$.  This large
temperature pushes the influence of this component almost into the Fe K band,
possibly interfering with the measurement of the red wing of the broad Fe
K$\alpha$ line by the {\tt kerrconv} inner disk reflection model.  Adopting
these assumptions, the authors cannot constrain the spin of the SMBH in
Fairall~9 using the dual reflector model, though they do achieve constraints
using a more phenomenological approach by modeling the broad Fe K$\alpha$ line
alone with a {\tt kerrdisk} component: $a=+0.67^{+0.10}_{-0.11}$.

\begin{figure}[hp]
\centerline{
\includegraphics[width=0.7\textwidth,angle=270]{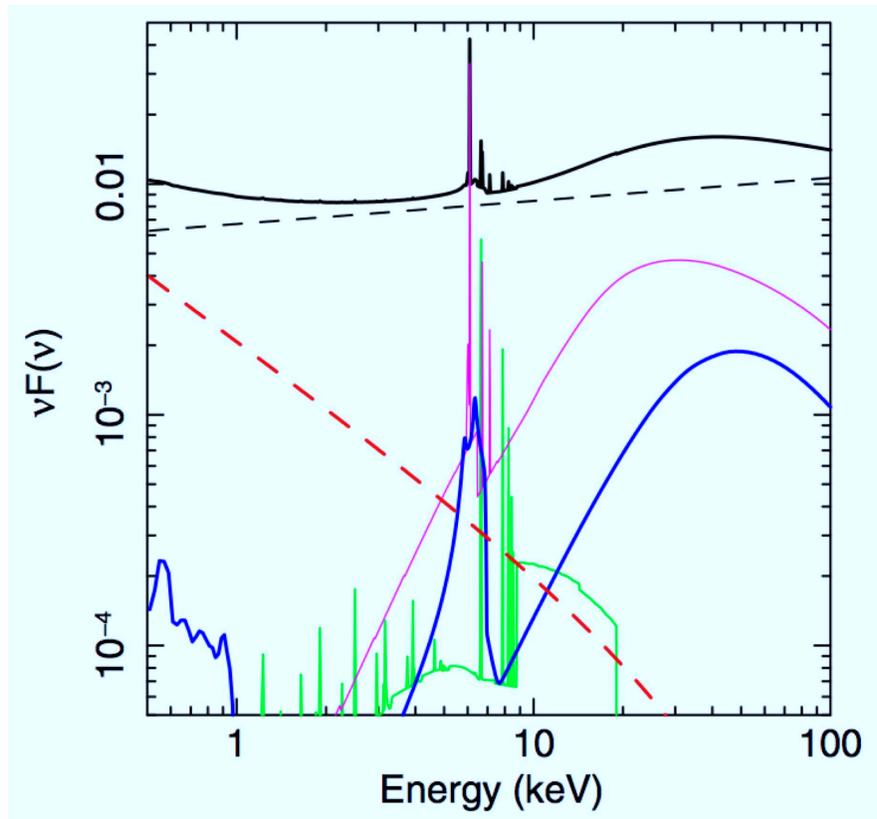}
}
\caption{\small{Best-fitting model components for the 2007 {\it Suzaku} data fit
by the L12 model including {\tt compTT} emission for the soft excess.  The total
model is the black solid line, the power-law is the black dashed line, the inner
disk reflector is in blue, the distant reflector in magenta, and the soft
Comptonization component in dashed red.  Figure is from Lohfink \etal
    2012, \ApJ, 758, 67.  Reproduced by permission of the AAS.}}
\label{fig:f9_eemo}
\end{figure}

A similar
approach was adopted by L12 to examine modeling degeneracies in the soft excess (see
Fig.~\ref{fig:f9_eemo}).  Comparing the L12 and P11 models fit to the {\it
  Suzaku} data, the addition of the {\tt compTT} component (here with comparable
optical depth, but higher temperature: $kT \sim 25 \keV$) drives the iron
abundance of the inner disk to a much higher value: ${\rm
  Fe/solar}=10^{+0}_{-2}$.  The ionization of the inner disk is also much
  smaller ($\xi \leq 70 \ergcmps$), since the inner disk no longer has to
  account for the soft excess emission on its own.  The inclination angle of the
  disk rises to $(48^{+6}_{-2}) \degmark$, while the spin of the black hole drops to
  $a=+0.52^{+0.19}_{-0.15}$. 

\begin{figure}[hp]
\centerline{
\includegraphics[width=0.75\textwidth,angle=0]{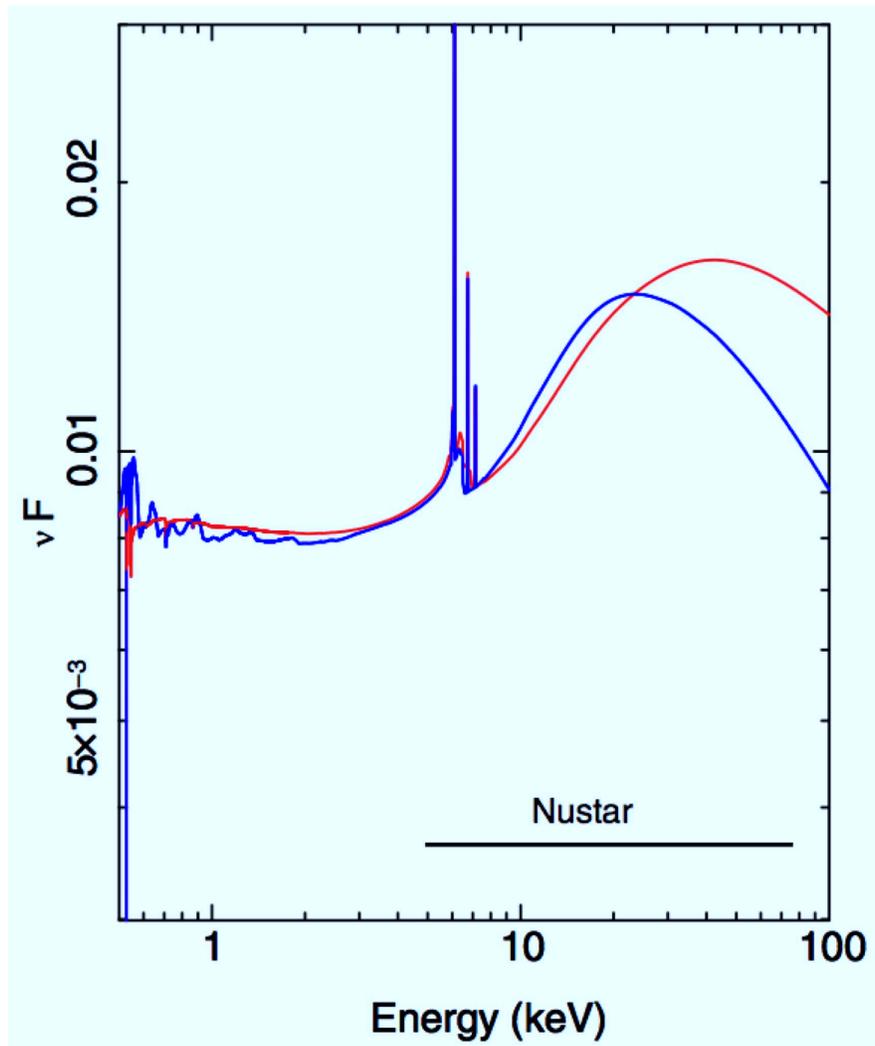}
}
\caption{\small{A comparison of the two best-fitting models to the 2007 {\it
      Suzaku} data for Fairall~9 presented in L12: the reflection-only model
    (blue; lower line above 10 keV) and the reflection plus Comptonization model
    (red; upper line above 10 keV).  Note the
    divergence of the two models above $10 \keV$, particularly.  This divergence
will be detectable with the high S/N spectra obtained from {\it
      NuSTAR}.  Figure is from Lohfink \etal
    2012, \ApJ, 758, 67.  Reproduced by permission of the AAS.}}
\label{fig:f9_nustar}
\end{figure}

In order to constrain the proper physical
components of the spectrum, we must await the high S/N achievable by
{\it NuSTAR}, which will be able to differentiate between the {\tt compTT}
(P11) and reflection-only (L12) models for the first time (see
Fig.~\ref{fig:f9_nustar}).  Given the differences between the various
modeling approaches used in L12 and
P11, however, it is somewhat surprising that the spin constraints achieved in each case
are consistent, within errors.  This could be an indication that the presence of
warm absorption is the greatest complicating factor in measuring spin, due to
the curvature it induces in the spectrum interfering with the isolation of the
red wing of the broad Fe K$\alpha$ line.  Alternatively (or perhaps in
addition to this point), the nature of the spin parameter space could
be playing a role in the similarity of the two measurements.  As seen
in Fig.~\ref{fig:isco}, the shape of the function relating spin to the
ISCO radius changes quite slowly in nearly linearly at moderate spin
values below $a \leq 0.9$, but changes much more rapidly above $a \geq
0.9$.  Therefore, differentiating between a spin of, e.g., $a=0.4$ and
$a=0.7$ is much more challenging, statistically, then differentiating
between spins of $a=0.9$ and $a=0.95$.  

This point was discussed at
some length in Walton \etal (2013; W13) as the authors analyzed the X-ray
spectra of 25 ``bare'' Seyfert AGN with {\it Suzaku}, specifically to
avoid the spectral complexities introduced by the presence of warm
absorbing gas along the line of sight to the nucleus.  In contrast to
Patrick \etal (2012; P12), in which a Comptonized soft excess is
assumed for the best fit to six bare
Seyferts, W13 do not employ a separate model component for the soft
excess, but rather allow it to be fit by the inner disk reflector.  Of
the five sources common to both the W13 and P12 samples, consistent
spin measurements (within errors) were found for Mrk~335 and Ark~120, but
not for Fairall~9, MCG--2-14-9 or NGC~7469.  The sixth source from P12,
SWIFT~J2127.4+5654, was not considered in the W13 sample, but the spin
derived in P12 is consistent with that measured by Miniutti \etal
(2009a), who employed a reflection modeling scheme for the soft excess
similar to W13.  The model used to parametrize the soft excess has a
noticeable impact on the derived spins in half of the small sample of
P12 bare Seyferts, then, underscoring the importance of establishing
the correct form of this spectral component.  It is also interesting
to note, however, that the three sources with consistent spin
measurements in spite of modeling differences all have spins of $a
\leq 0.85$, in keeping with the importance of the shape of the spin
parameter space.
 
\section{Measuring the SMBH Spin Distribution}

\subsection{Sources of Systematic Error}
\label{sec:systematics}

In previous sections we have noted the importance of both adequate data
(i.e., high S/N) and a physically self-consistent modeling approach to
constraining SMBH spins in AGN.  We have also stressed the importance of one very
critical assumption that must be made in order to calculate black hole spin: namely,
that the inner edge of the accretion disk truncates at the ISCO.  If the
optically-thick disk is
truncated further out, then any spin derived using this assumption and the
reflection modeling technique will be a lower limit.  Such truncated
disks may reside in radio-loud AGN; a disruption of the inner
accretion flow (using dips in the X-ray light curve as a proxy) is
thought to coincide roughly with the ejection of new
``knots'' of plasma into the radio jet (as observed in VLBI data).  This behavior has been noted
previous in, e.g., 3C~120 (Marscher \etal 2002) and 3C~111 (Kataoka \etal 2007).
By contrast, if there is significant emission
produced inside the ISCO, this
will lead to a systematic error on the black hole spin measurement
obtained via the reflection method that can be
$\geq20\%$ above the actual value of spin for non-spinning or retrograde black holes,
but is $\leq2\%$ higher than the real spin for black holes with
spins $a \geq +0.9$ (Reynolds \& Fabian 2008).  

The models currently used to represent both the accretion disk and the
relativistic smearing also have their inherent limitations and uncertainties.
The {\tt reflionx} and {\tt xillver} models both assume that the disk
is thin and can be well-approximated by a Novikov-Thorne formalism.
This is still an active topic of debate: disk thickness (e.g., Penna \etal 2010,
Noble \etal 2011) and disk
warping (e.g., Fragile
\& Anninos 2005) at small radii can have a substantial impact on measurements
of black hole spin. 
The current models also assume that the disk has a
constant density and ionization structure throughout, which cannot be the case,
physically.  
Portraying them as such is a necessary simplification,
computationally, and it is unclear whether even the highest-quality data can
differentiate between these simplified assumptions and more complex models that
have density and ionization varying as a function of radius and/or depth in the
disk.  

There is also some question about whether a
limb-brightening vs. limb-darkening algorithm should be used to represent the directionality of
the reflected emission from the disk when convolved with the smearing kernel
(Svoboda \etal 2010).  The nature of the disk emissivity profile itself is also an
active topic of research; though the disk is thought to dissipate energy as a
function of radius ($\epsilon \propto r^{-q}$), the emissivity index likely
varies as a function of radius as well (Wilkins \& Fabian 2011).  The
directionality of the coronal emission irradiating the disk also
impacts the observed reflection spectrum.  If the coronal photons
reflected back onto the disk are produced from a compact, localized
spot near the black hole spin axis and close to the disk surface,
light bending effects will focus the coronal emission preferentially
toward the center of the disk, resulting in an apparent enhancement of
the disk emissivity at small radii (corresponding to $q \geq 3$).  The degree to which
the emission is centrally concentrated in this scenario depends on the height $h$ of
the coronal active region above the disk (e.g., Miniutti \& Fabian
2004, Dauser \etal 2013).  We currently lack the ability to
characterize the physical properties of the corona in a given AGN,
however, which limits our ability to understand disk irradiation and
emissivity independently. 

Finally, when multiple detectors are involved in collecting the data
used to measure black hole spin, the cross-calibration uncertainty
between detectors can also contribute to the systematic error on the
spin constraint.  Given the need for high-S/N spectra across a wide
bandpass in X-rays, the use of multiple detectors is increasingly
necessary in order to achieve a reliable spin measurement.  For
example, B11 and W13 discuss how uncertainty in the cross-calibration
between the {\it Suzaku}/XIS and PIN instruments can affect
constraints on spin in NGC~3783 and a sample of 25 bare Seyfert AGN,
respectively.  The {\it NuSTAR} mission, too, is currently working to
improve the calibration between its two identical focal plane detectors, as well as its
cross-calibration with the instruments on both {\it XMM-Newton} and {\it
  Suzaku}.

\subsection{The Current Spin Sample}
\label{sec:sample}

\begin{figure}[hp]
\centerline{
\includegraphics[width=1.0\textwidth]{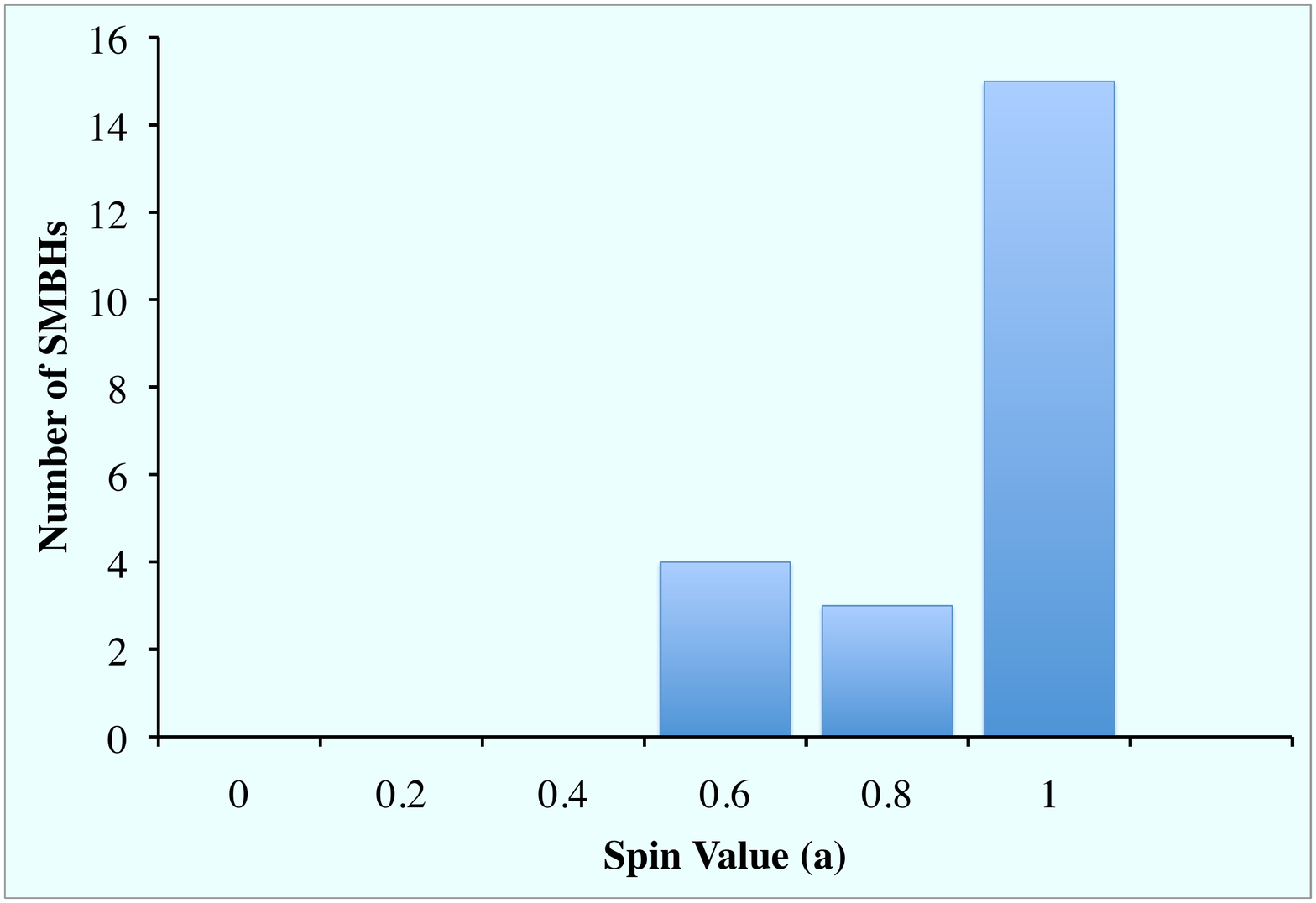}
}
\caption{\small{Distribution of the 20 SMBH spins measured so far, based on data from
    Table~\ref{tab:table2}.  Note the weighting toward large prograde spin values.}}
\label{fig:spin_dist}
\end{figure}

Taking all the caveats of \S\ref{sec:systematics} into account, one can begin to appreciate 
the challenge involved in obtaining precise, accurate spin constraints, and the
limitations of our sample size to bright, nearby AGN that are relatively
unobscured. 
For these reasons, there are currently only 22 AGN with
robust, published constraints for their
SMBH spins.  Here, I have defined a ``robust'' constraint in a manner similar to that of
Reynolds (2013), requiring that all other parameters of the accretion disk be
left free to vary during the fit (i.e., the disk inclination angle, iron
abundance and emissivity index, which must be itself constrained to $q \geq 2$
in order for the majority of the X-ray reflection to originate in the inner
disk).  The spins presented here meet these criteria and are single-valued.
These AGN, and
their properties, are listed in Table~1, and the histogram of spin values is
shown in Fig.~\ref{fig:spin_dist}.

{\small
\begin{table} 
\begin{tabular}{|ccccc|}\hline\hline
{\bf AGN} & {\bf a} & {\bf log M} & \textbf{$L_{\rm
    bol}/L_{\rm Edd}$} & {\bf Host}\\
\hline
MCG--6-30-15$^a$ & $\geq +0.98$ &
    $6.65^{+0.17}_{-0.17}$ & $0.40^{+0.13}_{-0.13}$ & E/S0 \\
\hline
Fairall~9$^b$ & $+0.52^{+0.19}_{-0.15}$ &
  $8.41^{+0.11}_{-0.11}$ & $0.05^{+0.01}_{-0.01}$ & Sc \\ 
\hline
SWIFT J2127.4+5654$^c$ & $+0.6^{+0.2}_{-0.2}$ & $7.18^{+0.07}_{-0.07}$ & $0.18^{+0.03}_{-0.03}$
    & --- \\ 
\hline
1~H0707--495$^d$ & $\geq +0.98$ & $6.70^{+0.40}_{-0.40}$ & $\sim 1.0_{-0.6}$ & --- \\
\hline
Mrk~79$^e$ & $+0.7^{+0.1}_{-0.1}$ &
$7.72^{+0.14}_{-0.14}$ & $0.05^{+0.01}_{-0.01}$ & SBb \\
\hline
Mrk~335$^f$ & $+0.70^{+0.12}_{-0.01}$ & $7.15^{+0.13}_{-0.13}$ & $0.25^{+0.07}_{-0.07}$
    & S0a \\
\hline
NGC~3783$^g$ & $\geq +0.98$ 
& $7.47^{+0.08}_{-0.08}$ & $0.06^{+0.01}_{-0.01}$ & SB(r)ab \\
\hline
Ark~120$^h$ & $+0.94^{+0.1}_{-0.1}$ &
$8.18^{+0.05}_{-0.05}$ & $0.04^{+0.01}_{-0.01}$ & Sb/pec \\
\hline
3C~120$^i$ & $\geq 0.95$ &
$7.74^{+0.20}_{-0.22}$ & $0.31^{+0.20}_{-0.19}$ &
    S0 \\
\hline
1~H0419--577$^j$ & $\geq +0.88$ & $8.18^{+0.12}_{-0.12}$ & $1.27^{+0.42}_{-0.42}$ & --- \\
\hline
Ark~564$^j$ & $+0.96^{+0.01}_{-0.06}$ & $\leq 6.90$ & $\geq 0.11$ & SB \\
\hline
Mrk~110$^j$ & $\geq +0.99$ & $7.40^{+0.09}_{-0.09}$ & $0.16^{+0.04}_{-0.04}$ & --- \\
\hline
SWIFT~J0501.9-3239$^j$ & $\geq +0.96$ & --- & --- & SB0/a(s) pec \\
\hline
Ton~S180$^j$ & $+0.91^{+0.02}_{-0.09}$ & $7.30^{+0.60}_{-0.40}$ & $2.15^{+3.21}_{-1.61}$ & --- \\
\hline
RBS~1124$^j$ & $\geq +0.98$ & $8.26$ & $0.15$ & --- \\
\hline
Mrk~359$^j$ & $+0.66^{+0.30}_{-0.54}$ & $6.04$ & $0.25$ & pec \\
\hline
Mrk~841$^j$ & $\geq +0.52$ & $7.90$ & $0.44$ & E \\
\hline
IRAS~13224-3809$^j$ & $\geq +0.995$ & $7.00$ & $0.71$ & --- \\
\hline
Mrk~1018$^j$ & $+0.58^{+0.36}_{-0.74}$ & $8.15$ & $0.01$ & S0 \\
\hline
IRAS~00521-7054$^l$ & $\geq +0.84$ & --- & --- & --- \\
\hline
NGC~4051$^m$ & $\geq +0.99$ & $6.28$ & $0.03$ & SAB(rs)bc \\
\hline
NGC~1365$^k$ & $+0.97^{+0.01}_{-0.04}$ & $6.60^{+1.40}_{-0.30}$ & $0.06^{+0.06}_{-0.04}$ & SB(s)b \\
\hline\hline 
\end{tabular}
\caption{\small{Summary of black hole spin measurements derived from
  relativistic reflection fitting of SMBH X-ray spectra.  All errors are
  quoted with $90\%$ confidence for one interesting parameter.  Data are taken
  with {\it Suzaku} except for 1H0707--495, which was observed with
  {\it XMM-Newton}; MCG--6-30-15, in which the data from {\it XMM-Newton}
  and {\it Suzaku} are consistent with each other; and NGC~1365, which
  was taken simultaneously with {\it XMM-Newton} and {\it NuSTAR}.  Spin ($a$) is
  dimensionless, as defined previously.  $M$ is the mass of the black hole in solar masses, and $L_{\rm
  bol}/L_{\rm Edd}$ is the Eddington ratio of its luminous output.
  Host denotes the galaxy host type. 
  All masses through 3C~120 are from Peterson \etal (2004) except MCG--6-30-15,
  1~H0707--495 and SWIFT~J2127.4+5654, which are taken from
  McHardy \etal (2005), Zoghbi \etal (2010) and Malizia \etal (2008),
  respectively.  All bolometric
  luminosities of these same objects are from Woo \& Urry (2002) except for the same three
  sources.  The same references for MCG--6-30-15 and SWIFT
  J2127.4+5654 are used, but host types for 1H0707--495 and SWIFT
  J2127.4+5654 are unknown.  Masses (bolometric luminosities) of the sources starting with
  1~H0419--577 are from, respectively: Fabian \etal 2005 (same),
  Collier \etal 2001 (Romano \etal 2004), Gliozzi \etal 2010 (same),
  Turner \etal 2002 (same), Miniutti \etal 2009b (Grupe \etal 2004), Zhou \&
  Wang 2005 (same), Zhou \& Wang 2005 (same),
  Kara \etal 2013 (Czerny \etal 2001), Bennert \etal 2011 (Woo \& Urry 2002), 
Zhou \& Wang 2005 (same), Risaliti \etal 2009b (Vasudevan
  \etal 2010).  References for each source are listed at the end of this work.}}
\label{tab:table2}
\end{table}
}

While it is difficult to draw any robust statistical inferences from a sample size of
22 objects, the trend toward higher spin values is obvious, even considering
that systematic error (which can lower measured spins as described above in
\S\ref{sec:systematics}) is not taken into account in the uncertainties quoted here.  There may be
selection biases in play which may make it more
likely that we measure higher spin values: AGN whose disks extend down
closer to the event horizon (i.e., those with large, prograde SMBH spins)
accrete more efficiently than those whose disks truncate farther from the SMBH,
provided that the disks in question conform to standard thin-disk profiles
(e.g., Novikov-Thorne). 
As such, an accreting, rapidly spinning black hole will be more luminous than an
accreting, slowing spinning black hole and hence will be over-represented in
flux-limited samples (B11).
The nature of the spin parameter space may also be playing a role
here, as discussed in W13: because of the rapid change in the shape of
the spin function vs. the ISCO radius at large prograde spin values,
it is easier to constrain spins with greater precision and accuracy
when they have spin values closer to $a=+1$.

Nonetheless, the pattern that is most readily apparent in Table~1 is that 15/22 AGN
have relatively high, prograde SMBH spins ($a \geq 0.8$), and no retrograde
spins have conclusively been measured (although the 90\% confidence lower bound
on the spin of the SMBH in of Mrk~1018 allows for retrograde spin).
Cowperthwaite \& Reynolds (2012) previously published a spin constraint of
$a \leq -0.1$ for 3C~120, but by taking multi-epoch, multi-wavelength data and the latest {\it
  Suzaku} calibrations into account, Lohfink \etal (2013) have revised this
measurement to $a \geq +0.95$.  

3C~120 is the one radio-loud galaxy with a
measured spin in the current sample, and is thus of great interest in terms of
probing the connection between black hole spin and jet production. 
Garofalo (2009) postulated that jet power is maximized for
rapidly-rotating retrograde black holes, though this idea is not without
controversy (e.g.,
Tchekhovskoy \& McKinney 2011).  More work needs to be done to constrain black hole spin and
jet power independently from observations in order to prove or disprove this
conjecture, and to place the rapid prograde spin measured for 3C~120 in context
with SMBH spins and jet luminosities for other radio-loud AGN.  It is worth
noting, however, that even the modest distribution in spin values seen in
Table~\ref{tab:table2} implies that black hole spin cannot be the primary driver
in determining whether an AGN possesses a relativistic jet. 

Narayan \& McClintock (2012) and Steiner \etal (2012) demonstrate two
examples of the beginnings
of such research in microquasars.  These authors report a correlation
between jet power and spin ($P_{\rm jet} \propto a^2$) in a sample of five stellar-mass black
holes, as expected based on the theoretical work of Blandford \& Znajek (1977).
There is some disagreement about this finding, however (e.g., see Fender,
Gallo \& Russell 2010), largely centered on how the jet
power is measured.  Daly (2011), meanwhile, has made a first effort at
measuring SMBH spins in 55 radio-loud AGN, finding a
distribution with an average close to $a=+0.5$, but with large
uncertainties on the individual spin values.  Precise measurements
of AGN jet magnetic fields are necessary in order to definitively
constrain the SMBH spins in these sources, however, and a $P_{\rm jet}
\propto a^2$ relation is assumed {\it a priori} in the work.  

\begin{figure}[H]
\vspace{-2.0cm}
\centerline{
\includegraphics[width=0.7\textwidth,angle=270,trim=3cm 0cm 3cm 0cm,clip=true]{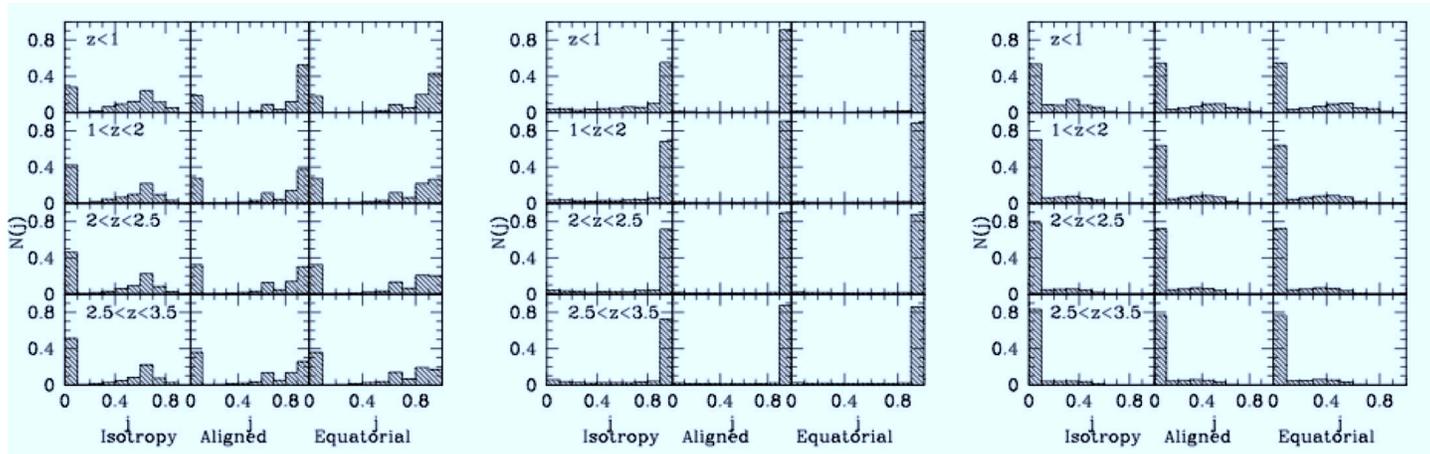}
}
\vspace{-2.0cm}
\caption{\small{Spin distribution as a function of redshift for the simulated
    SMBHs of Berti \& Volonteri (2008).  The left plot shows spin evolution
    driven by black hole mergers only, the middle plot shows mergers plus prolonged,
    prograde accretion, and the right plot shows mergers plus chaotic, random
    episodes of accretion.  Left-to-right columns in each plot show isotropic,
    aligned and equatorially-oriented mergers, respectively.  Figure is in Berti
\& Volonteri, 2008, \ApJ, 684, 822.  Reproduced by permission of the AAS.}}
\label{fig:BV08}
\end{figure}

If the trend toward large prograde spins continues to hold as our
sample size increases, we might ultimately infer that the growth of bright,
nearby AGN in recent epochs has been driven primarily by prolonged, prograde
accretion of gas.  If the overall distribution of SMBH spins in the local
universe begins to drift toward intermediate values, it is likely that the role
of mergers has been more significant than that of ordered gas accretion.
Similarly, if the distribution tends toward low values of spin, we can infer
that episodes of randomly-oriented accretion have been the dominant means of
SMBH and galaxy growth (Berti \& Volonteri 2008; see
Fig.~\ref{fig:BV08}).  Reynolds (2013) further note that both the most and least
massive SMBHs in Table~\ref{tab:table2} seem to have more moderate spin values
than their rapidly-spinning counterparts in the middle of the mass range.  If
this trend continues as the sample size of measured SMBH spins grows, it would
provide direct evidence for the increased role of chaotic accretion and/or major
mergers at these two extreme ends of the SMBH mass spectrum.

\section{Conclusions and Future Directions}
\label{sec:conclusions}

Measuring black hole spin is painstaking work, even with the best data from current
observatory-class missions such as {\it XMM-Newton}, {\it
  Suzaku} and {\it Chandra}.  Long observations ($\sim$hundreds of kiloseconds) of
bright AGN are needed, and multi-epoch, multi-instrument data
should be analyzed jointly whenever possible in order to assess the physical nature and
variability of all of the components in a given X-ray spectrum, knowing that the
spin value will not change over human timescales.  High
S/N across A broad energy
range is also desirable in order to constrain the properties of the continuum
and complex absorption, particularly, and to distinguish these components from
any signatures of inner disk reflection.  Only by isolating the broad Fe
K$\alpha$ line and its associated Compton hump can we measure black hole spin with the
accuracy and precision necessary to begin constructing a spin distribution for
local AGN.  We can then begin to draw inferences
regarding the dominant growth mechanism of these SMBHs over cosmic time, and to
understand the role of spin in jet production and AGN feedback.

Our current sample of 22 AGN with measured, published SMBH spins must be
extended in order to accomplish these goals.  The {\it Suzaku} Spin Survey has
recently been completed, and is providing rich legacy datasets that will benefit
this science for years to come.  Additionally, many datasets from the {\it XMM-Newton}
and {\it Suzaku} archives have recently been analyzed with an eye toward
measuring spin (e.g., P11, P12, W13).
{\it NuSTAR} will also play a vital role in this science, providing an
invaluable high-energy ($\sim3-79 \keV$)
complement to {\it XMM-Newton} and {\it Suzaku} spectra, particularly, when used simultaneously with
either observatory.  This high-energy capability will improve the accuracy of black hole
spin measurements, and will also enable improvements in precision in these
measurements by up to a factor of ten in some sources (e.g., Fig.~\ref{fig:mcg6_spin}).

{\it Astro-H},
scheduled for launch in 2015, will bring the science of
micro-calorimetry to X-ray astronomy with a spectral resolution of $\Delta E \sim 7 \eV$ over the
$0.3-12 \keV$ range.  Though the observatory will also fly a high-energy
detector capable of producing spectra up to $600 \keV$, the calorimeter will be
the unique strength of
this mission, enabling the broad and narrow Fe K emission and
absorption features to be definitively disentangled and the telltale signatures
of complex intrinsic absorption to be identified and modeled
correctly.  

In order to achieve the order of magnitude increase in sample size necessary to
begin assessing the spin distribution of SMBHs in the local universe from a
statistical perspective, future large-area ($\geq 1 \m^2$) X-ray observatories are
needed.  Proposed concepts such as {\it IXO/AXSIO} (White \etal 2010), {\it ATHENA}
(Barcons \etal 2012) and the {\it Extreme Physics Explorer (EPE)} (Garcia et
al. 2011) would all offer the necessary collecting area and superior
spectral resolution, allowing us to
extend our sample of measured SMBH spins to several hundred AGN using the
reflection modeling method.  

Additionally, such large-area observatories will also enable
the orbits of distinct ``blobs'' or ``hot spots'' of material within the
accretion disk to be measured via the periodicity of their emission, allowing
velocity to be charted as a function of
radius within the disk for tens of AGN.  Such measurements would provide an
independent check on the spin value
obtained from spectral fitting of the inner disk reflection signatures
averaged over many orbits, and would also yield important constraints on black hole
masses as well.  The Large Observatory For Timing ({\it LOFT}; Feroci
\etal 2012) is a proposed concept that, if funded, would provide the necessary effective area
($\geq 10\,{\rm m}^2$) to achieve this goal, coupled with moderate spectral
resolution ($\leq 260 \eV$) across a reasonably broad bandpass ($2-30 \keV$).
As discussed in Reynolds (2013), {\it LOFT} would also revolutionize the science of relativistic reverbertion
mapping, in which the lag time is measured between variations in the continuum emission from
the corona and variations in the response of the observed X-ray reflection
from the inner accretion disk.  The full, energy-dependent transfer function relating the changes
in these two X-ray spectral signatures encodes the spin of the black hole, among
other physical information about the inner accretion flow (Zoghbi \& Fabian
2011, de Marco \etal 2011, Fabian \etal 2012b, Kara \etal 2013).  Having an
instrument such as {\it LOFT} at our disposal would thus provide two more
methods to use in determining black hole spins.

The science of measuring the angular momenta of black holes is in its infancy.  Though the past
decade has seen great strides in our ability to constrain spin through long
X-ray observations coupled with detailed spectral
modeling, much work remains to be done in terms of improving the precision and
accuracy of these measurements, as well as the sample size.
The next decade will see an improvement in the
quality of black hole spin science, but a significant advance in the quantity of this
work in the decades beyond will depend critically on the amount of 
funding available to facilitate the international collaborations
necessary to build large-area X-ray spectroscopy missions, or on advances in
technology development that will allow such a large-area X-ray spectroscopic
mission to be flown for a fraction of the current cost.

\section{Epilogue: {\em NuSTAR} Validates Inner Disk
  Reflection in NGC~1365}
\label{sec:epilogue}

At the time of this writing, the {\it NuSTAR} X-ray observatory is undertaking
an ambitious campaign to observe several AGN simultaneously with either {\it
  XMM-Newton} or {\it Suzaku}.  These deep observations will yield the highest S/N spectra
from $0.2-79 \keV$ ever obtained, enabling the continuum, absorption and
reflection components of these AGN to be unambiguously disentangled.
As discussed in \S\ref{sec:mcg6}, deconvolving these spectral features will
allow black hole spin to be measured with greater precision and accuracy than
has ever been achieved in previous work.
 
The Seyfert 1.8 AGN NGC~1365 is the only AGN known to display, in addition to the
near-ubiquitous continuum and reflection from distant material, (1) extended
X-ray emission from a circumnuclear
starburst (Wang \etal 2009), (2) relativistic inner
disk reflection (Risaliti \etal 2009a, Walton \etal 2010, Brenneman \etal 2013),
(3) a warm absorber
(Risaliti \etal 2005b, Brenneman \etal 2013), and (4) a time-variable cold absorber that eclipses
the inner disk/corona (Risaliti \etal 2005a, Maiolino \etal 2010, Brenneman
\etal 2013).  It has been
the subject of over a dozen X-ray observations with
{\it XMM-Newton, Chandra} and {\it Suzaku} during the past decade. 
Most recently, Brenneman \etal (2013) jointly analyzed {\it Suzaku} spectra from three
different observations over a two-year period in order to maximize S/N in an
effort to separate the various spectral components.  A preliminary spin
constraint of $a=0.96 \pm 0.01$ was obtained using a {\tt relconv(reflionx)}
model for the inner disk reflection.  The limited S/N of the {\it Suzaku} data
above $10 \keV$ made it difficult to uniquely establish relativistic
reflection as the best-fitting model, however (e.g., vs. multiple complex
absorbers), calling into question the ability of the
data to truly constrain spin.

NGC~1365 was one of the first AGN observed by {\it NuSTAR} as
part of its science operations phase, and has now been the subject of four
separate observations taken simultaneously with {\it XMM-Newton}.  These observations
were taken in July and December 2012, and in January and February 2013, and
total nearly $500 \ks$ of simultaneous data from the two telescopes.  The
data from all four observations are currently being analyzed, and the first
results from spectral fitting of the July 2012 observation have now been
published (Risaliti \etal 2013).  

\begin{figure}[H]
\centerline{
\includegraphics[width=0.9\textwidth,angle=0]{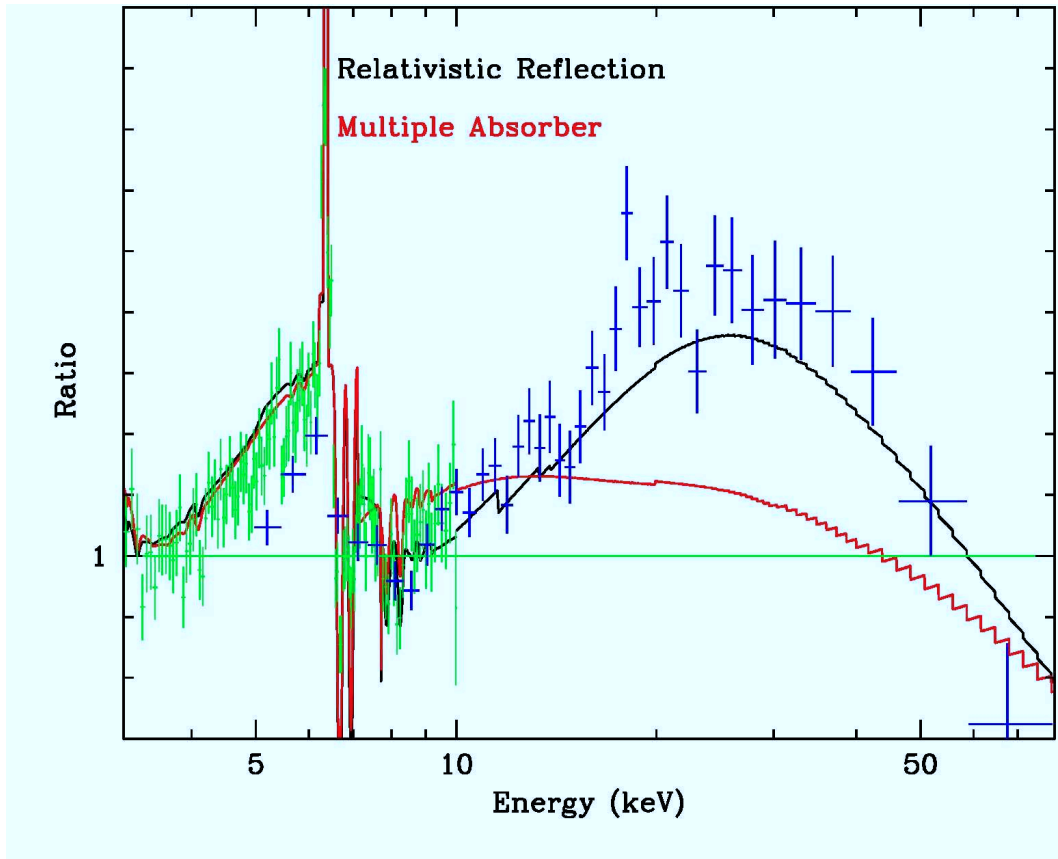}
}
\vspace{-4.0cm}
\caption{\small{The relativistic inner disk reflection model (black line)
    vs. the absorption-only model (red line) plotted against the {\it XMM-Newton}
    (green) and {\it NuSTAR} (blue) spectral data.  Both models are fit only
    below $10 \keV$ and extrapolated above this energy.  Note the strong
    resemblance of the data to the reflection model and divergence from the
    absorption model at high energies.  Both models fit the data equally well
    below $10 \keV$.  Credit: G. Risaliti, private communication.}}
\label{fig:n1365_refl_abs}
\end{figure}

\begin{figure}[H]
\centerline{
\includegraphics[width=1.0\textwidth,angle=0]{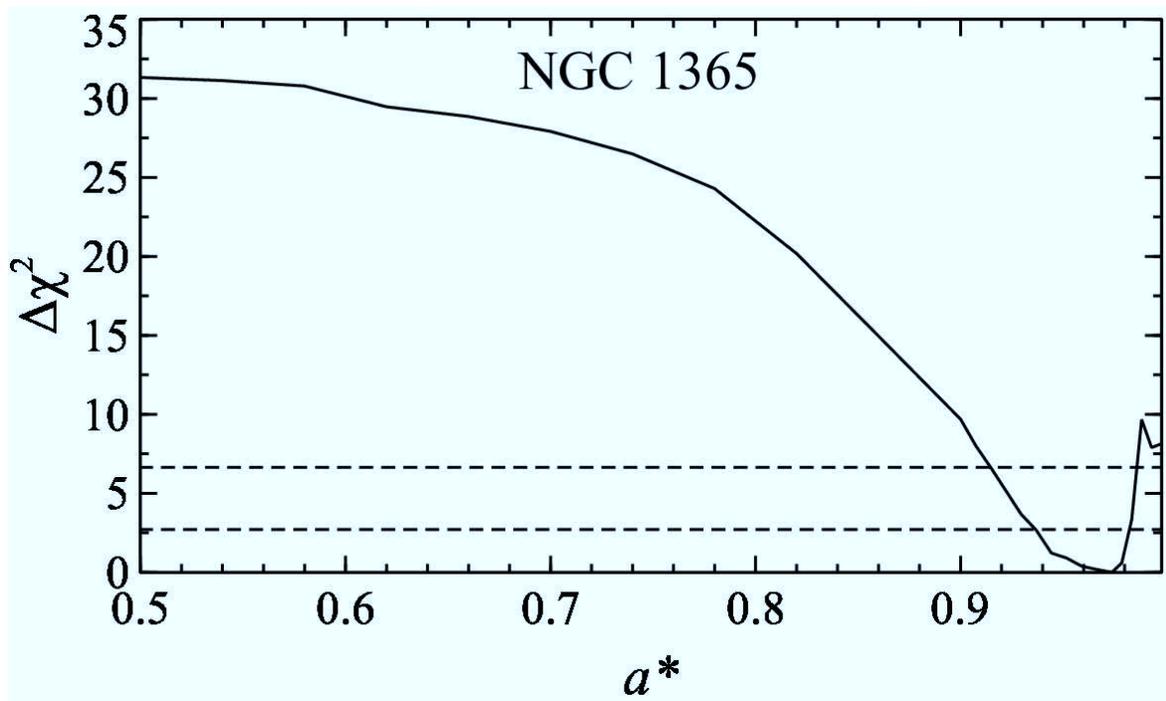}
}
\vspace{-4.0cm}
\caption{\small{Change in the global goodness-of-fit vs. spin for the reflection
    model, applied to data from the July 2012 NGC~1365 {\it NuSTAR+XMM-Newton}
    observation.  The dashed lines represent the $99\%$ (upper) and $90\%$
    (lower) confidence intervals.  Figure is from Risaliti \etal 2013, \Nat,
    494, 449.  Reprinted by permission from Macmillan Publishers, Ltd.}}
\label{fig:n1365_spin}
\end{figure}

Just as simulations for MCG--6-30-15 predicted
that the addition of {\it NuSTAR} data to that from {\it XMM-Newton} would enable the
reflection and absorption-only models to be conclusively disentangled
(Figs.~\ref{fig:mcg6_refl_abs}-\ref{fig:mcg6_spin}), the early {\it NuSTAR+XMM-Newton}
observations of NGC~1365 have conclusively demonstrated this capability.
Fig.~\ref{fig:n1365_refl_abs} shows the reflection model (black line) and the
absorption-only model (red line) fit to the July 2012 {\it XMM-Newton} data (green
points) below $10 \keV$.  The models are then extrapolated up to $79 \keV$ and
the {\it NuSTAR} data (blue points) are added, {\it without refitting}.  Though
the two models fit the data equally well below $10 \keV$, Note
the clear divergence of the two models above this energy, the striking
agreement between the {\it NuSTAR} data and the reflection model, and the clear
disagreement between the {\it NuSTAR} data and the absorption-only model.  The
data overwhelmingly support the presence of inner disk reflection signatures in
the data, in addition to both cold and warm absorption.  Applying a {\tt
  relconv(reflionx)} model to the spectra, a spin constraint of
$a=0.97^{+0.01}_{-0.04}$ is obtained, as quoted in Table~\ref{tab:table2}.  The
high S/N and broad-band spectral coverage of these data make this the most
statistically accurate, precise spin constraint achieved to date.
The robustness of this spin measurement is best appreciated through an examination
of the change in statistical goodness-of-fit with spin value, as shown in
Fig.~\ref{fig:n1365_spin}.  



\acknowledgments{LB thanks Chris Reynolds, Martin Elvis, Mike Nowak, Jon Miller,
Andy Fabian, Rubens Reis, Guido Risaliti, Dom Walton, Emanuele
Nardini, Giorgio Matt, Daniel Stern and Fiona Harrison for useful
discussions and collaborations that have contributed to many of the spin
measurements presented in this work.}

\bigskip \bigskip

\centerline{{\large REFERENCES}}
\bigskip

\noindent
{\bf TABLE 2:}

\noindent
$^a$BR06, Miniutti \etal (2007)\\ 
$^b$Schmoll \etal (2009), P12, L12, W13\\
$^c$Miniutti \etal (2009a), P12\\
$^d$Zoghbi \etal (2010), de La Calle P{\'e}rez \etal (2010),
    W13\\
$^e$Gallo \etal (2005, 2010)\\
$^f$P12, W13\\
$^g$B11, P11 \\
$^h$P12, Nardini \etal 2011 \\
$^i$Lohfink \etal 2013 \\ 
$^j$W13\\
$^k$Risaliti \etal 2009b, 2013; Brenneman \etal 2013\\
$^l$Tan \etal 2012\\
$^m$P12

\bigskip

\noindent
{\bf FULL TEXT:}

\noindent 
Agol, E. \& Krolik, J.H.: 2000, \ApJ, 528, 161.\\
Antonucci, R.: 1993, \ARAA, 31, 473. \\
Arnaud, K. \etal: 1985, \MN, 217, 105. \\
Barcons, X. \etal: 2012, arXiv:1207.2745.\\
Bardeen, J., Press, W. \& Teukolsky, S.: 1972, \ApJ, 178, 347.\\
Beckwith, K. \& Done, C.: 2005, \MN, 359, 1217.\\
Bennert, V. \etal: 2011, \ApJ, 726, 59.\\
Berti, E. \& Volonteri, M.: 2008, \ApJ, 684, 822.\\
Bianchi, S. \etal: 2009, \AaA, 495, 421.\\
Blandford, R.D. \& Znajek, R.L.: 1977, \MN, 179, 133.\\
Blandford, R.D. \& McKee, C: 1982, 255, 419.\\
Brenneman, L. \etal: 2011, \ApJ, 736, 103.\\
Brenneman, L. \etal: 2012, \ApJ, 744, 13.\\
Brenneman, L. \etal: 2013, \MN, 429, 2662.\\
Brenneman, L. \& Reynolds, C.: 2006, \ApJ, 652, 1028.\\
Brenneman, L. \& Reynolds, C.: 2009, \ApJ, 702, 1367.\\
Broderick, A. \etal: 2011, \ApJ, 735, 57.\\
Chayer, P., Fontaine, G. \& Wesemael, F.: 1995, \ApJS, 99, 189.\\
Chiang, C.Y. \& Fabian, A.: 2011, \MN, 414, 2345.\\
Collier, S. \etal: 2001, \ApJ, 561, 146.\\
Coppi, P.: 1999, \ASPC, 161, 375.\\
Cowperthwaite, P. \& Reynolds, C.: 2012, \ApJ, 752L, 21.\\
Crummy, J. \etal: 2006, \MN, 365, 1067.\\
Czerny, B. \etal: 2001, \MN, 325, 865.\\
Daly, R.: 2011, \MN, 414, 1253.\\
Dauser, T. \etal: 2010, \MN, 409, 1534.\\
Dauser, T. \etal: 2013, accepted by \MN, arXiv:1301.4922.\\
Davis, S. \etal: 2006, \ApJ, 647, 525.\\
de le Calle P\'{e}rez \etal: 2010, \AaA, 524, 50.\\
de Marco, B. \etal: 2011, \MN, 417, L98.\\
Di Matteo, T.: 2001, \AIPC, 599, 83.\\
Doeleman, S. \etal: 2008, \Nat, 455, 78.\\
Done, C. \etal: 2012, \MN, 420, 1848.\\
Dov\v{c}iak, M., Karas, V. \& Yaqoob, T.: 2004, \ApJS, 153, 205.\\ 
Elvis, M.: 2012, arXiv:1201.3520.\\
Emmanoulopoulos, D. \etal: 2011, \MN, 415, 1895.\\
Fabian, A.C. \etal: 1989, \MN, 238, 729.\\
Fabian, A.C. \etal: 2002, \MN, 335L, 1.\\
Fabian, A.C. \etal: 2005, \MN, 361, 795.\\
Fabian, A.C.: 2012, \ARAA, 50, 455.\\
Fabian, A.C. \etal: 2012b, \MN,  419, 116.\\
Fender, R., Gallo, E. \& Russell, D.: 2010, \MN, 406, 1425.\\
Ferland, G.J. \etal: 2013, \RMdAA, 49, 1.\\
Feroci, M. \etal: 2012, \ExA, 34, 415.\\
Ferrarese, L. \& Merritt, D.: 2000, \ApJ, 539L, 9.\\
Fragile, C.P. \& Anninos, P.: 2005, \ApJ, 623, 347.\\
Gallo, L.C. \etal: 2005, \MN, 363, 64.\\
Gallo, L.C. \etal: 2011, \MN, 411, 607.\\
Garcia, J. \& Kallman, T.: 2010, \ApJ, 718, 695.\\
Garcia, M. \etal: 2011, \SPIE, 8147E, 55.\\
Garofalo, D.; 2009, \ApJ, 699, 400.\\
Genzel, R. \etal: 2000, \MN, 317, 348.\\
George, I. \& Fabian, A.: 1991, \MN, 249, 352.\\
Ghez, A. \etal: 2000, \Nat, 407, 349.\\
Gierlinski, M. \etal: 2008, \Nat, 455, 369.\\
Gliozzi, M. \etal: 2010, \ApJ, 717, 1243.\\ 
Grupe, D. \etal: 2004, \AJ, 127, 156.\\
Guainazzi, M. \etal: 2006, \AN, 327, 1032.\\
G\"{u}ltekin, K. \etal: 2009, \ApJ, 698, 198.
Halpern, J.: 1984, \ApJ, 281, 90.\\
Harrison, F. \etal: 2013, accepted by \ApJ, arXiv:1301.7307.\\
Hawking, S.W.: 1974, \Nat, 248, 30.\\
Kallman, T. \& Bautista, M.: 2001, \ApJS, 133, 221.\\
Kara, E. \etal: 2013, \MN, 430, 1408.\\
Kataoka, \etal: 2007, \PASJ, 59, 279.\\
Kerr, R.P.: 1963, \PhRvL, 11, 237.\\
Krolik, J.: 1992, \AIPC, 254, 473.\\
Krongold, Y, \etal: 2005, \AIPC, 774, 325.\\
Laor, A.: 1991, \ApJ, 376, 90.\\
Lee, J.C. \etal: 2001, \ApJ, 554L, 13.\\
Lee, J.C.: 2010, \SSR, 157, 93.\\
Lohfink, A. \etal: 2012, \ApJ, 758, 67.\\
Lohfink, A. \etal: 2013a, arXiv:1301.4997.\\
Lohfink, A. \etal: 2013b, submitted to \ApJ.\\
Maiolino, R. \etal: 2010, \AaA, 517A, 47.\\
Malizia, A. \etal: 2008, \MN, 389, 1360. \\
Markoff, S., Nowak, M. \& Wilms, J.: 2005, \ApJ, 635, 1203.\\
Marscher, A.P. \etal: 2002, \Nat, 417, 625.\\
Matt, G. \etal: 1992, \AaA, 257, 63.\\
McHardy, I., Papadakis, I. \& Uttley, P.: 1999, \NuPhS, 69, 509.\\
McHardy, I. \etal: 2005, \MN, 359, 1469.\\
Miller, J.: 2007, \ARAA, 45, 441.\\
Miller, L., Turner, T. \& Reeves, J.: 2008, \AaA, 483, 437.\\
Miller, L., Turner, T. \& Reeves, J.: 2009, \MN, 399L, 69.\\
Miller, M. \& Colbert, E.: 2004, \IJMP, 13, 1.\\
Miniutti, G. \& Fabian, A.: 2004, \MN, 349, 1435.\\
Miniutti, G. \etal: 2007, \PASJ, 59S, 315.\\
Miniutti, G. \etal: 2009a, \MN, 398, 255.\\
Miniutti, G. \etal: 2009b, \MN, 401, 1315.\\
Murphy, K. \& Yaqoob, T.: 2009, \MN, 397, 1549.\\
Nandra, K. \etal: 2007, \MN, 382, 194.\\
Noble, S.C. \etal: 2011, \ApJ, 743, 115.\\
Narayan, R. \& McClintock, J.: 2012, \MN, 419L, 69.\\
Nardini, E. \etal: 2011, \MN, 410, 1251.\\
Patrick, A. \etal: 2012, \MN, 411, 2353.\\
Patrick, A. \etal: 2011, \MN, 416, 2725.\\
Penna, R.F. \etal: 2010, \MN, 408, 752.\\
Peterson, B. \etal: 2004, \ApJ, 613, 682. \\
P\'{e}tri, J.: 2008, \APSS, 318, 181.\\
Poutanen J., \& Svensson  R.: 1996, \ApJ, 470, 249.\\
Reis, R. \etal: 2012, \ApJ, 745, 93.\\
Remillard, R. \& McClintock, J.: 2006, \ARAA, 44, 49.\\
Reynolds, C.S.: 2013, arXiv:1302.3260.\\
Reynolds, C.S.: 1997, \MN, 286, 513.\\
Reynolds, C.S. \etal: 2012, \ApJ, 755, 88.\\
Reynolds, C.S. \& Fabian, A.: 2008, \ApJ, 679, 1181.\\
Reynolds, C.S. \& Nowak, M.: 2003, \PR, 377, 389.\\
Risaliti, G. \etal: 2005b, \ApJ, 630L, 129.\\
Risaliti, G. \etal: 2005a, \ApJ, 623L, 93.\\
Risaliti, G. \etal: 2009b, \ApJ, 696, 160.\\
Risaliti, G. \etal: 2009a, \MN, 393L, 1.\\
Risaliti, G. \etal: 2013, \Nat, 494, 449.\\
Romano, P. \etal: 2004, \ApJ, 602, 635.\\
Ross, R. \& Fabian, A.: 2005, \MN, 358, 211.\\
Schmoll, S. \etal: 2009, \ApJ, 703, 2171.\\
Schnittman, J. \& Krolik, J.: 2009, \ApJ, 701, 1175.\\
Seaton, M.: 1996, \APSS, 237, 107.\\
Shakura, N. \& Sunyaev, R: 1973, \AaA, 24, 337.\\
Silvestro, G.: 1974, \AaA, 36, 41.\\
Steiner, J. \etal: 2012, \ApJ, 745, 136.\\
Strohmayer, T.: 2001, \ApJ, 552L, 49.\\
Svoboda, J. \etal: 2010, \AIPC, 1248, 515.\\
Takahashi, T. \etal: 2010, \SPIE, 7732E, 27.\\
Tan, Y. \etal: 2012, \ApJ, 747, L11.\\
Tanaka, Y. \etal: 1995, \Nat, 375, 659.\\
Tchekhovskoy, A. \& McKinney, J.; 2012, \MN, 423L, 55.\\
Thorne, K.: 1974, \ApJ, 191, 507.\\
Titarchuk, L.: 1994, \ApJ, 434, 313.\\
Tombesi, F. \etal: 2010, \AaA, 521A, 57.\\
Tomsick, J. \etal: 2009, arXiv:0902.4238.\\
Turner, T.J. \etal: 2002, \ApJ, 568, 120.\\
Urry, C. \& Padovani, P,: 1995, \PASP, 107, 803.\\
Uttley, P., McHardy, I. \& Vaughan, S.: 2005, \MN, 359, 345.\\
Vasudevan, R.V. \etal: 2010, \MN, 402, 1081.\\
Volonteri, M. \etal: 2005, \ApJ, 620, 69. \\
Wassermann, D. \etal: 2010, \AaA, 524A, 9.\\
Walton, D.J., Reis, R.C. \& Fabian, A.C.: 2010, \MN, 408, 601.\\
Walton, D.J. \etal: 2013, \MN, 428, 2901.\\
Wang, J. \etal: 2009, \ApJ, 694, 718.\\
Watson, W. \& Wallin, B.: 1994, 432L, 35.\\
White, N. \etal: 2010, \AIPC, 1248, 561.\\
Wilkins, D. \& Fabian, A.: 2010, \MN, 414, 1269.\\
Wilms, J., Allen, A. \& McCray, R.: 2000, \ApJ, 542, 914.\\
Woo, J.H. \& Urry, C.M.: 2002, \ApJ, 579, 530.\\
Yaqoob, T. \& Padmanabhan, U.: 2004, \ApJ, 604, 63.\\
Zhou, X.-L. \& Wang, J.-M.: 2005, \ApJ, 618, L83.\\
Zoghbi, A. \etal: 2010, \MN, 401, 2419.\\
Zoghbi, A. \& Fabian, A.C.: 2011, \MN, 418, 2642.\\

\end{document}